\documentclass[preprint,3p,onecolumn,numbers]{elsarticle}

\usepackage{amsthm}
\usepackage{amsmath}
\usepackage{verbatim}
\usepackage{hyperref}
\usepackage{float}
\usepackage{bm,bbm}
\usepackage{subcaption}
\usepackage{url}
\usepackage{pifont}
\usepackage{amssymb} 
\usepackage{xcolor}

\journal{Journal of Economic Dynamics and Control}
\date{}

\begin{document}
\begin{frontmatter}

\title{Phase transitions in debt recycling}

\author[1]{Sabrina Aufiero}
\author[2]{Preben Forer}
\author[2]{Pierpaolo Vivo}
\author[1,3,4]{Fabio Caccioli}
\author[1]{Silvia Bartolucci}

\affiliation[1]{organization={Department of Computer Science, University College London},
            addressline={66-72 Gower Street}, 
            city={London},
            postcode={WC1E 6BT}, 
            country={UK}}
\affiliation[2]{organization={Department of Mathematics, King's College London},
            addressline={Strand}, 
            city={London},
            postcode={WC2R 2LS}, 
            country={UK}}
\affiliation[3]{organization={Systemic Risk Centre, London School of Economics and Political Sciences},
            city={London},
            postcode={WC2A 2AE}, 
            country={UK}}
            
             \affiliation[4]{organization={London Mathematical Laboratory},
addressline={8 Margravine Gardens}, 
            city={London},
            postcode={WC 8RH}, 
            country={UK}}

\begin{abstract}
Debt recycling is an aggressive equity extraction strategy that potentially permits faster repayment of a mortgage. While equity progressively builds up as the mortgage is repaid monthly, mortgage holders may obtain another loan  they could use to invest on a risky asset. The wealth produced by a successful investment is then used to repay the mortgage faster. The strategy is riskier than a standard mortgage-repayment plan since fluctuations in the house market and investment's volatility may also lead to a fast default, as both the mortgage and the liquidity loan are secured against the same good. The general conditions of the mortgage holder and the outside market under which debt recycling may be recommended or discouraged have not been fully investigated. In this paper, in order to evaluate the effectiveness of traditional monthly mortgage repayment versus debt recycling strategies, we build a dynamical model of debt recycling and study the time evolution of equity and mortgage balance as a function of loan-to-value ratio, house market performance, and return of the risky investment. We find that the model has a rich behavior as a function of its main parameters, showing strongly and weakly successful phases -- where the mortgage is eventually repaid faster and slower than the standard monthly repayment strategy, respectively -- a default phase where the equity locked in the house vanishes before the mortgage is repaid, signalling a failure of the debt recycling strategy, and a permanent re-mortgaging phase -- where further investment funds from the lender are continuously secured, but the mortgage is never fully repaid. The strategy's effectiveness is found to be highly sensitive to the initial mortgage-to-equity ratio, the monthly amount of scheduled repayments, and the economic parameters at the outset. The analytical results are corroborated with numerical simulations with excellent agreement.
\end{abstract}


\end{frontmatter}

\section{Introduction}\label{sec:intro}
In the contemporary financial landscape, the issuance of mortgages and the strategies for debt repayment are core elements that support the stability and growth of both individual finances and the global economy. The decision to take on a mortgage represents one of the most significant financial commitments for individuals and families, entailing long-term implications for their economic well-being \cite{LTVdef}. Consequently, the feasibility of debt repayments emerges as a critical area of study, offering essential insights into the sustainability of financial practices and the avoidance of over-leveraging, which can lead to distressing outcomes such as defaults and foreclosures. From an economic standpoint, mortgages play a central role in the housing market, influencing not only the dynamics of supply and demand, but also the overall economic cycle. The conditions under which mortgages are issued -- including the assessment of borrowers' ability to repay -- can significantly affect the health of the financial system. Lax lending standards can lead to increased default rates, with cascading effects on the housing market, financial institutions, and the broader economy, as vividly demonstrated by the 2008 financial crisis \cite{2008crises}. The study of debt repayment feasibility and mortgage issuance is thus imperative for crafting policies and practices that promote financial stability and responsible lending. It involves analyzing various factors, including interest rates, income stability, debt-to-income ratios, and economic conditions, to assess the risk profiles of borrowers and the sustainability of debt obligations. Moreover, this analysis is crucial for the development of innovative financial products and strategies that can enhance access to housing, while managing the risks associated with borrowing. Traditional mortgage schemes, while providing a structured path towards home ownership, inherently come with a significant limitation: the equity accumulated in the property remains largely inaccessible and non-productive until the mortgage is fully repaid. This standard feature of conventional mortgages may result in a sub-optimal use of homeowners' financial resources, as the equity tied up in the property does not yield any financial benefits or returns during the loan's tenure. Essentially, while the equity builds up in time, it remains dormant and does not contribute to the mortgage holder's broader financial well-being and/or investment capacity \cite{ER1}. This issue has prompted financial experts and institutions to design and propose alternative mortgage and debt management strategies. These innovative approaches aim to address the inefficiencies of traditional models by providing mechanisms through which homeowners can leverage their accumulated equity more effectively. The goals of these strategies are multifaceted and include accelerating the repayment process, enhancing the utility of locked equity, and potentially reducing the rates of default by improving the financial flexibility and resilience of borrowers against external stresses.
\\
In this paper, we will focus primarily on debt recycling, an aggressive strategy of equity extraction currently available only to Australian mortgage holders. Debt recycling leverages home equity building up in time to finance (risky) investments, effectively converting non-deductible mortgage debt into tax-efficient investment debt. This strategy offers a stark departure from traditional and more conservative financial strategies, which typically prioritize mortgage repayment before any investment activity is undertaken.
\\
The main idea is fairly simple: while the mortgage is being repaid, the home equity (essentially, the value of the fraction of the house owned thus far) is used to back a second loan, which in turn is invested into a (risky) asset. This way, the non-deductible Principal Place of Residence (PPOR) mortgage debt is converted into tax-efficient investment debt\footnote{The key to the success of debt recycling lies in the tax-deductibility of the interest on the loan used for investment purposes. The {\em Australian Taxation Office (ATO)} \cite{ATO} states when the interest on a loan is considered tax deductible: ``\textit{[...] If you borrow money to buy shares or related investments from which you earn dividends or other assessable income, you can claim a deduction for the interest you pay. Only interest expenses you incur for an income-producing purpose are deductible}."}, with the extra income hopefully generated by the investment becoming now available to repay the home mortgage faster. Recycling borrowed money therefore converts ``unproductive'' mortgage debt into additional cash flow that enjoys tax benefits, at the same time helping ease the financial burden on the house while it is being repaid. Other forms of less risky equity release are available in various markets, for example financial strategies to allow elderly citizens to extract liquid income from their property without having to relocate \cite{ER1} or debt swap strategies \cite{RBC}, which are reviewed in the next section.  
\\
The peculiar and rather extreme feature of the Australian debt recycling scheme, though, precisely lies in the interplay between \emph{two} (or more) loans that are backed by the \emph{same} good (the Principal Place of Residence, whose full ownership is being secured). While a good investment coupled with a booming house market can lead to increased locked-in equity being available, and in turn to a faster repayment of the mortgage, a bad investment or a depreciation of one's own house due to an economic downturn may increase the financial strain on the loans' holder to the point that they are no longer able to fulfill their obligations towards the lender, with potentially severe consequences. 
\\
Let us consider the case of a home valued at \$800,000 with a remaining mortgage of \$500,000. The owner may access \$100,000 from their equity via a line of credit. The expected annual return from the \$100,000 investment would be 7\%. The owner can then use the \$7,000 annual return to make additional payments on their home mortgage. Over time, these extra payments not only reduce the principal faster, but also decrease the amount of interest they would pay over the life of the mortgage. Moreover, the interest they pay on the \$100,000 line of credit, which could be around 4\% annually, is tax-deductible because it is tied to investment purposes. This deduction can offset the taxable income, providing further financial benefit. As their mortgage balance decreases and their home equity increases due to both repayments and potential appreciation of their home’s value, they could consider drawing additional funds for further investments. They might also re-invest dividends received, compounding their returns over time.
\\
The precise conditions of the borrower and the general economic climate that may lead to one outcome or the other are not described yet, and the literature on the topic is quite scarce, due to the novelty of the scheme, which flourished only recently due to rising home equity tied to booming real estate markets \cite{Bubble}. Financial advisors and planners began advocating for more sophisticated wealth-building strategies, including debt recycling, as part of a broader approach to financial management.
\\
In this paper, we provide a first dynamical model in discrete time of the joint evolution of home equity and mortgage, assuming that the mortgage holder follows a regular repayment schedule on top of converting a fraction of the equity to-date into a risky investment, whose performance is fluctuating randomly in time. The house market is also fluctuating randomly, leading to larger or smaller equity available to the investment. The average behavior in time of the equity and the remaining mortgage can be computed analytically as a function of the main parameters. We study the first \emph{hitting time}, namely the first time at which either the remaining mortgage hits zero (signaling that the house has been repaid in full), or the equity locked in the house hits zero (signaling that the house is fully mortgaged and the debt recycling strategy has failed). In the former case, we also monitor whether the full repayment (and consequent full ownership of the house) has materialized before or after what would have naturally happened if only the traditional monthly repayment strategy were adopted.
\\
We find that the complex interplay between house market and investment performance leads to a rich phase diagram, with regions of (strong or weak) success for the debt recycling strategy sharply separated from regions where the strategy fails, or where \emph{permanent re-mortgaging} may occur. In this scenario, securing additional investment funds from the lender creates a cycle where, despite ongoing investment losses, equity consistently exceeds the mortgage. This dynamic could potentially lead to an endless cycle of re-mortgaging. Although our model accommodates this theoretical possibility, it is unlikely to materialize in real-life settings. Our work therefore paves the way for a firmer quantitative understanding of the personal and environmental circumstances that make debt recycling a viable or non-viable strategy to repay one's mortgage faster.
\\
The plan of the paper is as follows. In Section \ref{Related Works}, we briefly review the existing literature on mortgages and risks of default, and strategies for equity release. In Section \ref{sec:Model Setup}, we set up the model and discuss the relevant variables and dynamical equations. In Section \ref{summary of results} we briefly summarize our results for the phase diagram and the main findings from analytical considerations and numerical simulations. In Section \ref{subsec:average} we describe the mathematical methods we employ to lead to the analytical solution of the model. In Section \ref{sec:Results} we report an extended discussion of results, including parameter dependence and different scenarios. \textcolor{black}{In Section \ref{sec:Policy}, we outline the policy implications derived from our model.} Finally, in Section \ref{sec:concl} we offer some concluding remarks and outlook for future research. \textcolor{black}{The Appendices are devoted to technical derivations, including the exact calculation of process fluctuations, and a discussion of the robustness of conclusions drawn from the outcome of the average process}.

\section{Related Works}\label{Related Works}
In this section we analyze the main literature related to (i) modelling and data analysis of mortgage defaults and (ii) assessment of debt recycling and other equity extraction strategies.

\subsection{Models and Analyses of Mortgage Default}

Credit default models are essential tools for understanding the dynamics that influence homeowners' decisions to default on mortgages. These analyses incorporate a variety of risk factors and economic variables to predict default behaviors under different circumstances. 

The life-cycle model illustrated in \cite{hatchondo} shows that households face income and house-price risk when buying homes with mortgages. Combining recourse mortgages and loan-to-value (LTV) limits effectively reduces default rates, and increases housing demand. The effect of monthly payment size on mortgage default is studied by Fuster et al. in \cite{payment size}. They find that payment reductions substantially lower mortgage defaults, even for borrowers that are deeply underwater: for instance, cutting the required payment in half reduces the delinquency hazard by about $55\%$. 

Campbell et al. in \cite{mortgage default} solve a dynamic model of households' mortgage decisions incorporating labor income, house price, inflation, and interest rate risk. The model quantifies the effects of adjustable versus fixed mortgage rates, LTV ratios, and mortgage affordability measures on mortgage premia and default. The model highlights the fact that the default decision depends not only on the extent to which a borrower has negative home equity, but also on the extent to which borrowers are constrained by low current resources. In the model, constraints shift the threshold at which a borrower optimally decides to exercise the irreversible option to default.

The impact of rising income inequality on the overall level of debt in the system as well as the default rates was studied by K{\"o}sem in \cite{Kosem}. By developing an equilibrium-based model of the mortgage markets, they show that rising income inequality actually leads to lower amount of mortgage debt, but higher mortgage default rates. This is due to a higher share of borrowers forced into risky low-value loans.

An empirical study into the effect of personal financial health on the probability of delinquency is done by Mocetti and Viviano in \cite{Mocetti}. Using datasets from Italian tax returns office and  the Bank of Italy, they examined how banks' selection policies and income shocks affect the rate of default. In particular, they show that the trend in default rates could be explained by the tightening of selection policies post 2008, and that individual financial health shocks have an out-sized impact on default rates. An income drop of 10\% caused default rates to go up by 5\%, while job losses double the likelihood of default.

Another study empirically looking at the link between LTV ratios and the risk of defaults was done by Gonz\' alez et al. \cite{Gonzalez}. Using regression models on data from mortgage loan portfolios in Spain between 2005-2008, they found a non-linear relationship between the two. Indeed, higher LTV ratios were associated with higher default rates, with a sharp increase seen for values greater than 80\%.

The study in \cite{LTVmodel} shows that real estate properties can be sold for more than their collateral values, leading to a significant bias in the LTV ratio that understates credit risk, especially with longer mortgage terms, higher LTV ratios, and aggressive lending products. This bias suggests that many mortgages, particularly those from the housing bubble peak, were already risky at the outset, questioning current underwriting and risk control practices.

The relationship between an individual’s mortgage repayment capacity and its impact on mortgage default risk was studied by O’Toole et al. \cite{OToole}. Using data from 2004-2007 Ireland, they used a household’s debt service ratio (i.e., the portion of net income that goes towards mortgage repayments), as a measure of a household’s repayment capacity. They found that a deterioration of the debt-service ratio led to an increase in default risk, regardless of the initial debt load. Additionally, in times of crisis this is worsened due to the presence of negative equity and liquidity constraints, essentially removing the household’s buffer.

\subsection{Equity Release}
A popular home-loan investment strategy in Europe and U.S. is the so-called \textit{equity release} or \textit{reverse mortgage} strategy. Equity release, often offered and considered by individuals aged 55 or above, represents a financial strategy enabling homeowners to access their property's equity without relocating. As eligibility criteria for older homeowners has become increasingly strict for traditional mortgages, using equity release to pay off a mortgage early has emerged as a popular way for homeowners in retirement to clear their existing mortgage debt \cite{ER1, ER2, ER3}. 
This mechanism permits the extraction of funds either as a lump sum or in increments, without necessitating immediate repayments. The schemes embed a \textit{No Negative Equity Guarantee} (NNEG), guaranteeing that the amount owed to the bank cannot be higher than the value of one's house when sold.
However, the compounded interest over time can significantly inflate the owed amount, potentially impacting the estate's value for inheritors.

In \cite{equityrelease}, a quantitative model is developed to study the differences between  equity release schemes from the perspective of a retired client: the reverse mortgages are the best option for prospective clients due to the higher lump sums paid out, as well as the protection against falling house prices provided via the NNEG.

The NNEG has been found by Sharma et al. \cite{ReleaseUK} to be one of the reasons for the disparity in use patterns of equity release schemes across the UK. They found that equity release schemes should only be used in areas of high house price due to the regional fluctuations of the cost of the NNEG that are not accounted for by market providers. Ignoring these regional fluctuations outside of areas of high house price growth leads to pre-mortgage equity release schemes being unprofitable to providers. The question of profitability of equity release schemes has also been studied by Hosty et al. \cite{hosty}. They investigate whether or not the competitiveness of the equity release market has pushed the sector towards unsustainable prices. To do so, they model the pricing of such products through `average' pricing assumptions that are meant to provide a benchmark to industry members. Another way of modelling the NNEG price was developed by Lie et al. in \cite{Siu-Hang Lie}. They propose a hedging strategy for the provider to use in order to hedge the house inflation risk. Using Monte Carlo simulations they could then estimate the hedging cost which allows them to price the NNEG.
An overview and comparison of the ways of modelling the NNEG, and thus pricing equity release programs, was done by Tunaru in \cite{Tunaru}: there is no best way of modelling the NNEG itself, and many different approaches have different benefits and drawbacks. The key to making each approach work is indeed accurate modelling of house prices, as well as testing the models for a variety of different scenarios. 

A qualitative study has also been undertaken by the \href{https://www.fca.org.uk}{Financial Conduct Authority} of the United Kingdom, aiming to understand the motivations and thought process of people who purchase reverse mortgages and equity release schemes \cite{Savanta}. They found that negative outcomes for the client were driven by the following key factors: the presence of a strong time constraint, being vulnerable to outside pressure, being financially unsophisticated, not seeking professional advice, and failing to explore other options.

More specifically, within strategies of debt conversion we also recall the \textit{debt swap} strategy \cite{RBC}. The homeowner disposes of their non-registered investments and uses the proceeds from the sale to pay off the mortgage (the non-deductible debt). They subsequently re-borrow the same amount (secured by the home) and use the proceeds to purchase a non-registered portfolio of income-producing assets. At this point, since they are then directly using the borrowed money for the purpose of earning income, the interest paid on the re-borrowing may be deductible for tax purposes.

Concerning the more aggressive debt recycling strategy, there is very little literature available. Most of the available information explain the risks and benefits associated with the strategy in a qualitative way. There are seemingly no quantitative analyses of the topic, as opposed to the extensive economic models developed for reverse mortgages and the NNEG discussed earlier. 

\section{Model Setup}\label{sec:Model Setup}
In this section, we introduce a discrete-time model for the evolution of mortgage and equity values via a debt-recyclying strategy incorporating the following update mechanisms: (i) \textcolor{black}{quarterly} loan repayments, accounting for the principal and interest components of each monthly mortgage payment arranged with the lender, (ii) fluctuations of the house market that affect the property value, (iii) dynamics of equity investment, as a fraction of their equity is invested via different routes.

At time $t=0,\ldots,T$ the \textit{house value} $H_t$ is simply the sum of the \textit{equity} owned $E_t$ and the \textit{mortgage} $M_t$, 
\begin{equation}
H_t = E_t + M_t \ .
\end{equation}
We assume that the {\em usable equity} at time $t$ -- the amount that can be used to back an investment -- is a fraction of the equity owned
\begin{equation}
U_t = \ell E_t \ , 
\end{equation}
where $\ell \in [0,1]$ is the so-called \textit{loan-to-value ratio} (LTV). The LTV ratio -- calculated by dividing the amount of the loan by the appraised value or purchase price of the property -- is used by financial institutions and other types of lenders to  assess the lending risk before approving a mortgage \cite{LTVdef}. Typically, loan assessments with high ($>80\%$) LTV ratios are considered higher-risk.
In this scenario, mortgage and equity values can vary depending on the (i) house market dynamics, and (ii) the performance of the investment.
At each time step, a fraction $I_t$ of the investment $U_t$ can be either earned or lost depending on the inherent risk associated with the investment
\begin{equation}\label{eq:I0}
I_t = \sigma _t \mu U_t \ ,
\end{equation}
where the risk factor $\mu \in [0,1]$ directly correlates to the investment's risk level. If the investment is successful, the wealth gained can be used to further cover part of the mortgage, leading in turn to an increase in equity owned. A higher risk factor $\mu$ signifies a riskier investment. The sequence of random variables  ${\bf\sigma} = \{\sigma_0, \dots, \sigma_T\}$ with $\sigma_t = \pm 1$ models the investments' positive or negative trend. 
The house market fluctuates as well, leading to an increase or decrease in the house value in time, therefore directly impacting the equity owned as well. In particular, at each time step a fraction $H_tr_t$ of the house value may be either lost or gained, depending on the house market trend. The random variable $r_t$ represents the percentage price change of the house market at time $t$. 

So, at a generic time step $t$, $E_t$ and $M_t$ are assumed to evolve according to the following set of coupled equations:
\begin{align}
E_t &= E_{t-1} + \pi_t + I_{t-1} + H_{t-1} r_t \ ,\label{eq:evo1}\\
M_t &= M_{t-1} - \pi_t - I_{t-1}\ .\label{eq:evo2}
\end{align}
The meaning of the evolution equations \eqref{eq:evo1},\eqref{eq:evo2} is as follows. The value of the equity increases at each step by an amount $\pi_t$ repaid regularly against the mortgage, which in turn decreases by the same amount. The equity may also increase (decrease) if the house market is growing (shrinking), with $r_t>0$ ($r_t<0$). A positive (negative) return on the investment $I_t$ leads respectively to an increase (decrease) in equity owned that can be re-invested, and a decrease (increase) in exposure towards the lender via the mortgage term.

Via simple manipulations, we can rewrite the process as follows:
\begin{align}
E_t &= \alpha_t E_{t-1} + \pi_t + r_t M_{t-1} \ ,\\ 
M_t &=  M_{t-1} -\pi_t +\delta_t E_{t-1}\ ,
\end{align}
where
\begin{align}
\alpha_t &= 1 + \ell \mu \sigma_{t-1} + r_t\ , \\
\delta_t &= - \ell \mu \sigma_{t-1}\ .
\end{align}

In matrix form
\begin{equation}
\begin{pmatrix}
 E_t \\ M_t
\end{pmatrix}
=
\begin{pmatrix}
\alpha_t & r_t \\ \delta_t & 1
\end{pmatrix}
\begin{pmatrix}
E_{t-1} \\ M_{t-1}
\end{pmatrix}
+
\begin{pmatrix}
\pi_t \\ -\pi_t  
\end{pmatrix}\ . 
\label{eq:processmatrix}
\end{equation}

The process in Eq. \eqref{eq:processmatrix} ends at the first \emph{hitting time} $t^\star=\min(t_E,t_M)$ (unless it continues indefinitely). The time $t_E$ characterizes the first time at which $E_{t_E}=0$, signaling that the homeowner's equity (seen as value of the fraction of the house owned) has been massively depleted:  the house is now entirely mortgaged, and the owner can only rely on monthly repayments to regain ownership of the house. In this extreme scenario, the debt recycling strategy has clearly failed. The time $t_M$ characterizes instead the first time at which $M_{t_M}=0$, showing that the borrower has managed to pay off their mortgage entirely, therefore now fully owning their house. In this scenario, we deem the strategy \textit{weakly successful} if the time taken to acquire the house through debt recycling, \(t^\star\), exceeds the time it would have taken to repay the house by standard monthly repayments, and as \textit{strongly successful} if the time required to own the house with the strategy is less than it would have been without recycling.

The question is therefore how to characterize $t^\star$, the \emph{first hitting time} of the stochastic process onto the absorbing boundaries $E=0$ or $M=0$, and consequently which boundary is reached first. \textcolor{black}{First-passage problems for coupled stochastic processes are notoriously hard to tackle exactly \cite{redner}. To make the problem more tractable, we are going to focus on the \emph{average} processes $\langle E_t \rangle$ and $\langle M_t \rangle$, where the averages are taken over the realizations of the random variables involved. We will also study analytically the fluctuations around the average processes and discuss the robustness of the outcome with respect to such fluctuations.}

In the following section, we are going to define the probability distributions and parameter assumptions of our model. Later, we are going to summarize the results for the hitting time of the average processes $\langle E_t \rangle$ and $\langle M_t \rangle$. 

\subsection{Assumptions}
\begin{itemize}

    \item We assume that the initial values of both equity and mortgage, denoted as \(E_0\) and \(M_0\) respectively, are fixed. Consequently, the average values are $\langle E_0 \rangle = E_0$ and $\langle M_0 \rangle = M_0$.

    \item The return of the investment $I_t$, defined in Eq.
    \eqref{eq:I0}, depends on the random variable $\sigma_t$. We assume that $\sigma_t$ at different times are independent Bernoulli variables
    \begin{equation}
    p(\sigma) = p \delta_{\sigma, +1} + (1-p) \delta_{\sigma, -1}\ , 
    \end{equation}
    accounting for the possibility of investment gain $(+1)$ or loss $(-1)$ with respective probabilities $p$ and $1 - p$.
    Consequently, the average investment outcome is: 
    \begin{equation}\label{eq:8}
    \langle \sigma \rangle = 2p-1\ ,
    \end{equation}
    where the parameter $p \in [0,1]$ denotes the probability of earning from the investment. A higher value of $p$ implies a greater likelihood of generating wealth from the investment, which can be funneled into the mortgage repayment. 

    \item We assume that the fluctuations in the house market value, introduced in Eq. \eqref{eq:evo1} with the term $H_{t-1} r_t$, are encoded in a normal distribution $p(r)$
    \begin{equation}\label{p(r)}
    p(r) \sim \mathcal{N}(s, \phi^2)
    \end{equation}
    with average\footnote{The choice to let \(s\) vary within the range \([-4\%, +4\%]\) is based on data regarding the percentage change in the Residential Property Price Index (RPPI). The RPPI aggregates the Established House Price Index (HPI) and the Attached Dwelling Price Index (ADPI) to measure changes in residential dwelling prices across Australia's eight Greater Capital City Statistical Areas (Sydney, Melbourne, Brisbane, Adelaide, Perth, Hobart, Darwin, and Canberra). This index, published by the \href{https://www.abs.gov.au/statistics/economy/price-indexes-and-inflation/residential-property-price-indexes-eight-capital-cities/dec-2021}{Australian Bureau of Statistics} (ABS) assesses  the residential property price inflation, covering all residential properties, regardless of ownership or the occupants' tenure.} $s \in [-4\%, +4\%]$ and variance $\phi^2$. A higher \(s\) value indicates a positive market trend, leading to an increase in house value; instead, a negative \(s\) value implies poor market performance, resulting in a loss in house value.

    \item The scheduled repayment to the bank \(\pi_t\), introduced in Eq. \eqref{eq:evo1}, is assumed to be non-negative ($\pi_t$ \textcolor{black}{$\geq$} $0$) at every time step \(t\), indicating consistent payments to the bank. However, our model allows for the rare event of a few installments of the repayment being skipped (with probability $q$). Otherwise, with probability $1-q$ each installments has value $\pi^\star$.
\begin{equation}
p(\pi)=q\delta(\pi) + (1-q)\delta(\pi-\pi^\star) \  ,\label{pofpi}
\end{equation}

implying an average installment of 
\begin{equation}\label{avg pi}
\langle\pi\rangle=(1-q)\pi^\star\ .
\end{equation}
\end{itemize}

In summary, the parameters of the models are: 

\begin{itemize}
    \item $\ell \in [0, 1]$: {LTV ratio};
    \item $\mu \in [0, 1]$: risk factor of the investment;
    \item $p \in [0,1]$: probability that the investment produces a positive income (gain);
    \item $s \in [-4\%, +4\%]$:
    average of the fluctuations in the house market value;
    \item $q \ll 1$: probability of skipping an installment of the repayment to the lender;
    \item $\pi^\star$: value of the monthly installment;
    \item $E_0, M_0$: initial conditions of equity and mortgage respectively.
\end{itemize}    
Given the availability of official housing market data, which are typically reported on a quarterly basis, we define our model's time steps as quarters, equating to 3-month periods. 

\section{Summary of Results}\label{summary of results}

In this section, we are going to present a concise summary of our findings. We then provide the methodology in Section  \ref{subsec:average}, and a more detailed discussion of the results in Section \ref{sec:Results}.

In Section \ref{subsec:average}, we will find that the evolution in time of the equity $E$ and mortgage $M$ is described by a $2\times 2$ matrix. Since we are interested in the average processes $\langle E_t \rangle$ and $\langle M_t \rangle$ -- where the averages are taken over the realizations of the random variables involved -- we focus on the average of the matrix that describes the evolution. 
We find that the relevant variables to describe our processes are the combinations $\lambda_1 = s+1$ and $\lambda_2 = \ell \mu (2p-1) +1$, which correspond to the eigenvalues of the average matrix. 

To further fix the setup, we now consider $\ell=0.5$ and $\mu=0.5$, corresponding to a scenario where the LTV ratio is below the suggested maximum value of $\sim 80\%$, and the fraction earned or lost is half of the investment.

\begin{figure}[h]
    \centering
    \includegraphics[width=0.8\textwidth]{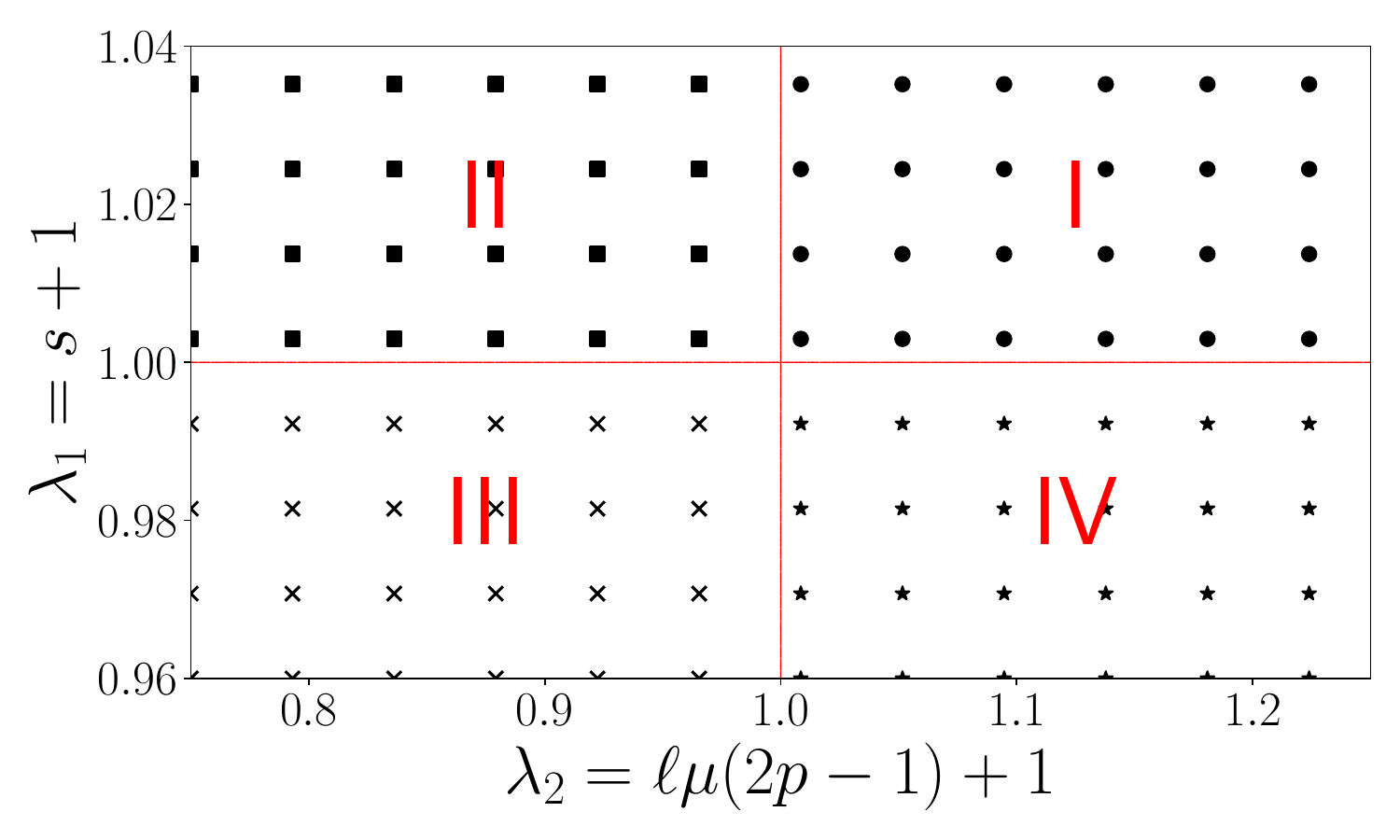}
    \caption{$\lambda$-space for $\ell=0.5$, $\mu = 0.5$. \\ Legend: \\
    \MakeUppercase{\romannumeral 1} quadrant ($\bullet$): $\lambda_{1,2} >1$ \\
    \MakeUppercase{\romannumeral 2} quadrant ($\blacksquare$): $\lambda_1 >1; 0< \lambda_2 <1$ \\
    \MakeUppercase{\romannumeral 3} quadrant ($\times$): $0< \lambda_{1,2} <1$ \\
    \MakeUppercase{\romannumeral 4} quadrant ($\bigstar$): $\lambda_2 >1; 0<\lambda_1 <1$ \\ }
    \label{fig:lambda space}
\end{figure}

In Fig. \ref{fig:lambda space}, we present a schematic summary of the space of possible values taken up by the eigenvalues. The markers indicate the four possible combinations of their magnitudes -- reflecting all potential outcomes, given that \(\lambda_1\) and \(\lambda_2\) cannot be negative, regardless of parameter choices. The red lines represent the transitions between different zones. Alterations in the parameters \(\ell\) and \(\mu\) affect only the range within which \(\lambda_2\) varies, but the transition thresholds are consistent: at \(s=0\%\), where \(\lambda_1\) exceeds $1$, and at \(p=0.5\), where \(\lambda_2\) exceeds $1$.

The four regions depicted in the $\lambda$-space in Fig. \ref{fig:lambda space} correspond to the four dynamical scenarios summarized in Table \ref{tab:dynamic outcomes}, which in turn end up in three possible outcomes of the debt recycling strategy.
The strategy is deemed \textit{successful} when the average process reaches the \(M=0\) boundary first, a result achievable with all four potential combinations of eigenvalue magnitudes. Whether the strategy is weakly successful -- if the time \(t^\star\) needed to acquire the house via debt recycling exceeds the time needed without the strategy -- or strongly successful -- if \(t^\star\) is smaller than the time needed without debt recycling -- depends on the specific scenario and the values of the parameters involved. The strategy \textit{defaults} (fails) when the process hits the \(E=0\) boundary first, an occurrence seen in the \MakeUppercase{\romannumeral 3} and \MakeUppercase{\romannumeral 4} quadrants of the \(\lambda\)-space. Lastly, the strategy may fall into a state of \textit{permanent re-mortgaging} where neither process ever hits the absorbing boundary: as the house value increases, securing additional bank funding for investment creates a virtuous cycle of equity extraction. Despite sustained investment losses, equity remains above the mortgage, a situation observed in the \MakeUppercase{\romannumeral 2} quadrant. In practice, this theoretical outcome will eventually be halted by the lender's decision to discontinue funding at some point. 
We will now discuss the possible dynamical scenarios in details.

\begin{itemize}
    \item \underline{\MakeUppercase{\romannumeral 1} quadrant ($\bullet$): $\lambda_{1,2} > 1$; Strong Success.} \\
    
        The parameters significantly favor the mortgage holder due to a positive trend in the housing market (\(s>0\)) combined with high likelihood of the investment yielding a positive return (\(p>0.5\)). The average equity displays an increasing exponential trend, whereas the average mortgage rapidly decreases, reaching the absorbing boundary at zero within a few time steps. Consequently, the strategy proves to be strongly successful.
    
    \item \underline{\MakeUppercase{\romannumeral 2} quadrant ($\blacksquare$): $\lambda_1 >1$, $0< \lambda_2 <1$; Weak Success or Permanent Re-Mortgaging.} \\
    
        The housing market shows a positive trend (\(s>0\)), this time paired with a low probability of the investment yielding a positive return (\(p<0.5\)). This configuration leads to two possible outcomes for the strategy, determined by the value of the product \(\ell\mu\): either the average equity increases while the average mortgage decreases and eventually hits zero, marking the strategy as weakly successful, or neither process ever hits the absorbing boundary, displaying either exponential growth or stabilizing at a constant value, determined by the magnitude of $s$. This dynamics happens because, in scenarios where the house keeps increasing in value (\(s>0\)), obtaining additional funds from the bank for investment initiates a cycle where, despite persistent investment losses, equity consistently exceeds the mortgage at each time step \(t\). This cycle leads to further lending, potentially resulting in a never-ending sequence of re-mortgaging. While our model theoretically allows for this scenario, this is unlikely to occur in practice, indicating the need for external mechanisms to suppress it (typically, a time cap in the lender's willingness to provide credit). 
    
    \item \underline{\MakeUppercase{\romannumeral 3} quadrant ($\times$): $0<\lambda_{1,2}<1$; Success or Default.} \\
    
        In this region, the values of the parameters should discourage the mortgage holder from undertaking a debt recycling strategy due to a negative trend in the housing market (\(s<0\)), and a low probability of the investment yielding a positive return (\(p<0.5\)). While the average mortgage consistently decreases, the average equity reaches an absolute minimum in the early stages of the process before following an increasing trend. For a given \(p\), but with increasing \(s\), the average equity curves move upwards, thereby creating the possibility of an early default if the minimum of the equity is negative, but also the possibility of a success (both strong and weak) if the minimum is positive. In the majority of scenarios, though, due to unfavorable parameters, the time required for the strategy to be successful is considerably high, suggesting that a simple monthly repayment without recycling is the better strategy.
        In Fig. \ref{fig:translation s} the equity for fixed $p=0.4$ and increasing $s$ is shown.

        \begin{figure}[h]
        \centering \includegraphics[width=0.6\textwidth]{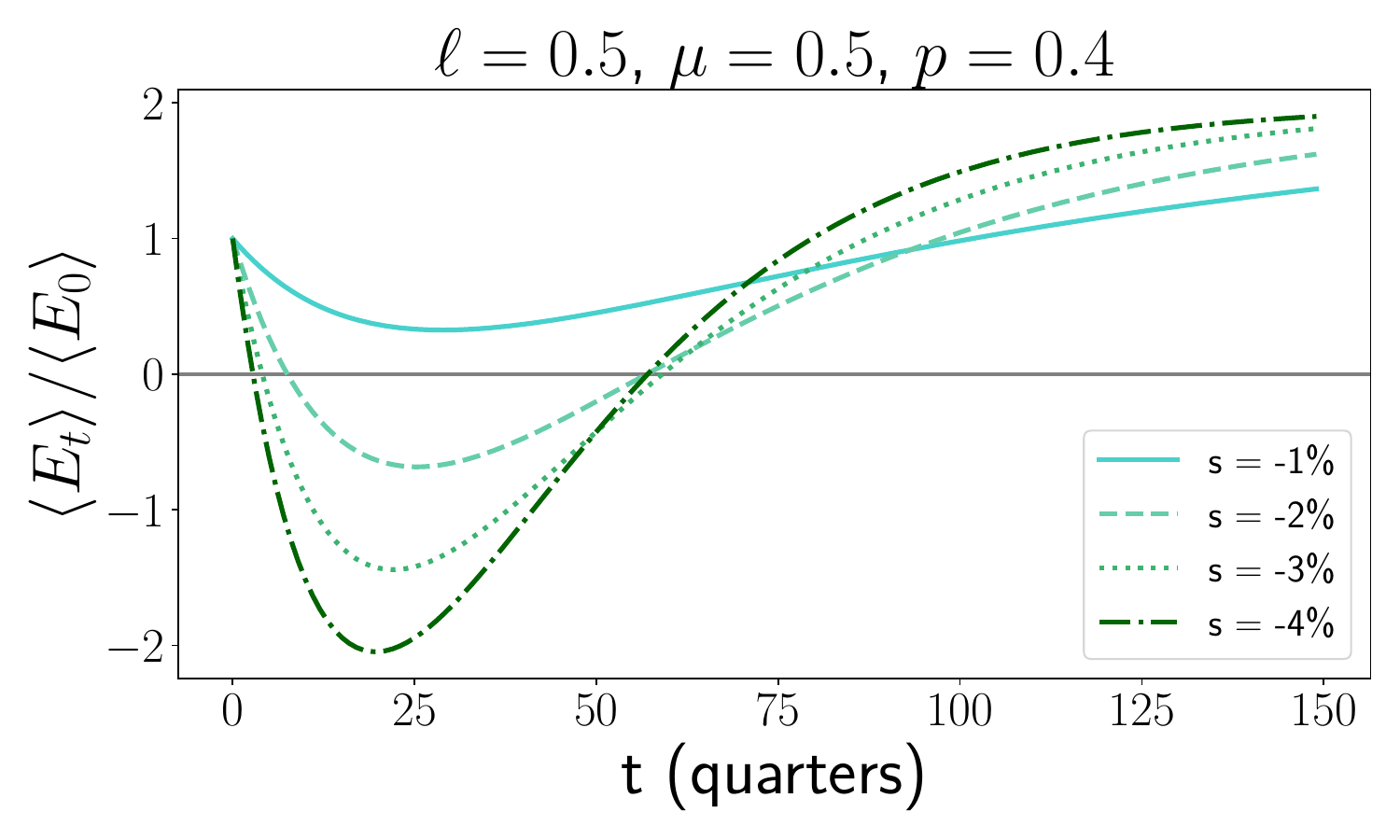}
        \caption{Average equity dynamics for increasing $s$ and fixed $p$, in the \MakeUppercase{\romannumeral 3} quadrant ($\times$) of the $\lambda$-space. }
        \label{fig:translation s}
        \end{figure}
    
    \item \underline{\MakeUppercase{\romannumeral 4} quadrant ($\bigstar$): $\lambda_2 >1$, $0<\lambda_1 <1$; Success or Default.} \\
    
        There is a negative trend in the housing market  ($s<0$), but a high probability of the investment yielding a positive return ($p > 0.5$). The outcome of the strategy can either be success (both strong and weak), with the average equity showing an increasing exponential trend and the average mortgage steadily decreasing, or default if the trends are respectively decreasing and increasing. When the strategy is successful, it is almost always strongly successful, as a \(s<0\) value implies that, with each time step, the total value of the house decreases, thereby also lowering the paid equity threshold required to achieve ownership of the house.

    \end{itemize}

In the following Table, we summarize the various outcomes of the strategy in different regions of the phase diagram, and the typical process dynamics in each case.

\begin{table}[H]
\centering
\resizebox{\columnwidth}{!}{%
\begin{tabular}{|c|c|c|}

\hline
& Strategy Outcome 
& Process Dynamics 
\\ \hline 

\begin{tabular}[c]{@{}c@{}}
    $\bullet $\\ 
    $\lambda_{1,2} >1$
\end{tabular} &
\begin{tabular}[c]{@{}c@{}}
\begin{minipage}{6cm}
    \textcolor{orange}{\\ Strong Success} \\ The average equity exhibits an increasing exponential trend; the average mortgage shows a decreasing exponential pattern, until it is eventually extinguished.
\end{minipage} 
\end{tabular} &
\begin{tabular}[c]{@{}c@{}}
    \includegraphics[width=4cm, height=2.4cm]{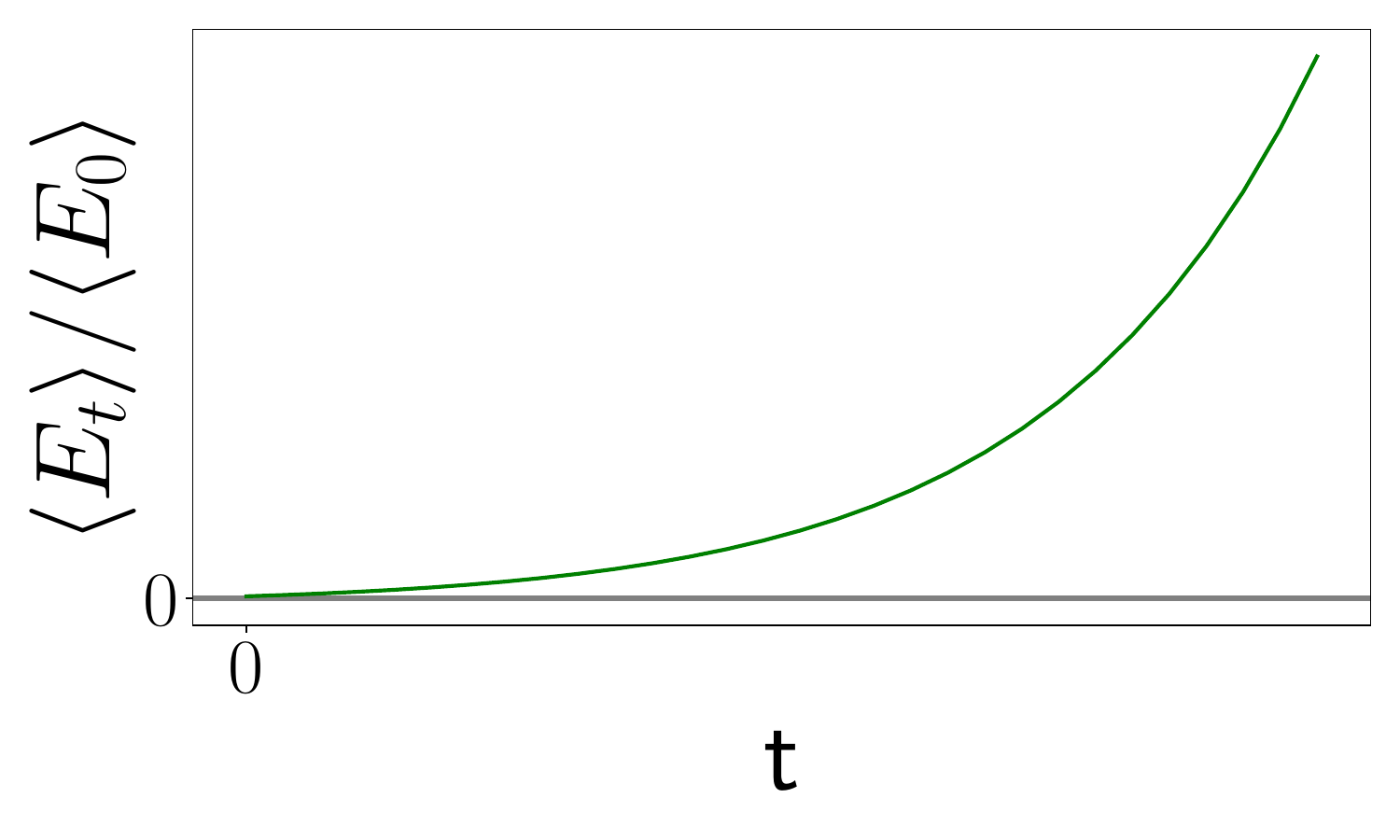} \includegraphics[width=4cm, height=2.4cm]{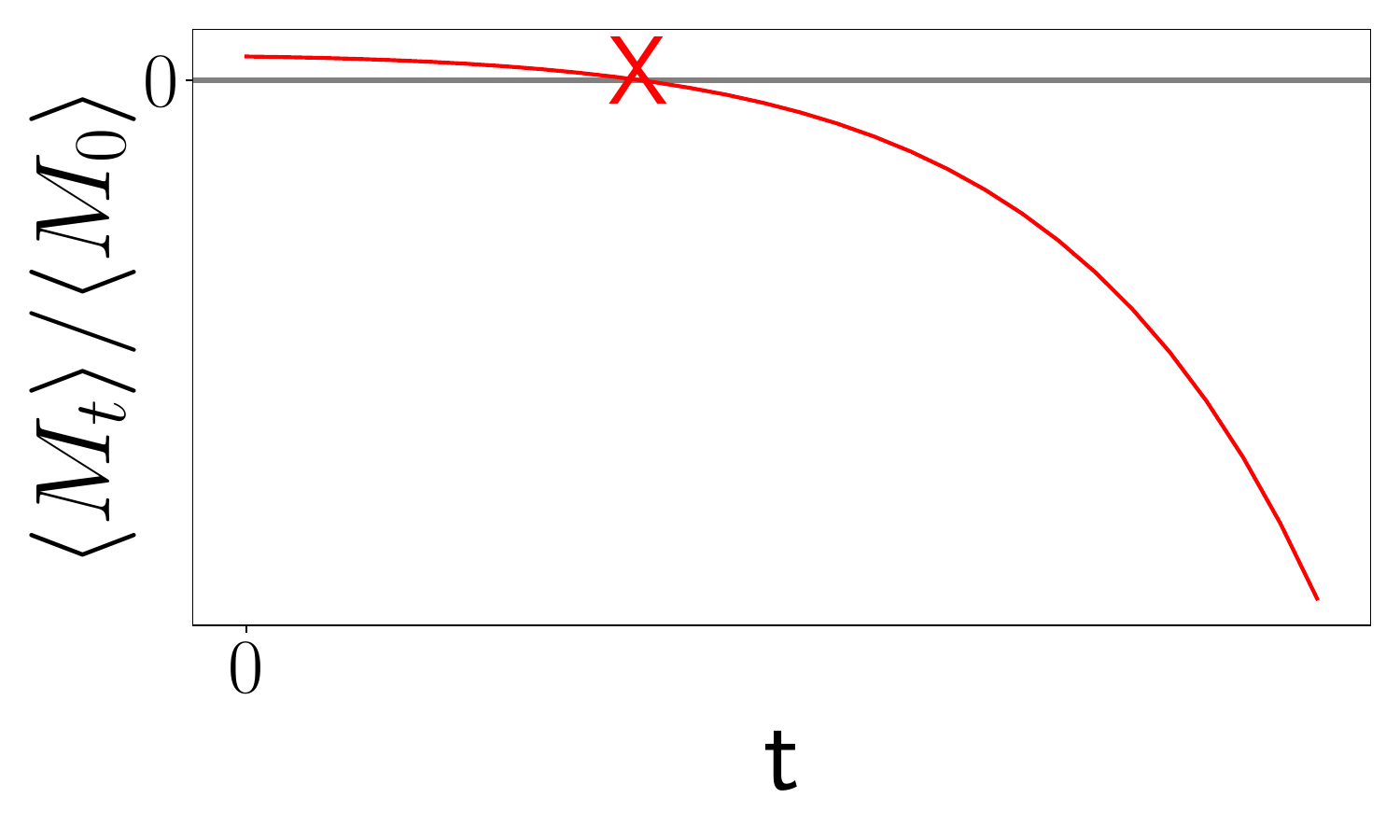}  
\end{tabular}
\\ \hline
   
\begin{tabular}[c]{@{}c@{}}
    $\blacksquare $\\
    $\lambda_1 >1$\\ 
    $0< \lambda_2 <1$
\end{tabular} &
\begin{tabular}[c]{@{}c@{}} 
\begin{minipage}{6cm}
    \textcolor{orange}{\\ Weak Success (low $\ell \mu$)} \\ The average equity exhibits an increasing trend; the average mortgage shows a decreasing pattern, until it is eventually extinguished.
\end{minipage} \\ 
\rule{6cm}{0.01mm} \\ 
\begin{minipage}{6cm}
    \textcolor{gray}{Permanent Re-Mortgaging \\ (high $\ell \mu$)} \\ Neither process ever hits the absorbing boundary: they either display increasing behavior or converge to a constant. \\
\end{minipage} 
\end{tabular} &
\begin{tabular}[c]{@{}c@{}}
    \includegraphics[width=4cm, height=2.4cm]{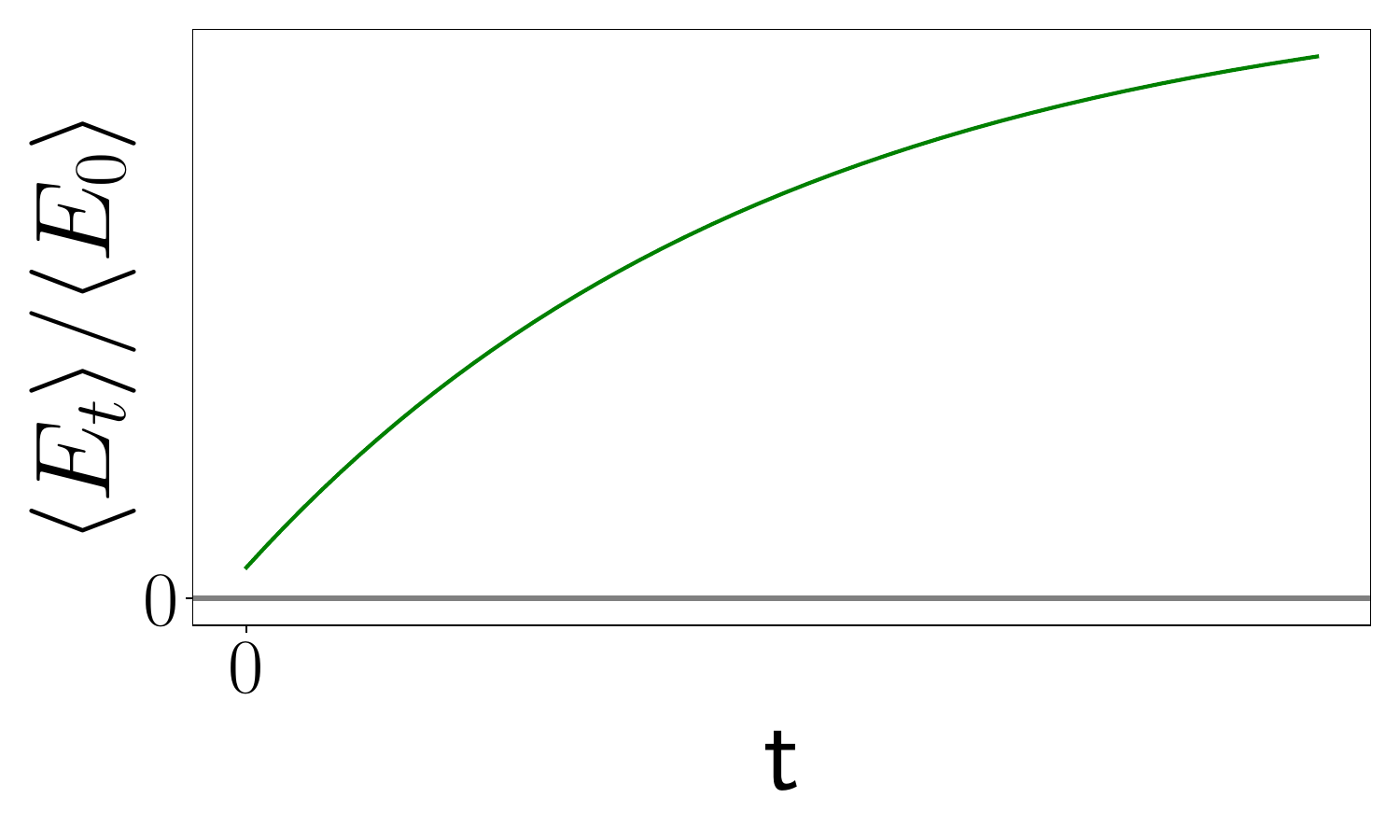}  \includegraphics[width=4cm, height=2.4cm]{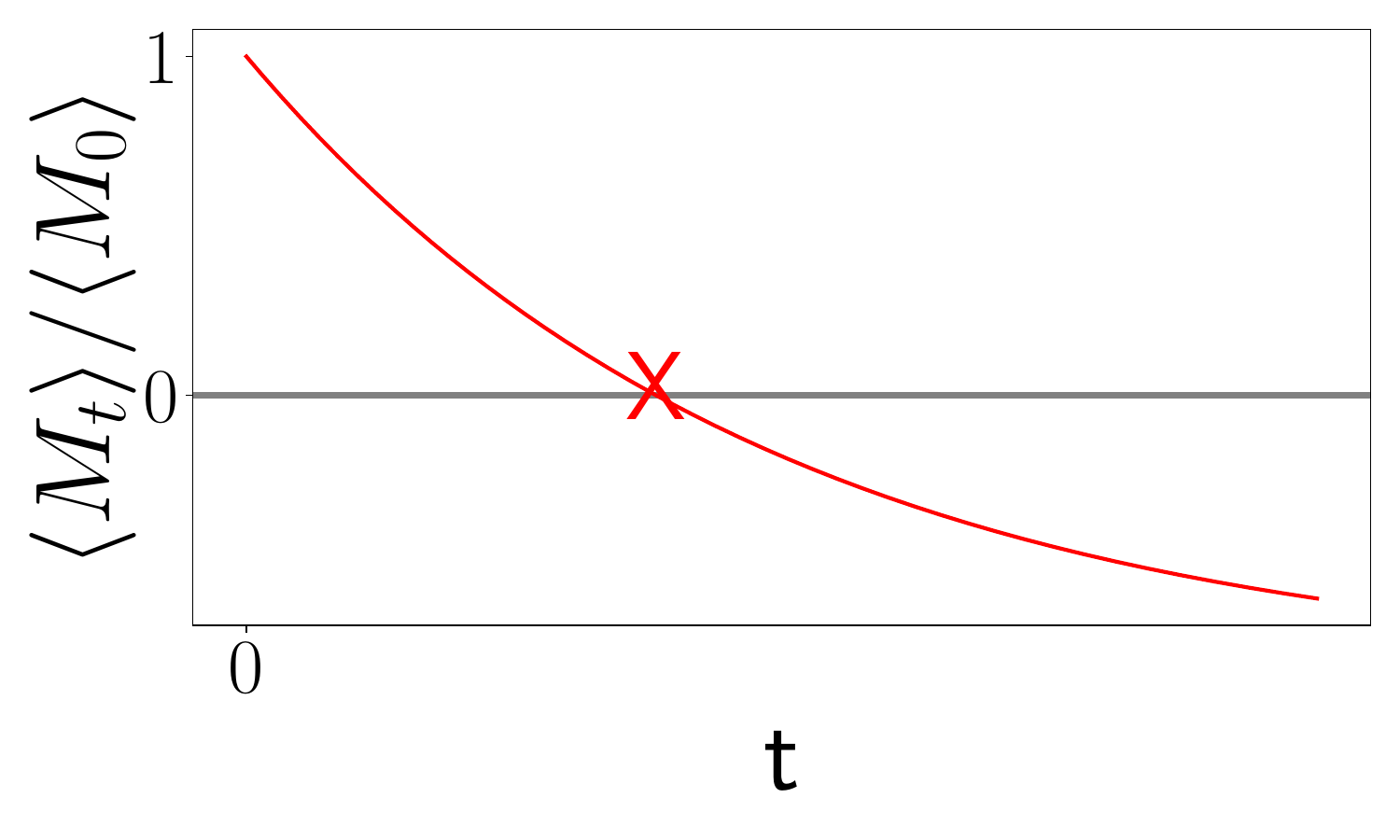} \\
    \rule{8cm}{0.01mm} \\
    \includegraphics[width=4cm, height=2.4cm]{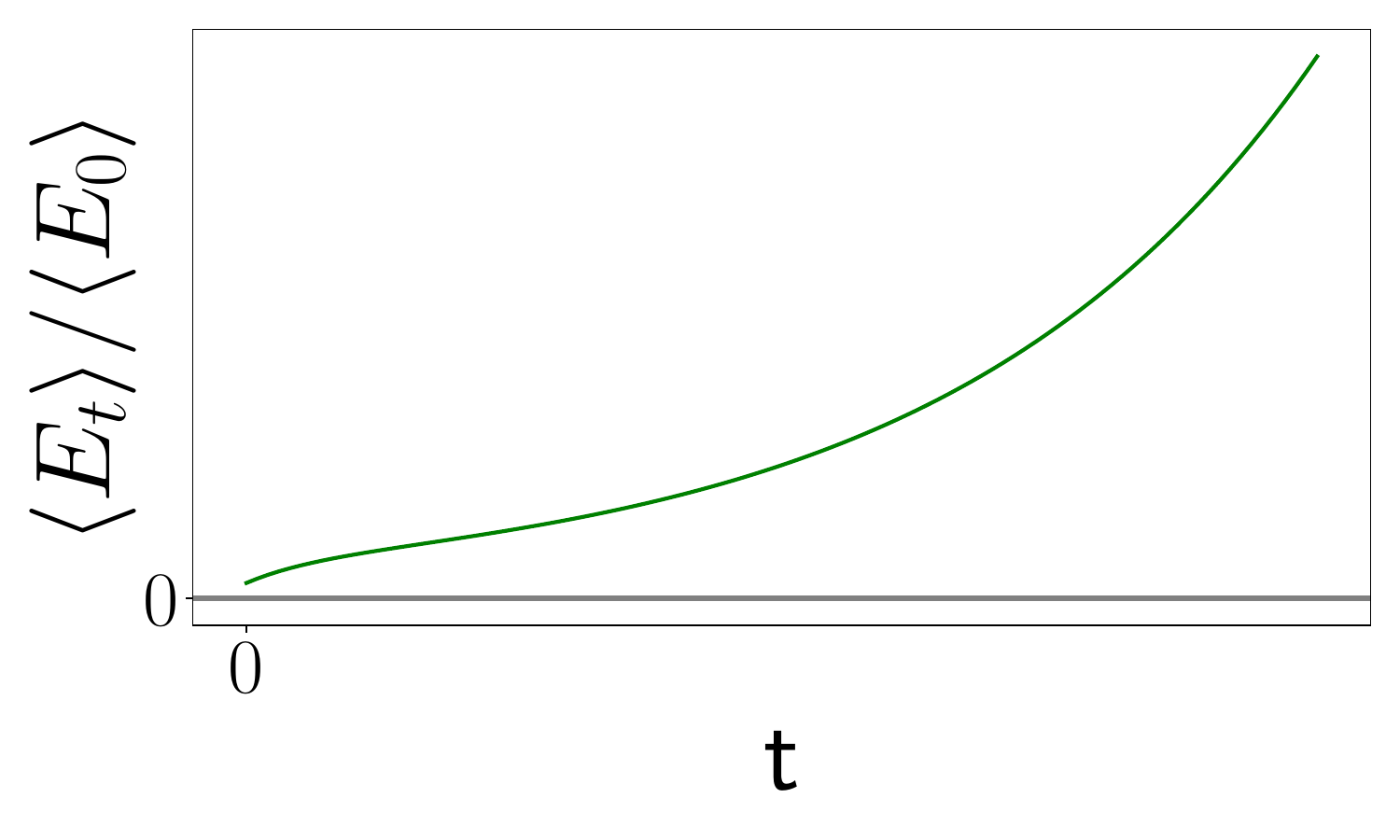} \includegraphics[width=4cm, height=2.4cm]{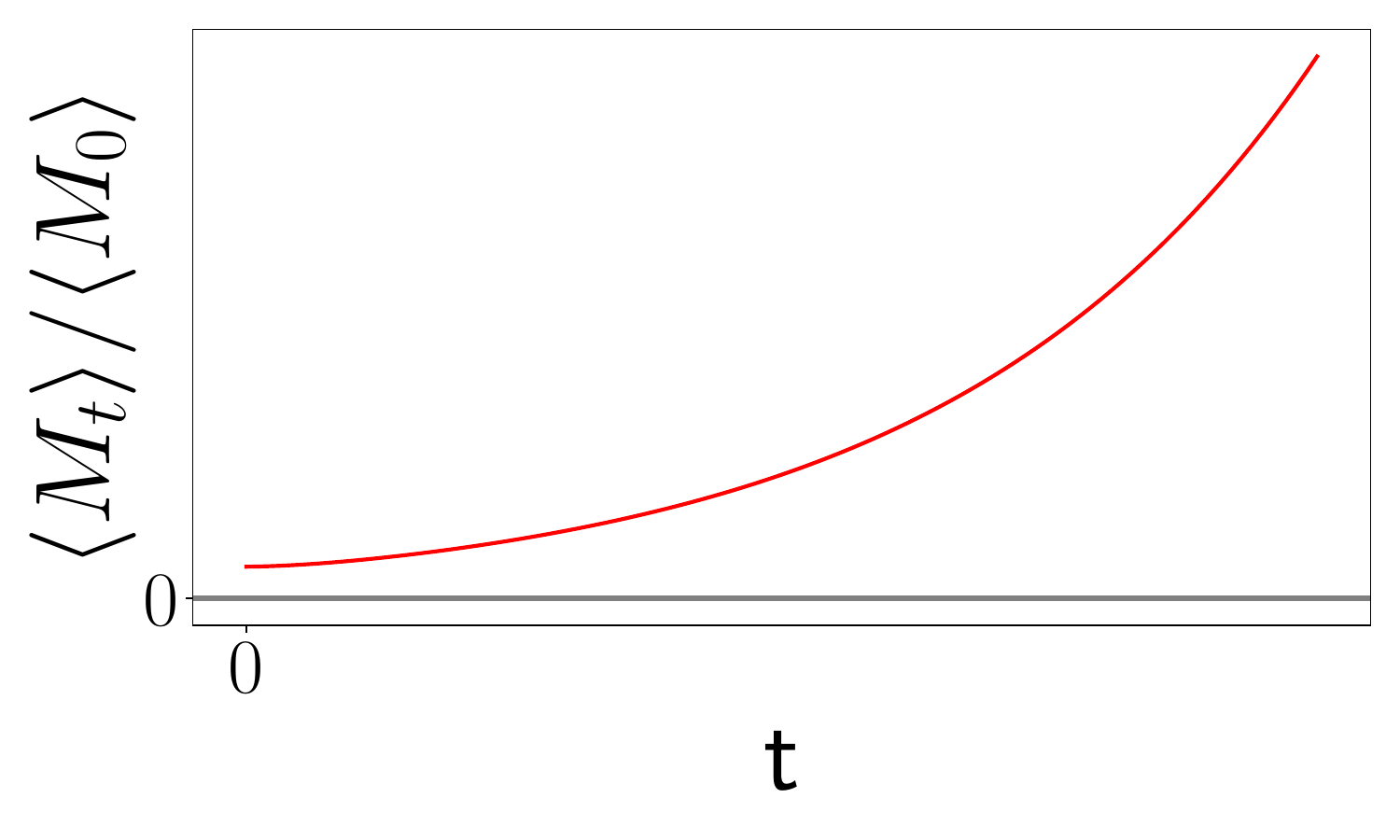} 
    \end{tabular}
\\ \hline
   
\begin{tabular}[c]{@{}c@{}}
    $\times $\\ 
    $0< \lambda_{1,2} <1$
\end{tabular} &
\begin{tabular}[c]{@{}c@{}}
\begin{minipage}{6cm}
    \textcolor{orange}{\\ Success (high $s$)} \\ The average equity displays a positive minimum in the initial phase, then follows an increasing trend; the average mortgage shows a decreasing pattern, until it is eventually extinguished.
\end{minipage} \\ 
\rule{6cm}{0.01mm} \\ 
\begin{minipage}{6cm}
    \textcolor{violet}{Default (low $s$)} \\ The average equity rapidly reaches a negative minimum in the initial phase; the average mortgage shows a decreasing pattern, but slower. The strategy defaults.\\
\end{minipage} 
\end{tabular} &
\begin{tabular}[c]{@{}c@{}}
        \includegraphics[width=4cm, height=2.4cm]{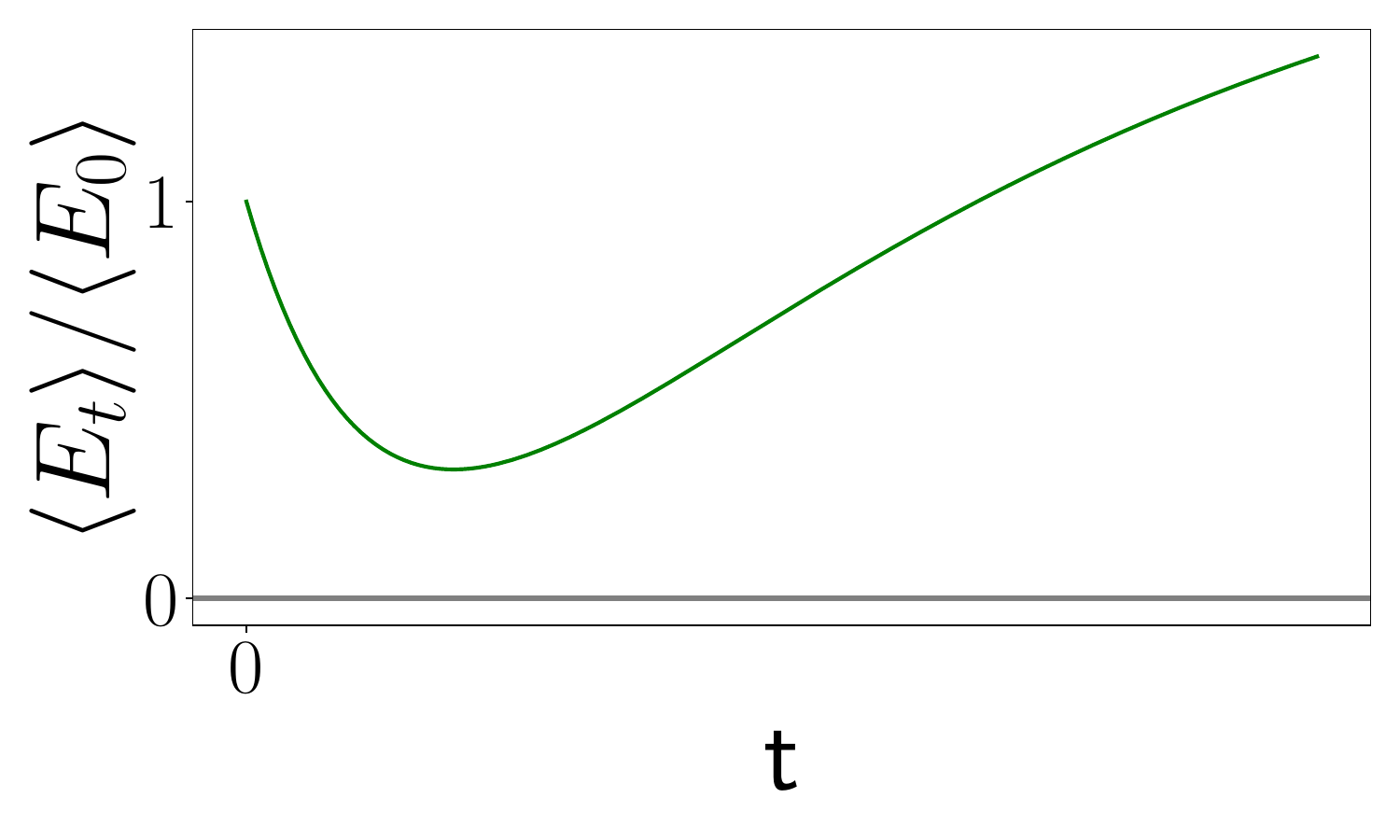} \includegraphics[width=4cm, height=2.4cm]{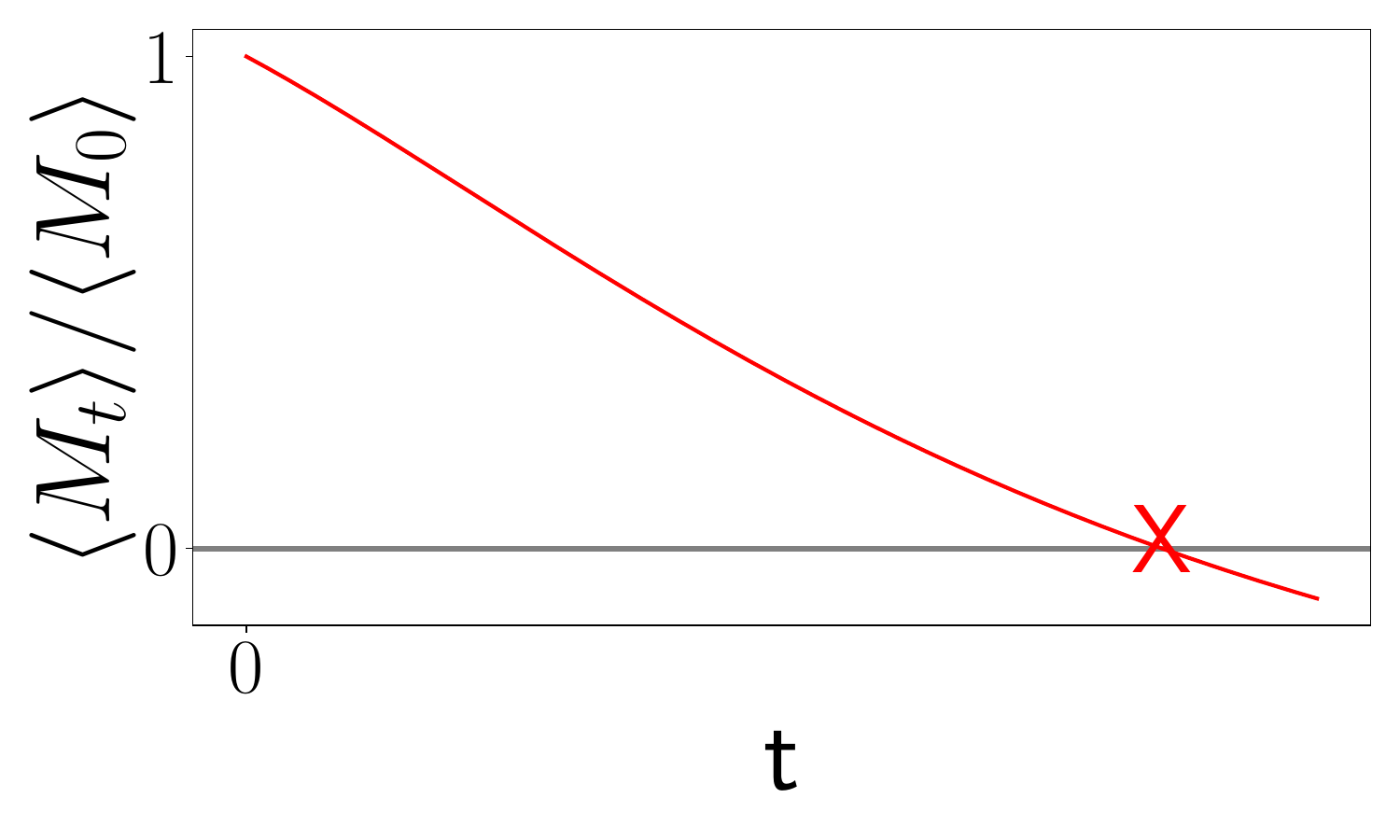} \\ \rule{8cm}{0.01mm} \\ 
        \includegraphics[width=4cm, height=2.4cm]{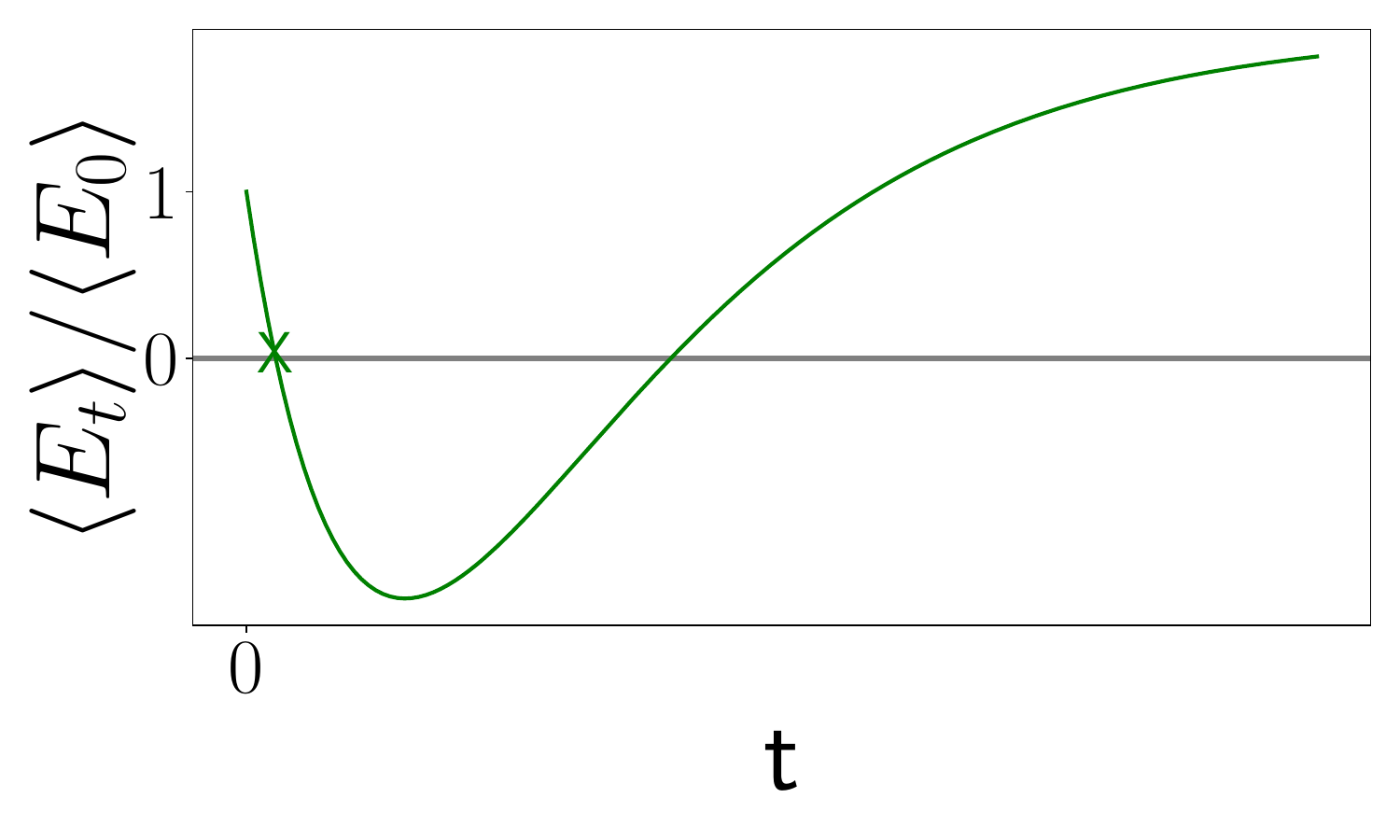} \includegraphics[width=4cm, height=2.4cm]{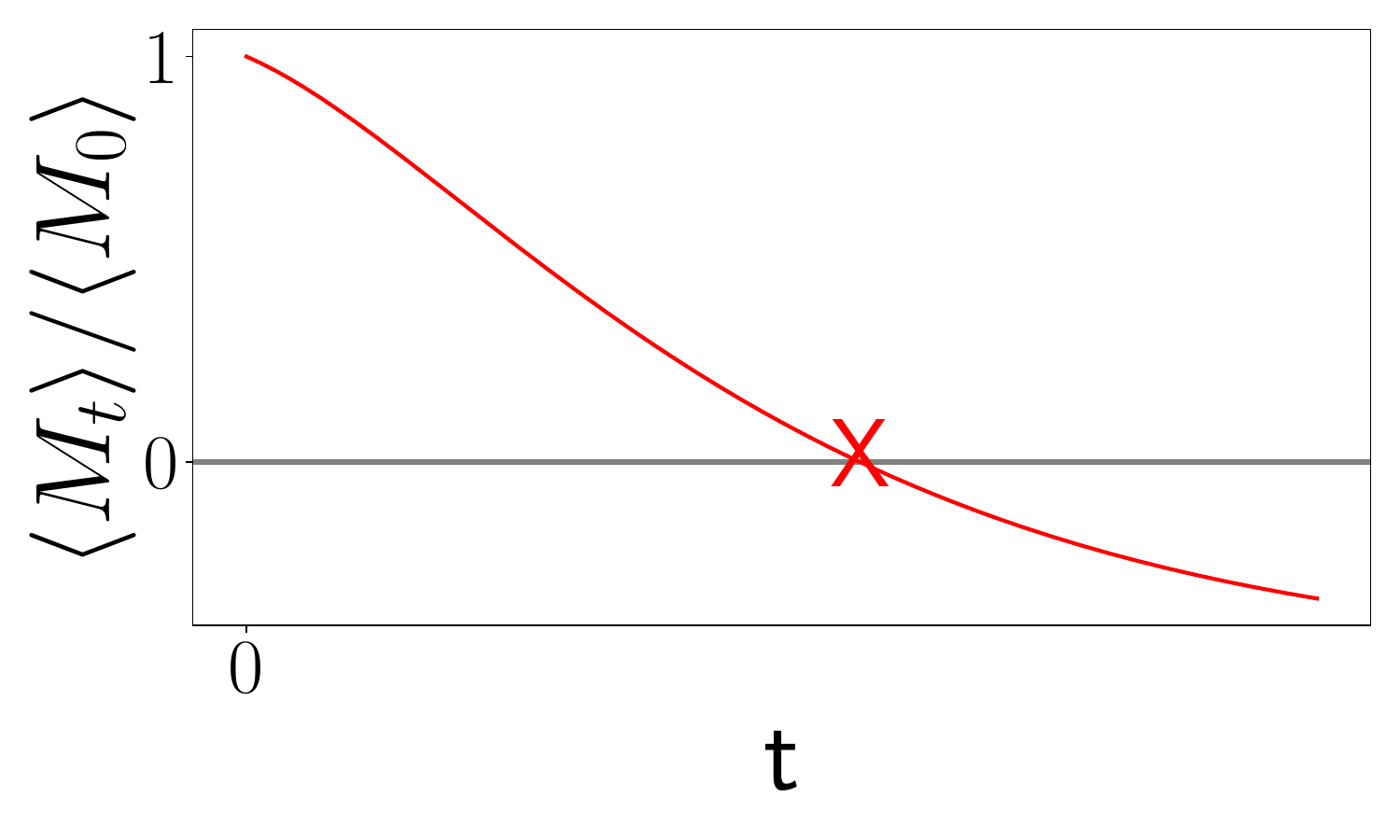}
\end{tabular}
\\ \hline
   
\begin{tabular}[c]{@{}c@{}}
$\bigstar$\\ 
$\lambda_2 >1$\\ 
$0<\lambda_1 <1$
\end{tabular} &
\begin{tabular}[c]{@{}c@{}}
\begin{minipage}{6cm}
    \textcolor{orange}{\\ Success} \\ The average equity exhibits an increasing exponential trend; the average mortgage shows a decreasing exponential pattern, until eventually it is extinguished.
\end{minipage} \\
\rule{6cm}{0.01mm} \\ 
\begin{minipage}{6cm}
    \textcolor{violet}{Default} \\ The average equity exhibits a decreasing exponential trend, hitting the absorbing boundary early on; the average mortgage shows an increasing exponential pattern. The strategy defaults.
\end{minipage}
\end{tabular} &
\begin{tabular}[c]{@{}c@{}}
    \includegraphics[width=4cm, height=2.4cm]{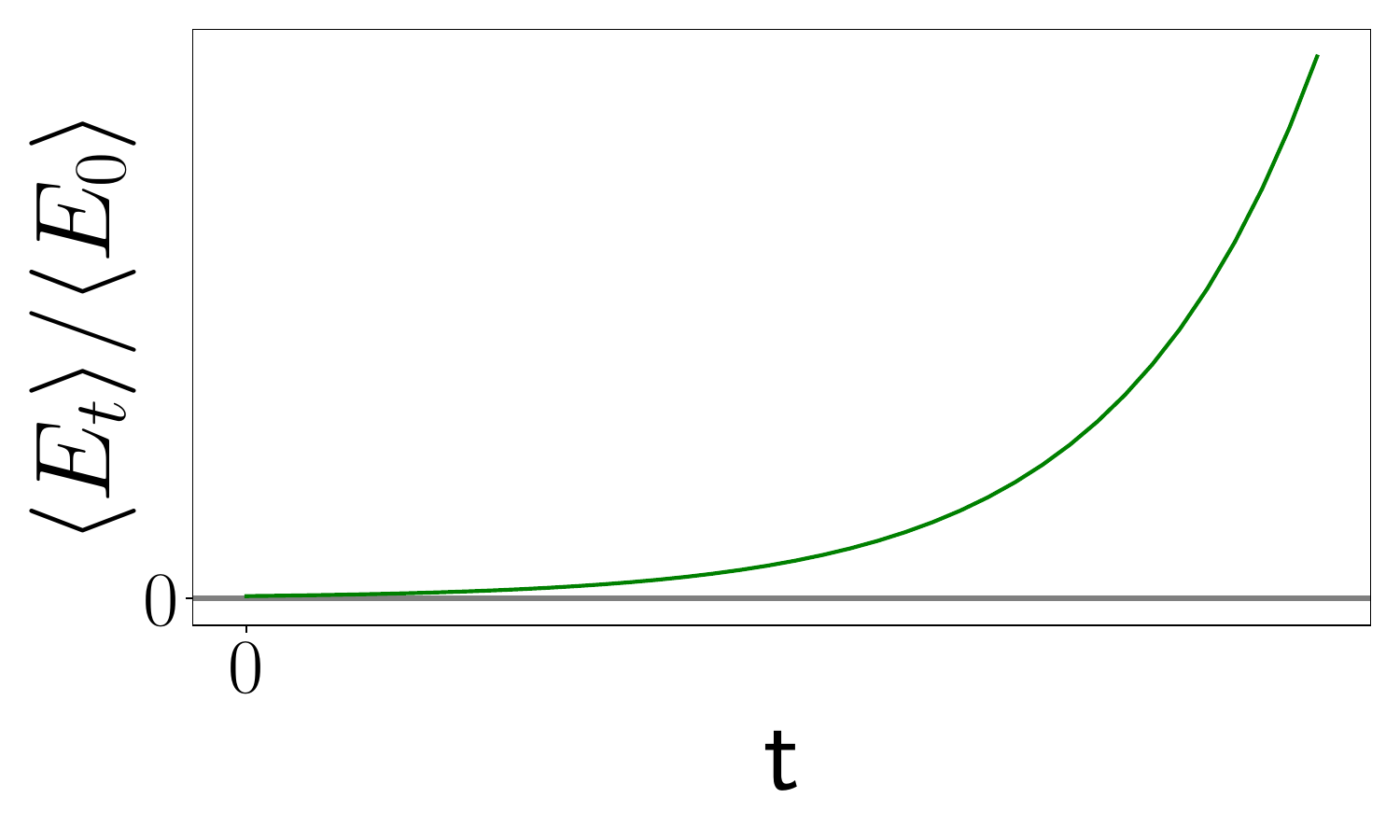} \includegraphics[width=4cm, height=2.4cm]{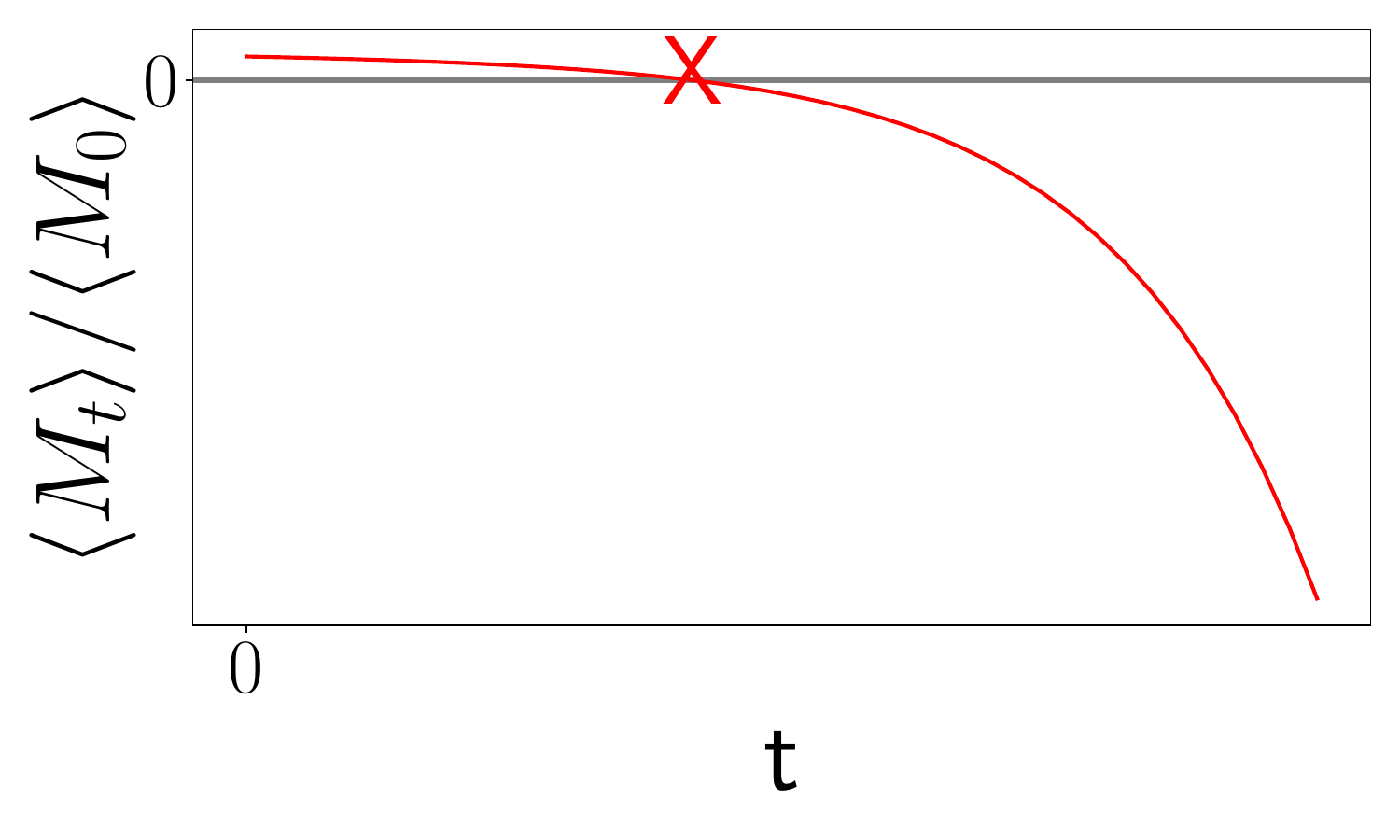} \\ \rule{8cm}{0.01mm} \\
    \includegraphics[width=4cm, height=2.4cm]{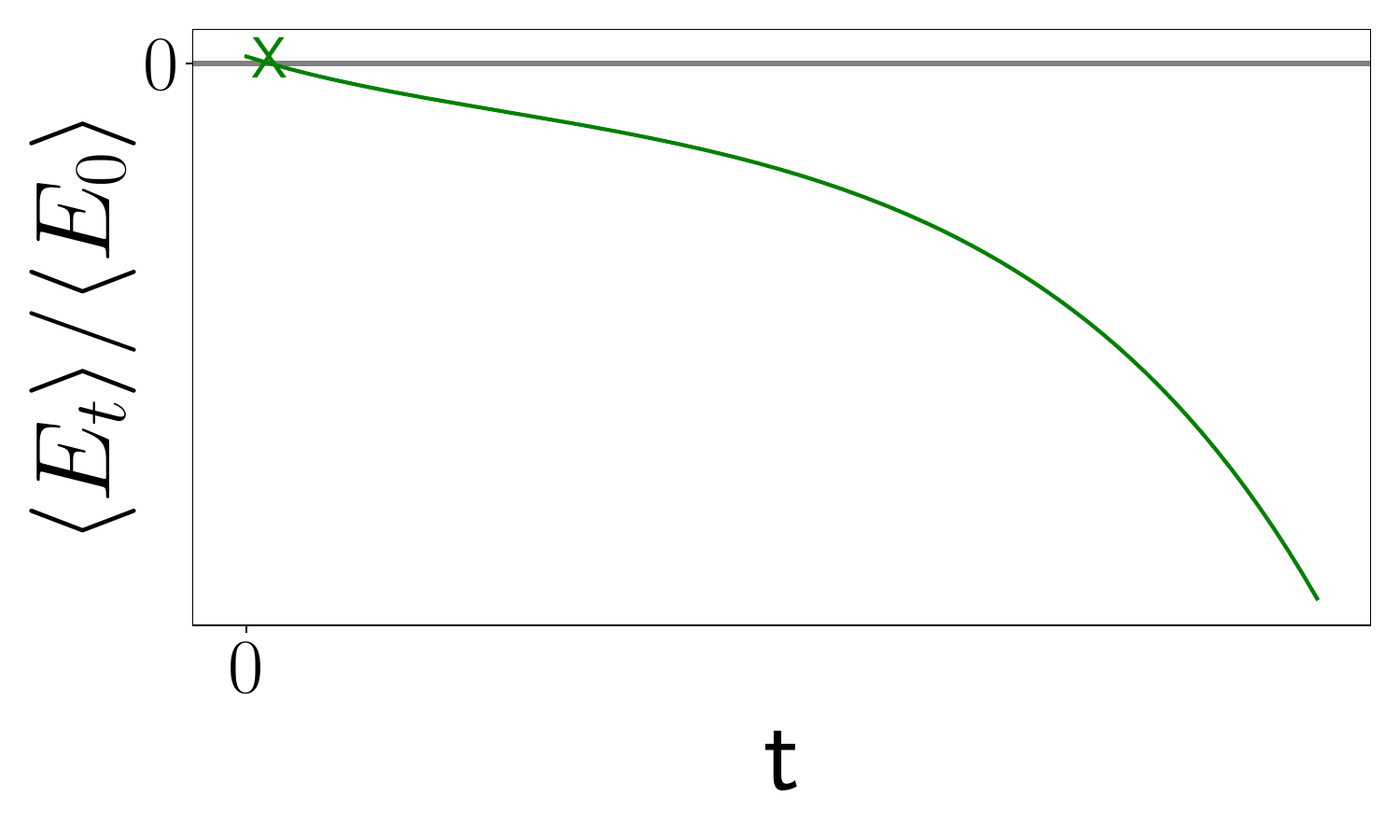}\includegraphics[width=4cm, height=2.4cm]{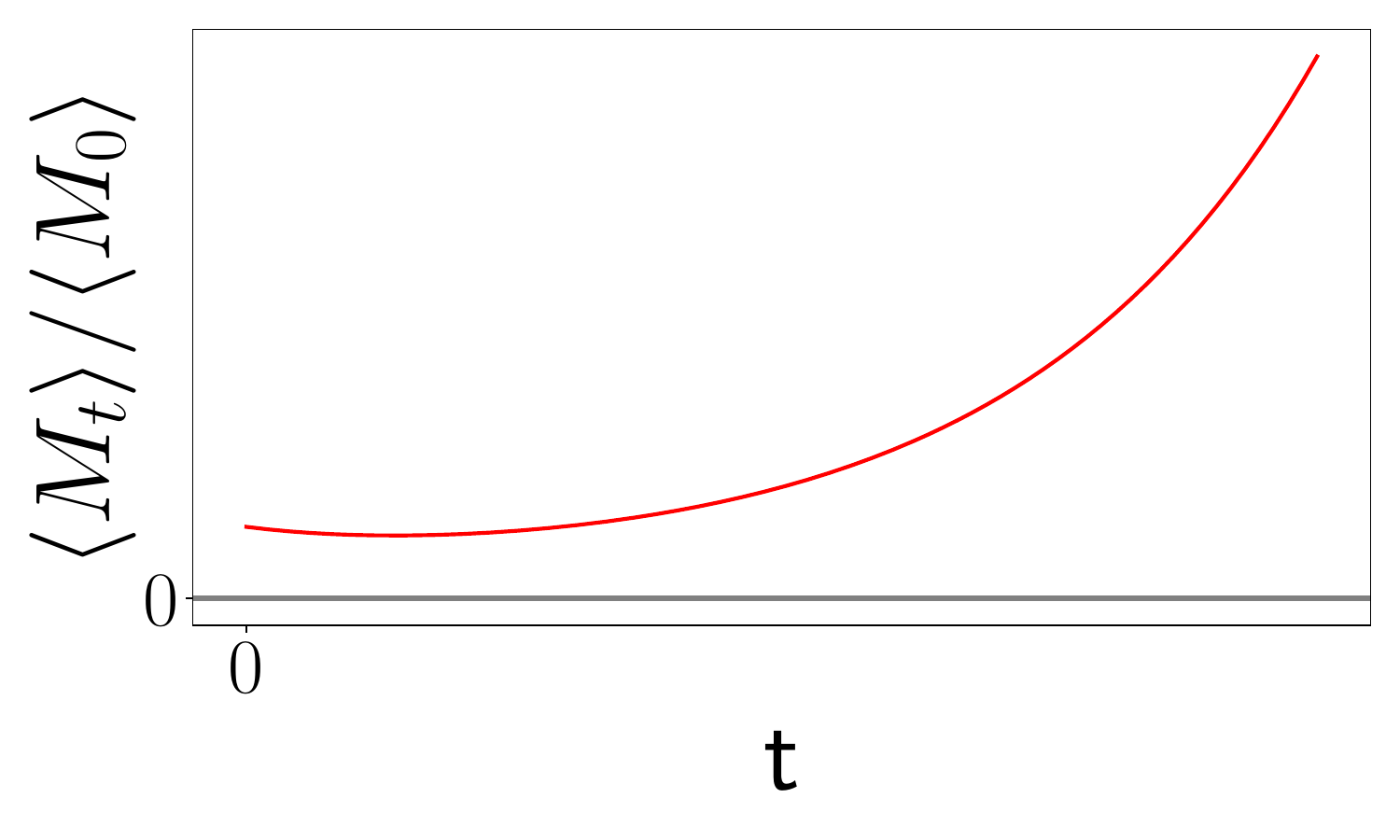}
\end{tabular}
\\ \hline
\end{tabular}%
}
\caption{Average equity and mortgage dynamics, based on the value of the eigenvalues $\lambda_1$ and $\lambda_2$.}
\label{tab:dynamic outcomes}
\end{table}

\newpage
We can now construct a phase diagram in the $(p,s)$ plane, keeping the other parameters fixed. 

\begin{figure}[H]
    \centering
    \includegraphics[width=1\textwidth]{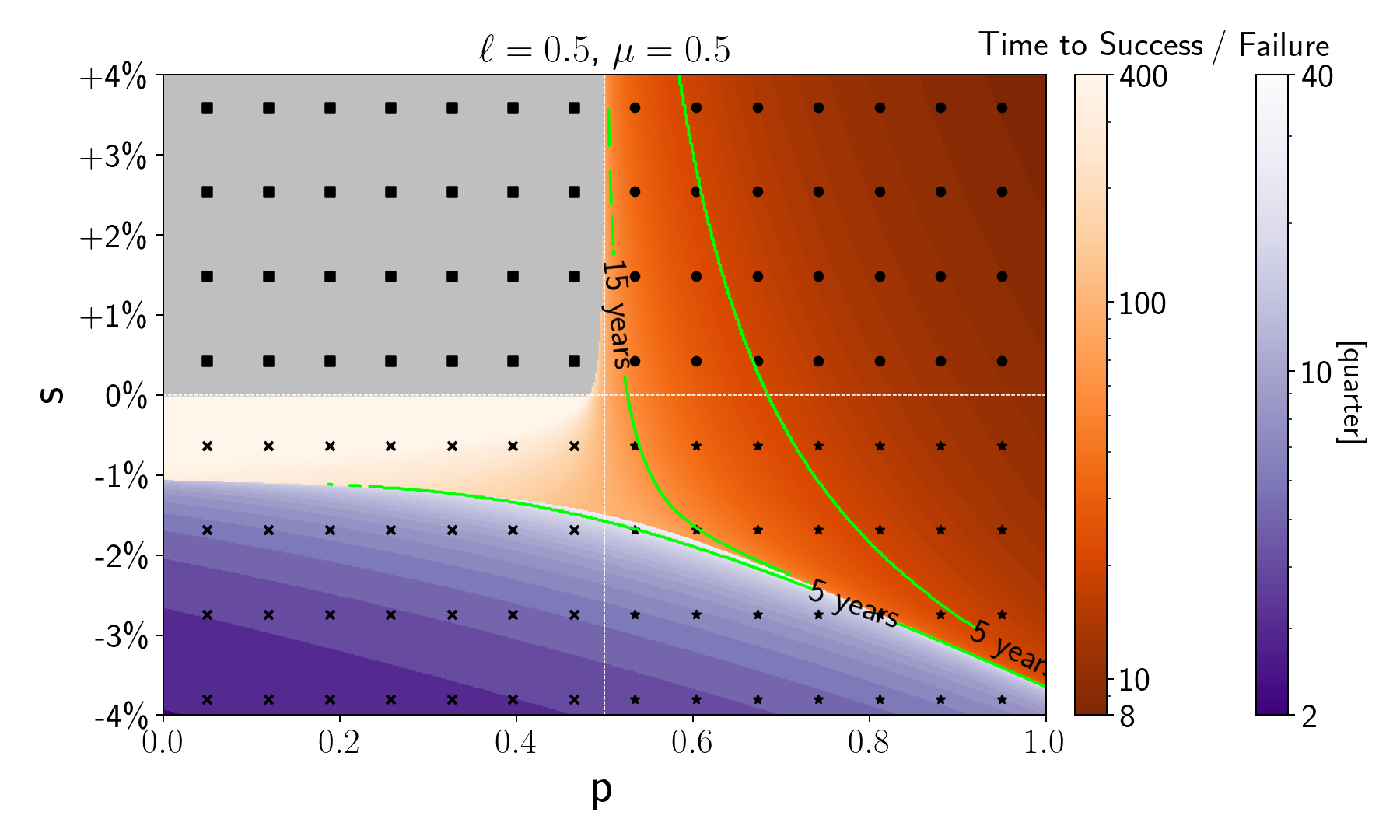}
    \caption{Here and in the following phase diagrams, the plot is constructed as follows: area in orange represents where debt recycling is (either weakly or strongly) successful; area in purple represents regions in the plane where debt recycling fails; area in gray denotes the zone of permanent re-mortgaging. The intensity of the orange (purple) directly correlates with the speed at which the strategy succeeds (fails). The colorbar is measured in quarters. Lime lines are contour lines, and the markers on the plot represent the values of the eigenvalues of the process at that point, with white lines marking the transition between one area and another. The markers follow the legend: \\
    ($\bullet$): $\lambda_{1,2} >1$ \\
    ($\blacksquare$): $\lambda_1 >1; 0< \lambda_2 <1$ \\
    ($\times$): $0< \lambda_{1,2} <1$ \\
    ($\bigstar$): $\lambda_2 >1; 0<\lambda_1 <1$ \\}
    \label{fig:phase diagram 1}
\end{figure}

In Fig. \ref{fig:phase diagram 1} we present the phase diagram of the debt recycling strategy with the following parameters: average initial equity \(\langle E_0 \rangle = \$30,000\),
i.e., the amount that the homeowner paid at the beginning of the mortgage agreement, or equivalently the value of the portion of the house that the borrower owns totally from the beginning, free from any liabilities or mortgages; average initial mortgage \(\langle M_0 \rangle = \$300,000\), i.e., the part of the property's purchase price that is financed through the loan; value of the installment of the repayment to the bank \(\pi^\star = \$3,000\); \(q = 1 \%\) chance of skipping an installment; LTV ratio $\ell=0.5$; risk factor of the investment $\mu =0.5$.

Within the parameter space, we analyze the average equity and mortgage, coloring each point of the phase diagram in orange, purple, or gray to represent the three potential outcomes of the strategy detailed in Table \ref{tab:dynamic outcomes}. Orange indicates a successful strategy (both weakly and strongly), where the mortgage holder is generally able to fully repay the mortgage and achieve complete ownership of the house. Purple signifies a default scenario, where the house value is entirely mortgaged, suggesting that a standard monthly repayment strategy (without recycling) should have rather been undertaken. Gray is used for scenarios leading to permanent re-mortgaging.

The color intensity directly correlates with how fast the final outcome is achieved: darker orange (purple) indicates faster acquisition (loss) of house ownership. For visual clarity, the maximum intensity of orange is limited to $t=400$ quarters, or $100$ years, with points depicting weakly successful debt recycling beyond $100$ years shown in beige. The beige region is, obviously, unphysical. 

Markers on the diagram align with the legend in Fig. \ref{fig:lambda space} or the first column of Table \ref{tab:dynamic outcomes}, showing the possible combinations of the magnitudes of eigenvalues $\lambda_1$ and $\lambda_2$.

Lime lines represent contour lines, which are the locus of points on the plane $(p,s)$ that have the same first hitting time \(t^\star\). We choose to illustrate the points of the plane where the process concludes at 5 and 15 years, aligning with the guideline that debt recycling is best suited for individuals who can sustain the cycle for a minimum of 5 years. This approach is shared by many financial advisory firms and mortgage consultancy companies that provide guidance on loans and accounting services.

\begin{figure}[h]
    \centering
    \includegraphics[width=0.8\textwidth]{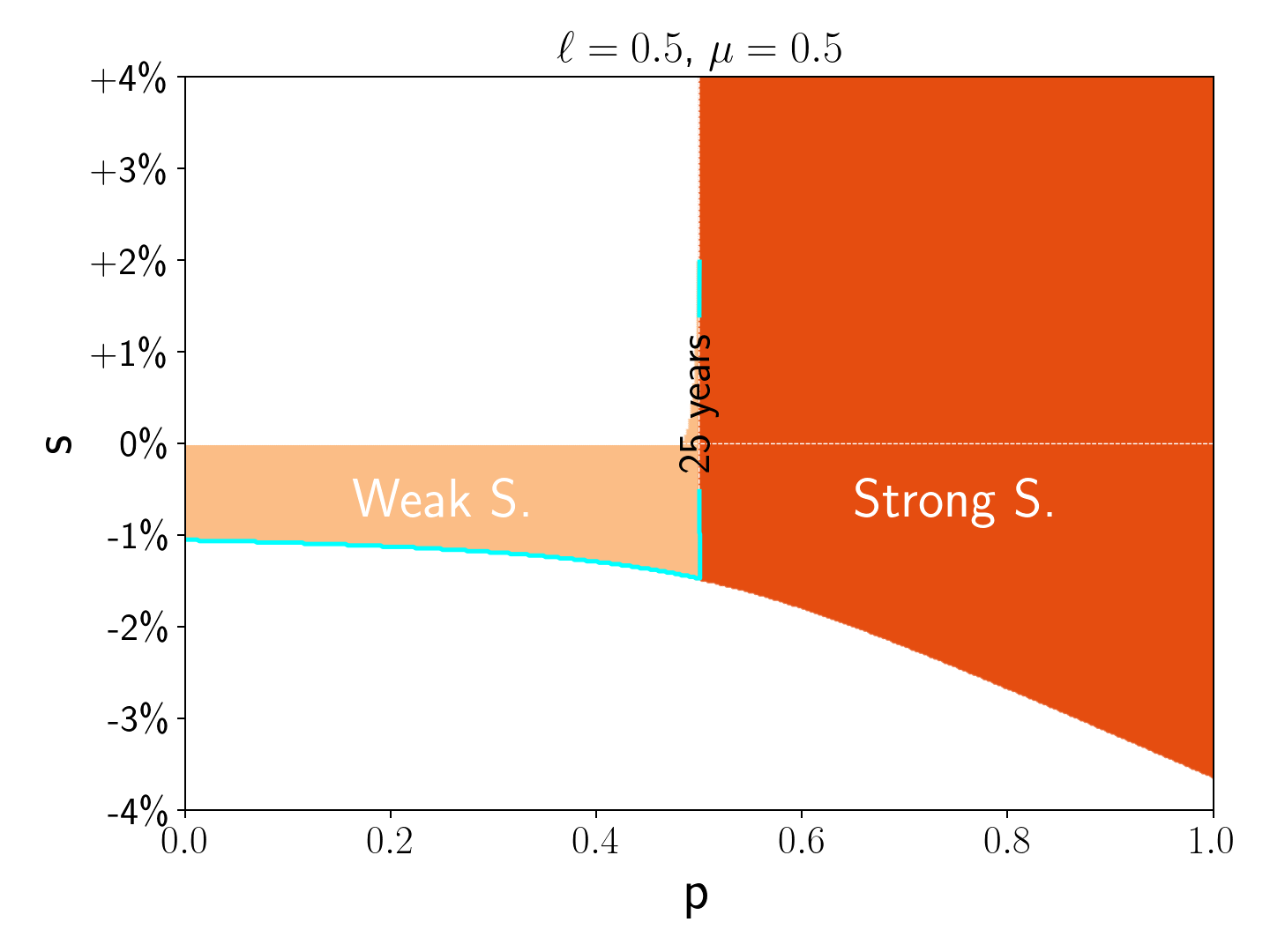}
    \caption{Another representation of the phase diagram in Fig. \ref{fig:phase diagram 1}, with focus on the strong (dark orange) and weak (light orange) success of the strategy. In cyan, the contour line of $t^* = t_{\text{ no recycling}} = \langle M_0 \rangle /\pi^\star = 25$ years for our choice of parameters $\langle M_0 \rangle = \$300,000$ and $\pi^\star = \$3,000$.} 
    \label{fig:phase diagram SW 1}
\end{figure}

In Fig. \ref{fig:phase diagram SW 1}, we further distinguish between areas of strong success (in dark orange), where the time to achieve house ownership through debt recycling is shorter than without any strategy, and areas of weak success (in light orange), where the strategy requires more time than the conventional monthly repayment approach, suggesting the latter as the more advisable option. The cyan line is also a contour line, similar to the ones in lime shown in the previous Fig. \ref{fig:phase diagram 1}. It is still calculated as the locus of points on the plane where the first hitting time takes on the same value $t_{\text{ no recycling}}$, the time required to achieve house ownership without any debt recycling strategy.

Strong success is consistently observed in the \MakeUppercase{\romannumeral 1} quadrant and weak success in the \MakeUppercase{\romannumeral 2} quadrant. In the majority of standard scenarios, where parameters are within typical ranges, the \MakeUppercase{\romannumeral 3} quadrant typically exhibits weak success, and the \MakeUppercase{\romannumeral 4} quadrant strong success, with \(p=0.5\), serving as the separating threshold between the two. In rare cases involving extreme parameter values, specifically when the total value of the house is exceptionally high (worth millions) or the initial mortgage-to-equity ratio is significantly skewed towards a high level of debt with very little initial equity (over-collateralized situations), strong success can manifest in the \MakeUppercase{\romannumeral 3} quadrant or weak success in the \MakeUppercase{\romannumeral 4} quadrant. 

The outcome of the debt recycling strategy is primarily influenced by the mortgage-to-equity ratio and the monthly repayment schedule. Raising the LTV ratio, the risk parameter, or both can accelerate the process and shift outcomes from default to success in the \MakeUppercase{\romannumeral 4} quadrant, where the housing market has a negative trend. This shift is more pronounced when the total house value is lower, showing that the strategy's effectiveness is also significantly impacted by risk levels.

For a more detailed and extended discussion of the results, please refer to Section \ref{sec:Results}. In the next section, we are going to describe the mathematical methods used to solve the joint stochastic process $(E_t,M_t)$ defined in Eq. \eqref{eq:processmatrix}.

\section{Methods}\label{subsec:average}

In this section, we derive the solution for the average behaviour of the equity and mortgage process. Introducing the $2\times 2$ matrix $A_t$
\begin{equation}\label{eq: A_t}
A_t=
\begin{pmatrix}
\alpha_t & r_t \\ \delta_t & 1
\end{pmatrix}
=
\begin{pmatrix}
1 + \ell \mu\sigma_{t-1} +r_t & r_t \\ 
- \ell \mu \sigma_{t-1} & 1
\end{pmatrix}\ ,
\end{equation}
the coupled dynamical process \eqref{eq:processmatrix} can be iterated to yield
\begin{align}\label{eq:5}
\nonumber \begin{pmatrix}
 E_t \\ M_t
\end{pmatrix}
&=
A_t A_{t-1} \cdots A_1
\begin{pmatrix}
E_0 \\ M_0
\end{pmatrix}
+
A_t A_{t-1} \cdots A_2
\begin{pmatrix}
\pi_1 \\ -\pi_1
\end{pmatrix}
+
A_t A_{t-1} \cdots A_3
\begin{pmatrix}
\pi_2 \\ -\pi_2
\end{pmatrix}
\\
&+ \ldots +
\begin{pmatrix}
\pi_t \\ -\pi_t
\end{pmatrix}\ ,
\end{align} 
where $E_0$ and $M_0$ are the initial conditions of equity and mortgage respectively, and $\pi_t$ is the scheduled repayment to the bank at the time $t$.
Isolating $E_t$ and $M_t$, and taking the average over the disorder, we may use independence at different times, and the fact that the matrix entries are identically distributed to deduce the following for product of operators
\begin{equation}
    \Big\langle\prod_{t'=k}^t A_{t'}\Big\rangle=\prod_{t'=k}^t \langle A_{t'}\rangle=\bar{A}^{t-k}\ ,
\end{equation}
where $\bar{A}$ is the average matrix
\begin{equation}\label{eq:A avg}
\bar{A} = 
\begin{pmatrix}
\langle 1 + \ell \mu\sigma +r \rangle & \langle r \rangle \\ 
\langle - \ell \mu \sigma \rangle & 1
\end{pmatrix}
=
\begin{pmatrix}
1 + \ell \mu (2p-1) + s & s \\ 
- \ell \mu (2p-1) & 1
\end{pmatrix}\ ,
\end{equation}
and $\bar{A}^{t-k}$ stands for matrix power. The average matrix depends on four main parameters: $\ell$ (\textit{LTV ratio}), $\mu$ (\textit{risk factor}), $p$ (\textit{probability of gaining from the investment}), and $s$ (\textit{average value of the market's fluctuations}).
 Therefore from Eq. \ref{eq:5} we obtain
\begin{multline}\label{eq:11}
    \langle E_t \rangle = (\bar{A} ^t)_{11} \langle E_0 \rangle + (\bar{A} ^t)_{12} \langle M_0 \rangle + \\
    (\bar{A}^{t-1})_{11} \langle \pi \rangle + (\bar{A}^{t-1})_{12} \langle - \pi \rangle +
    (\bar{A}^{t-2})_{11} \langle \pi \rangle + (\bar{A}^{t-2})_{12} \langle - \pi \rangle + \ldots + \langle \pi \rangle\ .
\end{multline}
Now, the matrix $\bar A$ can be decomposed as follows
\begin{equation}
    \bar A = U\Lambda U^{-1}\ ,
\end{equation}
where 
\begin{equation}\label{eq: autovalori}
\Lambda = 
\begin{pmatrix}
s+1 & 0\\
0 & \ell\mu(2p-1)+1
\end{pmatrix}
\end{equation}
is the diagonal matrix of (distinct) eigenvalues, and
\begin{align}
U &=\begin{pmatrix}
 -\frac{s}{\ell \mu  (2p-1)} & -1 \\
 1 & 1 
\end{pmatrix} \label{U}\\
U^{-1} &=\begin{pmatrix}
\frac{l\mu(2p - 1)}{l\mu(2p - 1) - s} & \frac{l\mu(2p - 1)}{l\mu(2p - 1) - s} \\
 -\frac{l\mu(2p - 1)}{l\mu(2p - 1) - s} & -\frac{s}{l\mu(2p - 1) - s} \\
\end{pmatrix}\ \label{Uminusone}
\end{align}
are the matrix of the eigenvectors and its inverse, respectively.
It should be noted that \(\lambda_1 =s+1 > 0\) is always true, since \(s\) represents the percentage change in market fluctuations, and \(\lambda_2 = \ell \mu (2p-1)+1 > 0\) is always true as well, as \(\ell\), \(\mu\), and \(p\) are defined to be within the range \([0,1]\).
Eq. \ref{eq:11} can therefore be rewritten as
\begin{equation}\label{eq:master E}
   \frac{\langle E_t \rangle}{\langle E_0 \rangle} = (U \Lambda^t U^{-1})_{11} + c_1  (U \Lambda^t U^{-1})_{12} + c_2 \sum_{k=1}^{t} \left[ (U \Lambda^{t-k} U^{-1})_{11} - (U \Lambda^{t-k} U^{-1})_{12} \right]\ ,
\end{equation}
with
\begin{equation}\label{c1,c2}
    c_1 = \frac{\langle M_0 \rangle}{\langle E_0 \rangle};\qquad c_2 = \frac{\langle \pi \rangle}{\langle E_0 \rangle}\ .
\end{equation}
Following the same procedure, we can find the evolution equation for the average mortgage value 
\begin{multline}\label{eq:12}
 \langle M_t \rangle = (\bar{A} ^t)_{21}  \langle E_0 \rangle + (\bar{A} ^t)_{22} \langle M_0  \rangle + \\
(\bar{A}^{t-1})_{21} \langle \pi  \rangle + (\bar{A}^{t-1})_{22} \langle - \pi \rangle +
(\bar{A}^{t-2})_{21} \langle \pi \rangle + (\bar{A}^{t-2})_{22} \langle - \pi \rangle + \ldots - \langle \pi \rangle \ ,
\end{multline}
or, equivalently:
\begin{equation}\label{eq: master M}
\frac{\langle M_t \rangle}{\langle M_0 \rangle} = c_3 (U \Lambda^t U^{-1})_{21} + (U \Lambda^t U^{-1})_{22} + c_4 \sum_{k=1}^{t} \left[ (U \Lambda^{t-k} U^{-1})_{21} - (U \Lambda^{t-k} U^{-1})_{22} \right]\ ,
\end{equation}
with
\begin{equation}\label{c3,c4}
c_3 =\frac{1}{c_1}= \frac{\langle E_0 \rangle}{\langle M_0 \rangle};\qquad c_4 = \frac{\langle \pi \rangle}{\langle M_0 \rangle}\ .
\end{equation}
We now wish to extract from Eq. \ref{eq:master E} and Eq. \ref{eq: master M} the first time $t^\star$ at which either of these processes hits zero. After some lengthy algebra (see \ref{app:Calculation Methods} for details) we manage to extract the $t$-dependence exactly as
\begin{align}
\frac{\langle E_t \rangle}{\langle E_0 \rangle} &= \mathcal{A} \lambda_1^t + \mathcal{B} \lambda_2^t +\mathcal{C}\label{avproc1}\ ,\\
\frac{\langle M_t \rangle}{\langle M_0 \rangle} &= \mathcal{D} \lambda_1^t + \mathcal{E} \lambda_2^t +\mathcal{F}\label{avproc2}\ ,
\end{align}
where the constants are given explicitly as
\begin{align}
\mathcal{A} &= - \frac{s (c_1 + 1)}{\ell \mu (2p-1) -s} = \frac{(c_1+1)(\lambda_1 - 1)}{\lambda_1 - \lambda_2} \label{eq: A} \ , \\ 
\nonumber\mathcal{B} &= \frac{-c_2 s + \ell \mu (2p - 1) (c_1 s + c_2 + \ell \mu (2p - 1))}{\ell \mu (2p - 1) (\ell \mu (2p - 1) - s)} =\\
&=\frac{c_2\left(\lambda_1-1\right)-\left(\lambda_2-1\right)\left(c_1\left(\lambda_1-1\right)+c_2+\lambda_2-1\right)}{\left(\lambda_1-\lambda_2\right)\left(\lambda_2-1\right)}  \ ,\label{eq: B} \\
\mathcal{C} &= - \frac{c_2}{\ell \mu (2p-1)} = - \frac{c_2}{\lambda_2-1}\label{eq: C} \ ,\\
\mathcal{D} &= \frac{(c_3+1) \ell \mu (2p-1)}{\ell \mu (2p-1)-s} = 
-\frac{\left(c_3+1\right)\left(\lambda_2-1\right)}{\lambda_1-\lambda_2}
\label{eq: D} \ , \\
\nonumber\mathcal{E} &= \frac{c_4 s - \ell \mu (2p-1) (c_3 \ell \mu (2p-1) + c_4 +s)}{\ell \mu (2p-1) (\ell \mu (2p-1) -s)} =\\ 
&=\frac{-c_4\left(\lambda_1-1\right)+\left(\lambda_2-1\right)\left(c_3\left(\lambda_2-1\right)+c_4+\lambda_1-1\right)}{\left(\lambda_1-\lambda_2\right)\left(\lambda_2-1\right)} \label{eq: E}  \ ,\\
\mathcal{F} &= \frac{c_4}{\ell \mu (2p-1)} = \frac{c_4}{\lambda_2-1} \label{eq: F} \ ,
\end{align}
while the eigenvalues $\lambda_1$ and $\lambda_2$ are given in Eq. \eqref{eq: autovalori}.
It should be noted that the parameters \(\ell\) and \(\mu\) never occur independently of each other: it is the product \(\ell\mu \in [0,1]\) that acts as the critical factor influencing the amplitude of the fluctuations in \(E_t\) and \(M_t\). In the next section we present a more detailed and extended discussion of our results.

\textcolor{black}{We refer to \ref{app:variance} for the calculation of the fluctuations (variances) of the processes $E_t$ and $M_t$ at all times. The fluctuations will be shown to be fundamentally incapable of changing the conclusions drawn on the basis of the average processes alone, and summarized in the table and phase diagrams.}

\section{Results}\label{sec:Results}
We recall that the parameters involved in the simulations are the following:
\begin{itemize}
    \item $\ell \in [0, 1]$: LTV ratio;
    \item $\mu \in [0, 1]$: risk factor of the investment;
    \item $p \in [0,1]$: probability that the investment generates a gain;
    \item $s \in [-4\%, +4\%]$:
    average value of the market's fluctuations;
    \item $q \ll 1$: probability of skipping an installment of the repayment to the lender;
    \item $\pi^{*}$: value of the regular installment;
    \item $E_0, M_0$: initial conditions of equity and mortgage respectively.
\end{itemize} 

We have derived the evolution equation for the average equity process \(\langle E_t \rangle\) in Eq. \eqref{avproc1}, and for the average mortgage process \(\langle M_t \rangle\) in Eq. \eqref{avproc2}. We are interested in the time $t^\star=\min(t_E,t_M)$, where $t_E$ is the first time at which \(\langle E_{t} \rangle = 0\), while $t_M$ is the first time at which \(\langle M_{t} \rangle = 0\). If $t^\star=t_E$, the homeowner's equity (or value of the fraction of the house owned) has been massively depleted. In this scenario, the borrower would need to rely uniquely on monthly repayments to regain ownership, indicating that the strategy has not succeeded. Instead, \(t_M\) signifies that the homeowner has managed to pay off the entire mortgage, achieving full ownership of the house. If $t^\star=t_M$ then the strategy has been successful. In this case we distinguish between \emph{strong} success,  where the time to achieve house ownership through debt recycling is shorter than without any strategy, and \emph{weak} success, where the strategy requires more time than the conventional monthly repayment approach, suggesting that the latter would still be the more advisable option -- even if eventually full home ownership is reached anyway. 

The question is therefore which of the two scenarios occurs first. To shed some light on the question, we first conducted a few numerical simulations of the joint dynamical process $(E_t,M_t)$ described by Eq. \eqref{eq:processmatrix} for different values of the model parameters. Next, we present a more detailed investigation of phase diagrams for this problem.

\subsection{Simulations of trajectories for mortgage balance and equity}\label{repayment traj}
\begin{figure}[H]
    \begin{minipage}{0.5\textwidth}
        \centering \includegraphics[width=\textwidth]{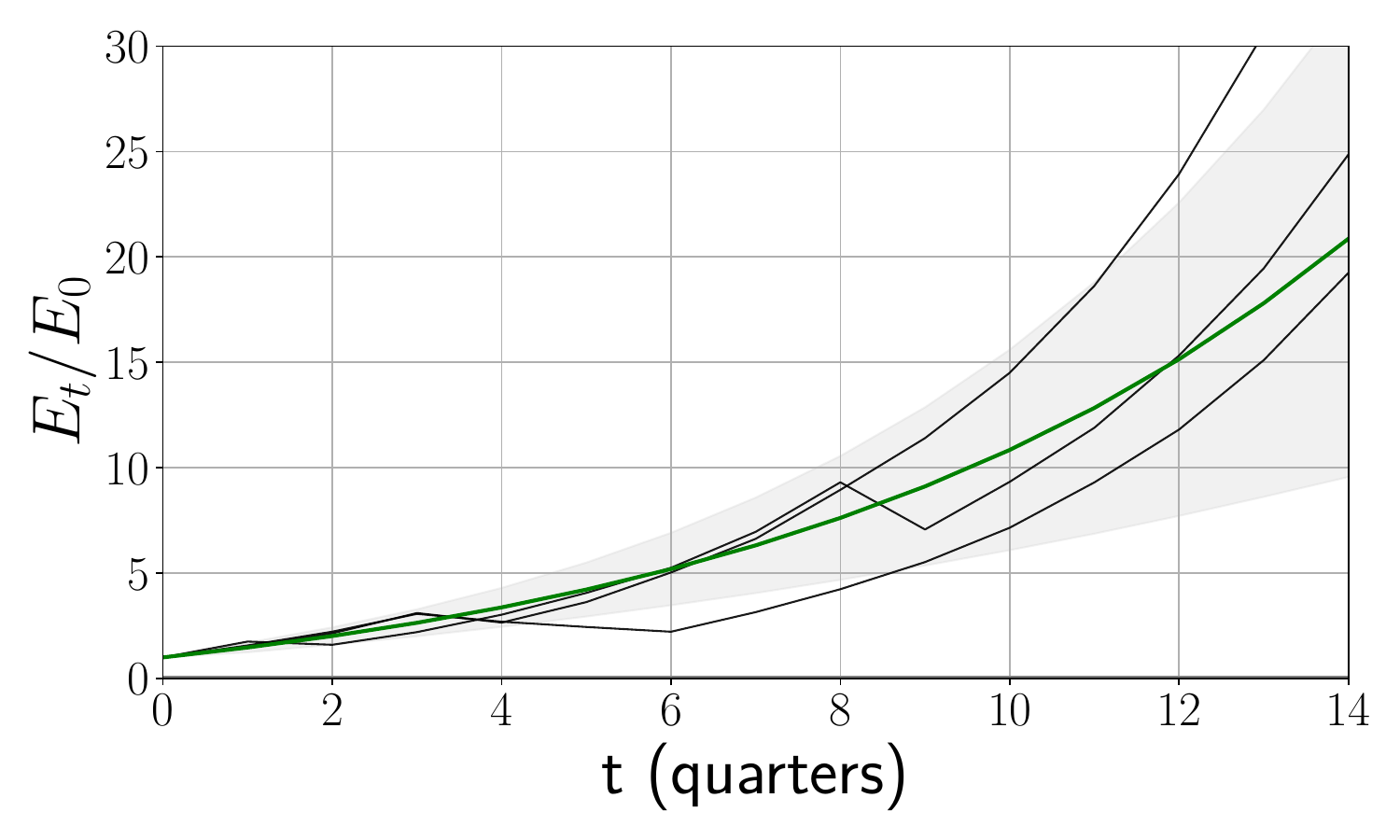}
        \subcaption{Equity over time. The average process, Eq. \eqref{avproc1}, is illustrated by the solid green line; some simulations of the process, Eq. \eqref{eq:processmatrix} \textcolor{black}{with a standard deviation $\phi = 1\%$ of the fluctuations in the house market value}, are shown in black; the standard deviation band, calculated using $N=10^5$ simulations, is displayed in gray. 
        } \label{fig:trajectory1}
    \end{minipage}
    \hspace{10pt}
    \begin{minipage}{0.5\textwidth}
        \centering \includegraphics[width=\textwidth]{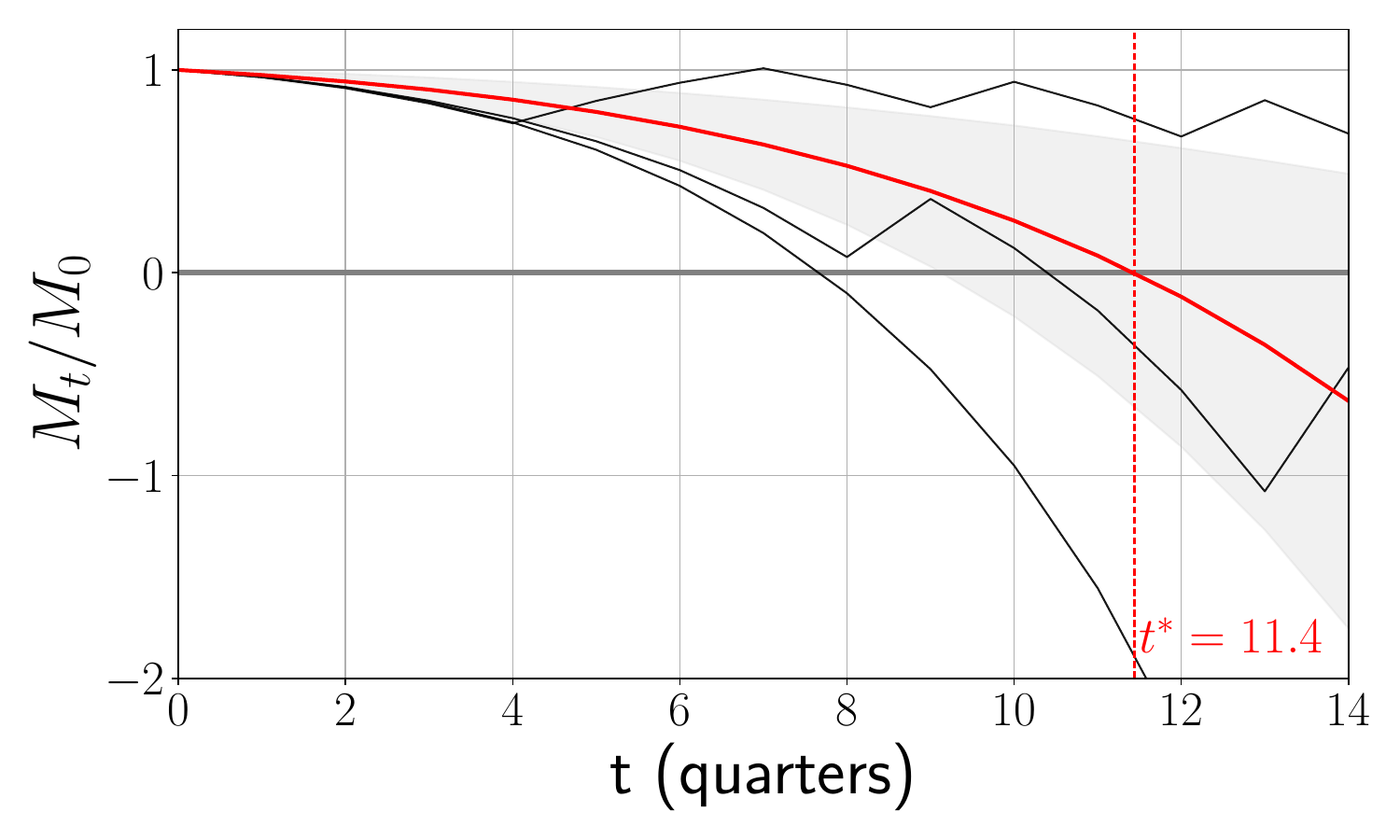} 
        \subcaption{Mortgage over time. The average process, Eq. \eqref{avproc2},  is illustrated by the solid red line; some simulations of the process, Eq. \eqref{eq:processmatrix} \textcolor{black}{with a standard deviation $\phi = 1\%$ of the fluctuations in the house market value}, are shown in black; the standard deviation band, calculated using $N=10^5$ simulations, is represented in gray. The vertical line at $t^\star=11.4$ signals the moment the average mortgage balance hits the absorbing wall at zero.}\label{fig:trajectory2}
    \end{minipage}
    \caption{Example of time evolution of the equity and mortgage balance processes with the debt recycling strategy: the equity increases steadily, while the mortgage balance decreases steadily until it hits the absorbing boundary within $3$ years, marking the strategy as strongly successful. 
    Referring to Tab. \ref{tab:dynamic outcomes}, we are in the scenario \((\bullet): \lambda_{1,2}>1\).} 
    \label{fig: trajectories}
\end{figure}

In Fig. \ref{fig: trajectories} we display the time evolution of the equity $E_t$ (in units of the initial equity $E_0= \$ 30,000 $) [left panel] and the mortgage balance $M_t$ (in units of the initial mortgage $M_0= \$ 300,000 $) [right panel] in a scenario where the parameters are chosen to be firmly in favor of the homeowner being able to successfully repay their mortgage: \(\ell = 0.5\), \(\mu =0.5\), \(p = 0.8\), \(s = 2\%\), $q= 1\%$, and $\pi^{*}= 3,000\$$. In particular, we have a high probability $p=0.8$ that the asset being invested on will generate a positive income, and we have a positive trend $s=2\%$ of the house market leading to the possibility to extract more equity.  

In line with our expectations, most of the equity trajectories as well as the average process (solid green line) are steadily growing in time, while the mortgage balance curves and their average (solid red line) are steadily decreasing in time, until at time $t^\star=11.4$ the mortgage holder has (on average) successfully managed to acquire full ownership of the house.
From Fig. \ref{fig:phase diagram SW 1}, we know that \((p,s) = (0.8, 2\%)\) corresponds to a point in the phase diagram indicating strong success of the strategy. Indeed, the time required to acquire the house without any strategy, calculated as \(t = M_0 / \langle \pi \rangle \sim 101\) quarters ($\sim 25$ years), is significantly higher than \(t^*\) (equal to nearly $3$ years).

\begin{figure}[H]
    \begin{minipage}{0.5\textwidth}
        \centering \includegraphics[width=\textwidth]{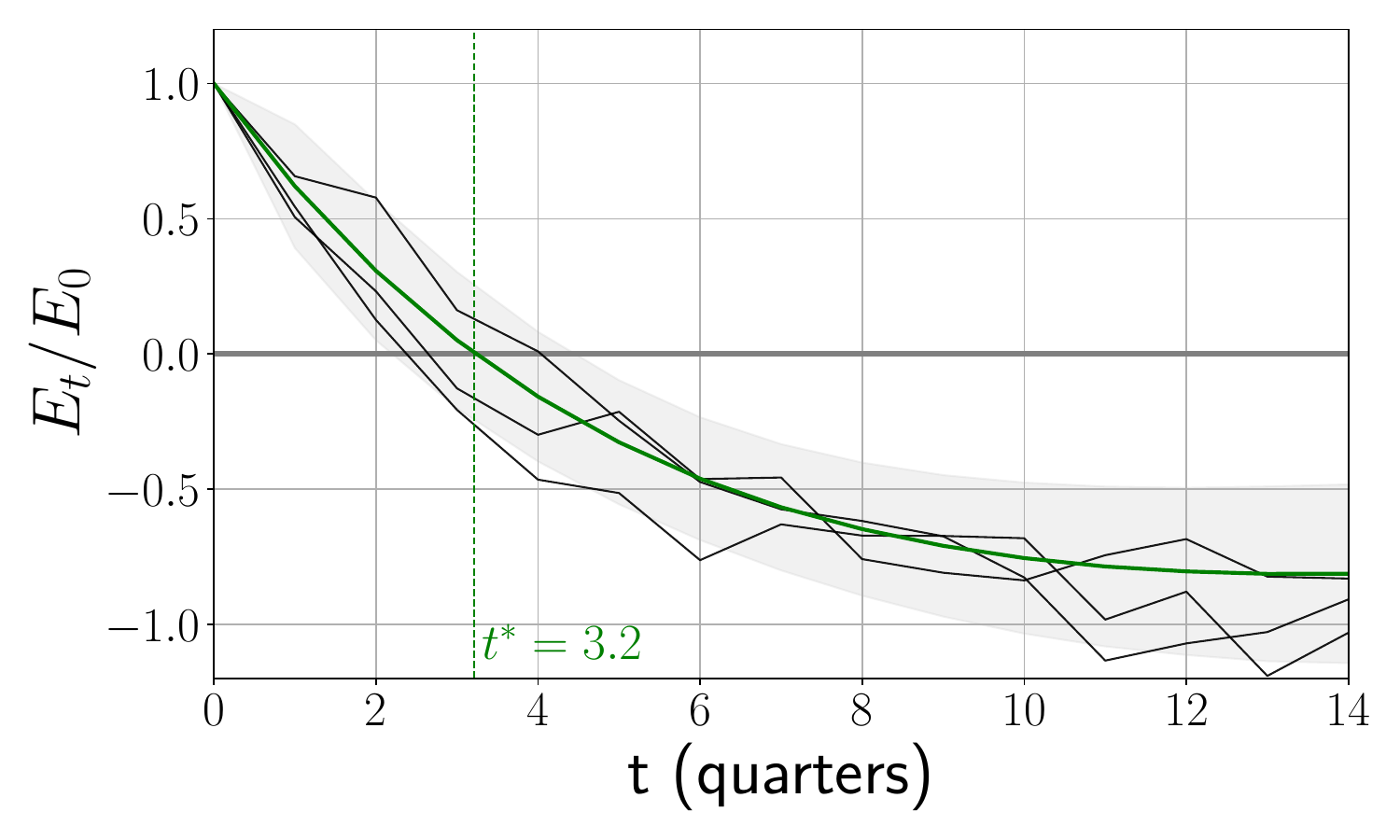} \subcaption{Equity over time. The average process, Eq. \eqref{avproc1}, is illustrated by the solid green line; some simulations of the process, Eq. \eqref{eq:processmatrix} \textcolor{black}{with a standard deviation $\phi = 1\%$ of the fluctuations in the house market value}, are shown in black; the standard deviation band, calculated using $N=10^5$ simulations, is displayed in gray. The vertical line at $t^*=3.2$ signals the moment the average equity hits the barrier at zero.} \label{fig:trajectory3}
    \end{minipage}
    \hspace{10pt}
    \begin{minipage}{0.5\textwidth}
        \centering \includegraphics[width=\textwidth]{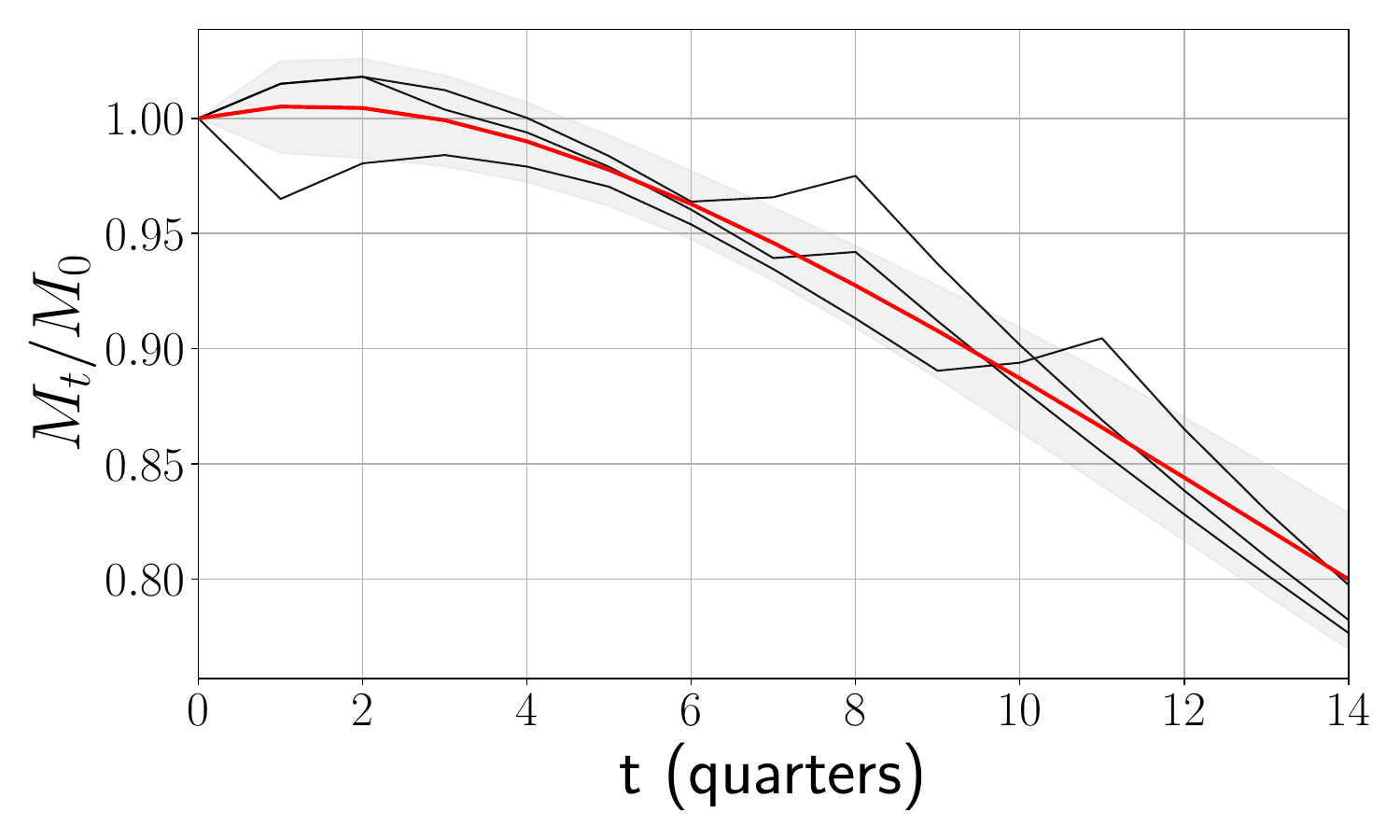} \subcaption{Mortgage over time. The average process, Eq. \eqref{avproc2}, is illustrated by the solid red line; some simulations of the process, Eq. \eqref{eq:processmatrix} \textcolor{black}{with a standard deviation $\phi = 1\%$ of the fluctuations in the house market value}, are shown in black; the standard deviation band, calculated using $N=10^5$ simulations, is displayed in gray. }\label{fig:trajectory4}
    \end{minipage}
    \caption{Example of time evolution of the equity and mortgage balance processes with the debt recycling strategy: they both decrease, but the equity does it faster and hits the absorbing boundary within $9$ months, marking the strategy as failing. 
    Referring to Tab. \ref{tab:dynamic outcomes}, we are in the scenario $(\times): 0<\lambda_{1,2}<1$. Note that the equity trajectory in the left panel is  limited to the initial few time steps, as we aim to emphasize the first hitting time. As a result, the  increasing trend potentially observed \emph{after} the hitting time (see scenario $(\times)$ in Tab. \ref{tab:dynamic outcomes}) is not shown.} 
    \label{fig: trajectories2}
\end{figure}

In Fig. \ref{fig: trajectories2}, we instead present a scenario where the parameters are unfavorable to the mortgage holder: here, we have a low probability \(p=0.2\) that the investment will result in a positive gain and an unfavorable housing market with \(s=-3\%\). All other parameters remain unchanged. Both equity and mortgage balance exhibit a decreasing trend (due to the negative value of \(s\)), but the equity curves (as well as the average process) hit the \(E=0\) absorbing wall in just over $3$ time steps, indicating a default situation where the house value is entirely eroded by the mortgage.

\subsection{Phase Diagrams and dependence on parameters}\label{subsec:phase diagram}

In this section, we present more phase diagrams of the debt recycling process, for different initial conditions. 
Starting from individual trajectories of the average process, which depend on the parameters (\(\ell\), \(\mu\), \(p\), \(s\)) as seen in Sec. \ref{repayment traj}, we explore the parameter space \((p,s)\) and build the phase diagrams, dependent on the parameters \(\ell\) and \(\mu\). 
Every point in the \((p,s)\) space is colored in orange if the average process trajectories correspond to a successful strategy outcome (see example in Fig. \ref{fig: trajectories}), in purple if they correspond to a strategy failure (see example in Fig. \ref{fig: trajectories2}), in gray if they lead to permanent re-mortgaging.
These processes follow an underlying dynamic that is determined by the magnitudes of the eigenvalues $\lambda_1$ and $\lambda_2$ at that point in parameter space (i.e. the markers on the phase diagram relate to the dynamical situation at that point). Based on the combination of the parameters $\lambda_1=s+1$ and $\lambda_2 = \ell \mu (2p-1) +1$, the dynamics of the processes is described by one of the cases showed in Tab. \ref{tab:dynamic outcomes}. For each combination of symbols and colors within the parameter space, the dynamics of \(\langle E \rangle\) and \(\langle M\rangle \) is the same. The results are of course consistent for different initial conditions.

The strategy is always strongly successful in the \MakeUppercase{\romannumeral 1} quadrant, as suggested by a high chance of investment's success (\(p>0.5\)) and a growing housing market (\(s > 0\)). The equity exhibits an increasing exponential trend, while the mortgage balance shows a decreasing exponential trend, reaching the \(M=0\) barrier early in the process. The lime-colored flow lines in the phase diagram presented in Fig. \ref{fig:phase diagram 1} illustrate that progressing towards the upper right corner of the \MakeUppercase{\romannumeral 1} quadrant shortens the timeframe for acquiring house ownership. In the \MakeUppercase{\romannumeral 2} quadrant, the strategy can be weakly successful with lower values of the product \(\ell \mu\), or it may lead to permanent re-mortgaging with higher values of \(\ell \mu\), in which case neither process ever hits the absorbing boundary. The \MakeUppercase{\romannumeral 3} and \MakeUppercase{\romannumeral 4} quadrants are where default can potentially occur. The other possible outcome is success (both strong and weak).

As stated in Eqs. \eqref{c1,c2} and \eqref{c3,c4} the parameters leading the evolution of the equity and the mortgage balance are: 
\begin{equation}
    c_1 = \frac{\langle M_0 \rangle}{\langle E_0 \rangle};\qquad c_2 = \frac{\langle \pi \rangle}{\langle E_0 \rangle};\qquad c_3 =\frac{1}{c_1}= \frac{\langle E_0 \rangle}{\langle M_0 \rangle};\qquad c_4 = \frac{\langle \pi \rangle}{\langle M_0 \rangle}\ .
\end{equation}

\begin{figure}[H]
    \centering
    
    \begin{subfigure}{.5\textwidth}
        \centering \includegraphics[width=\linewidth]{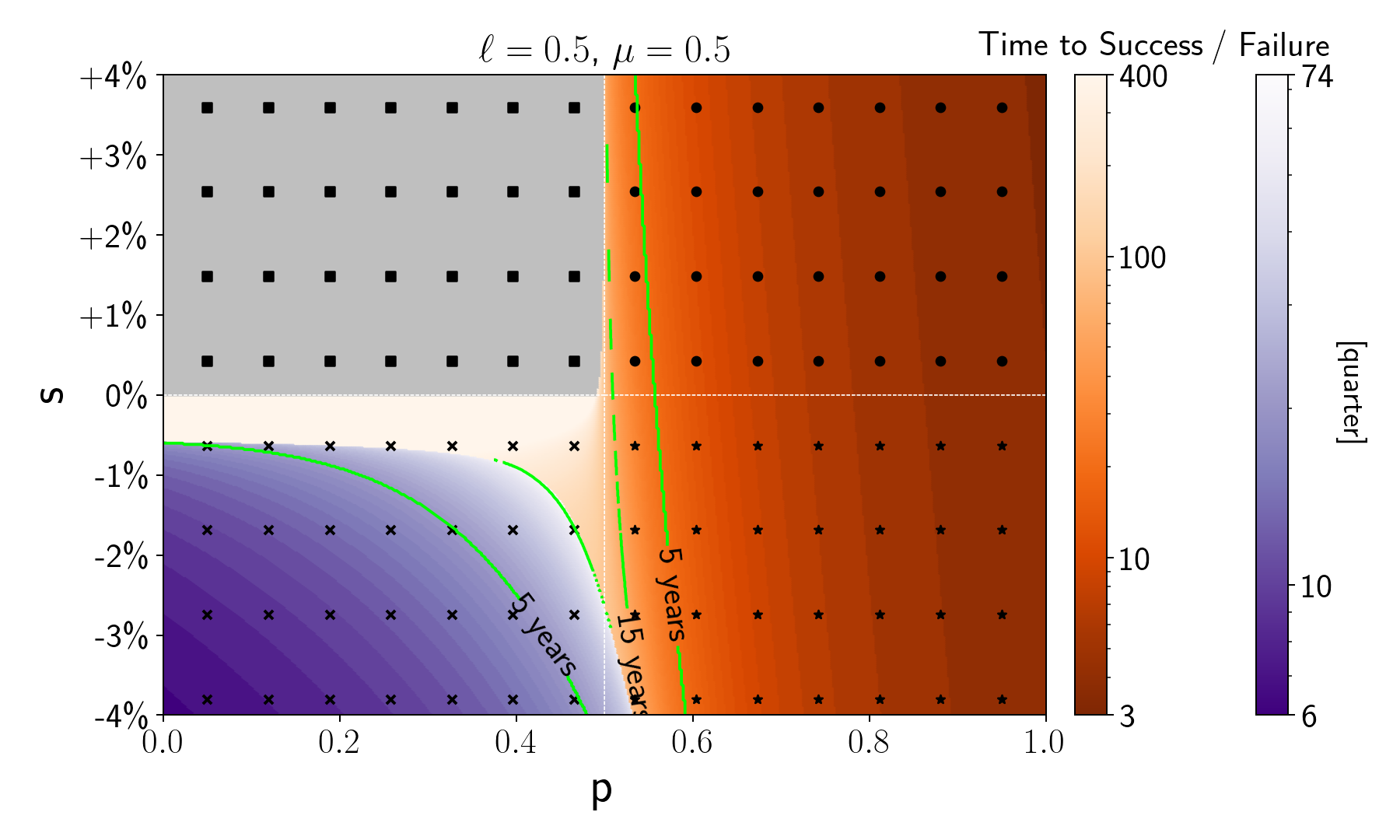}
        \subcaption{Case $1a$: $\langle E_0 \rangle = \$300,000 $; $\langle M_0 \rangle = \$300,000 $; \\ \textcolor{black}{$c_1=1$},$\langle \pi \rangle = \$3,000 $}
        \label{fig:sub1}
    \end{subfigure}\hfill
    \begin{subfigure}{.5\textwidth}
        \centering
        \includegraphics[width=\linewidth]{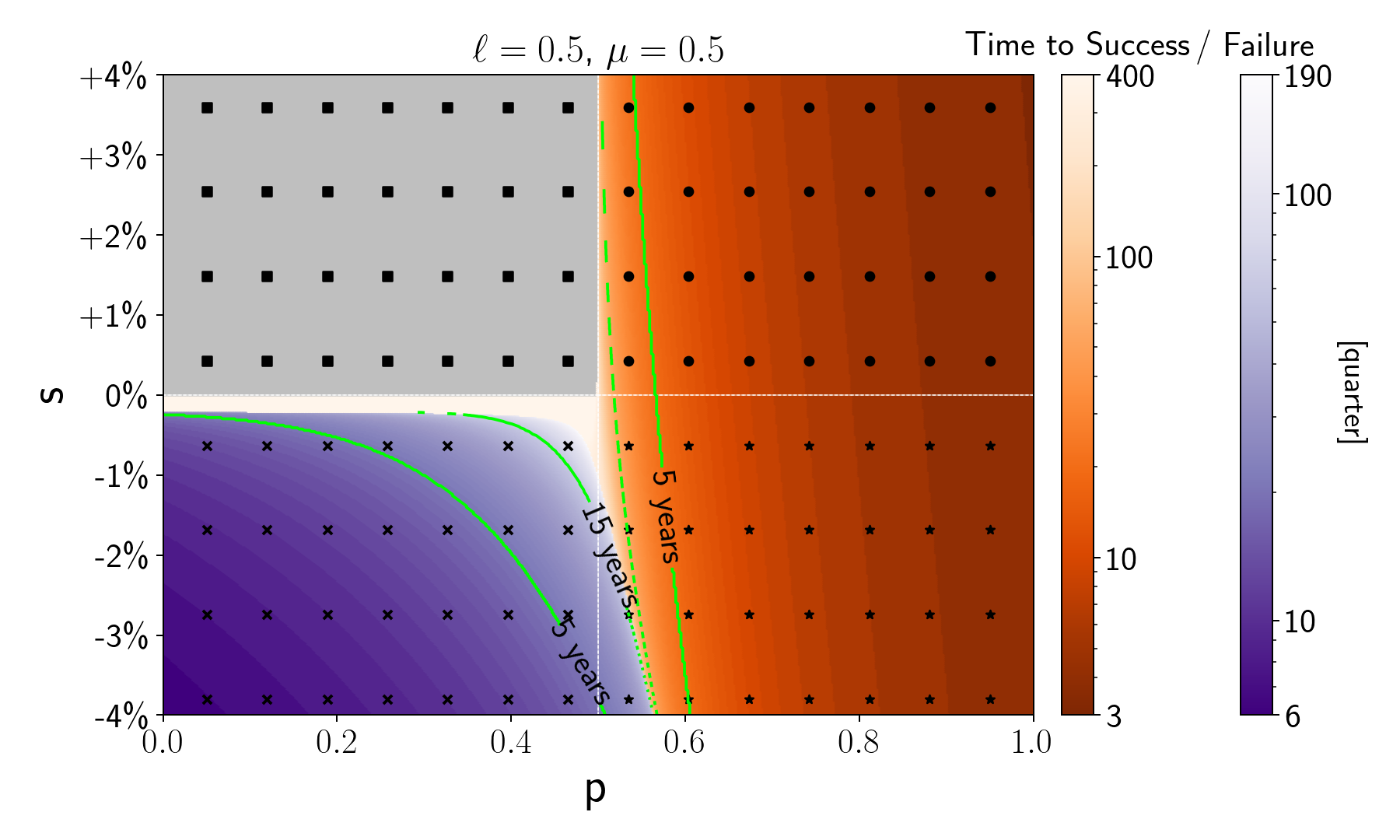}
        \subcaption{Case $1b$: $\langle E_0 \rangle = \$800,000 $; $\langle M_0 \rangle = \$800,000 $; \\ \textcolor{black}{$c_1=1$}, $\langle \pi \rangle = \$3,000 $}
        \label{fig:sub2}
    \end{subfigure}
    
    \vspace{5mm} 
    
    \begin{subfigure}{.5\textwidth}
        \centering
        \includegraphics[width=\linewidth]{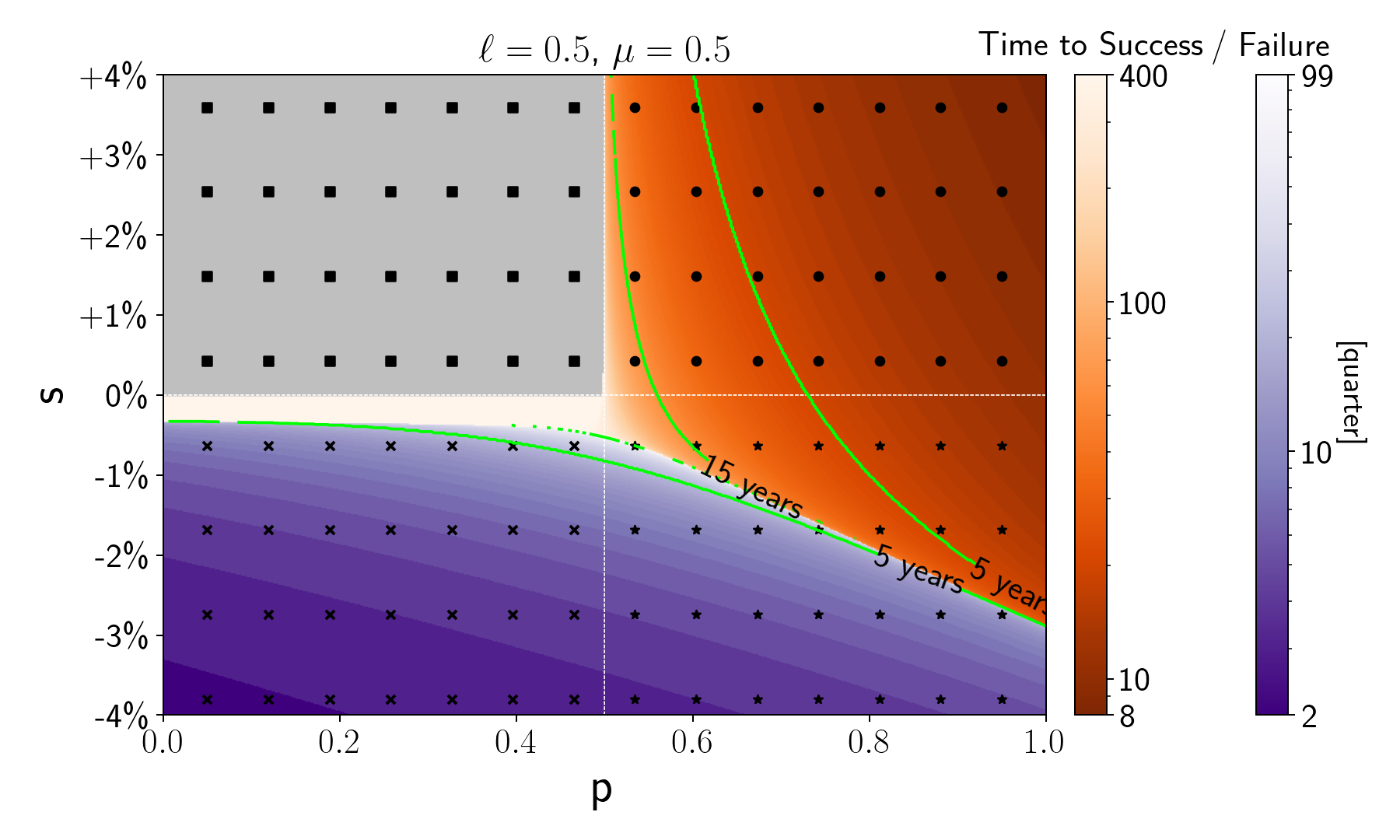}
        \subcaption{Case $2a$: $\langle E_0 \rangle = \$90,000 $; $\langle M_0 \rangle = \$900,000 $; \\\textcolor{black}{$c_1=10$}, $\langle \pi \rangle = \$3,000 $}
        \label{fig:sub3}
    \end{subfigure}\hfill
    \begin{subfigure}{.5\textwidth}
        \centering
        \includegraphics[width=\linewidth]{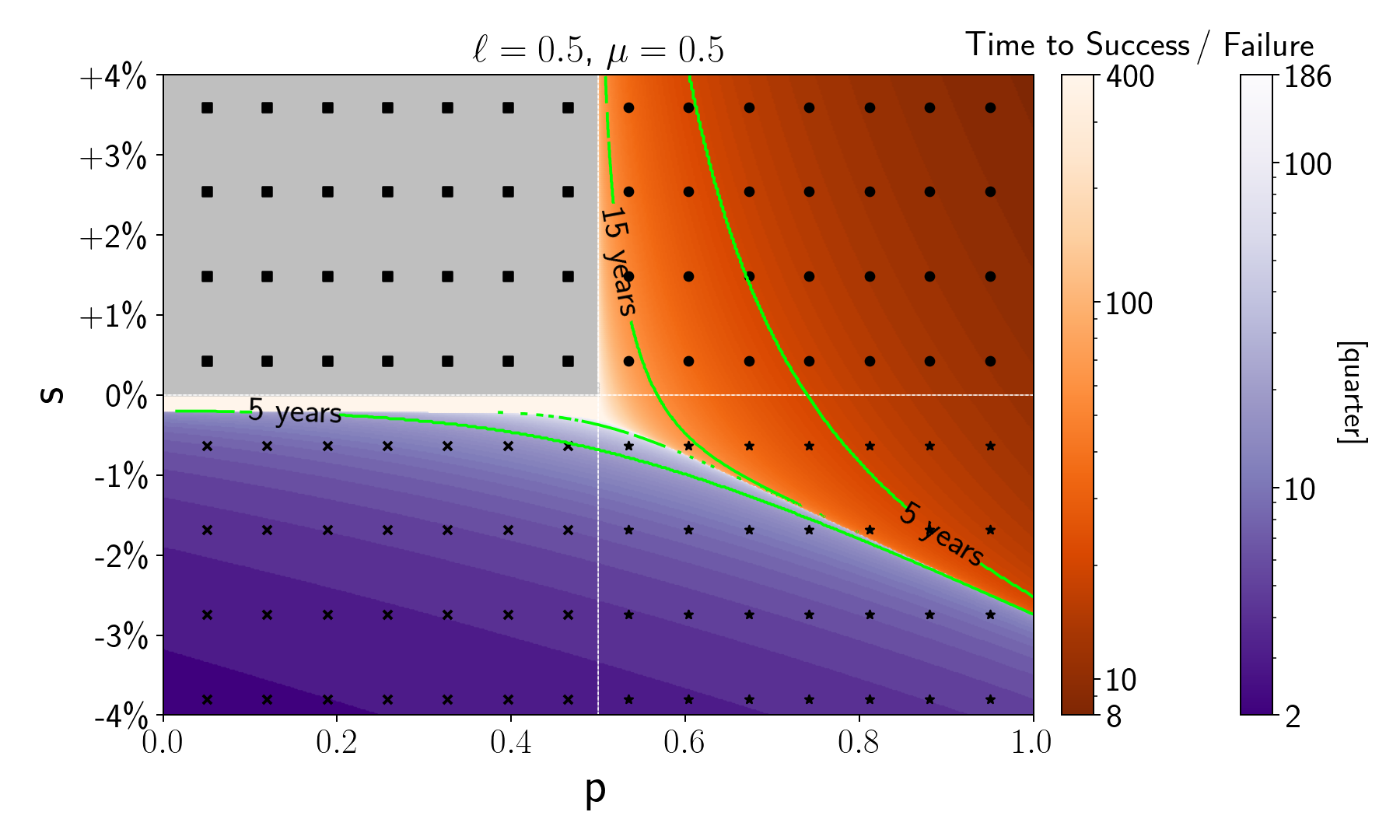}
        \subcaption{Case $2b$: $\langle E_0 \rangle = \$150,000 $; $\langle M_0 \rangle = \$1,500,000 $; \\\textcolor{black}{$c_1=10$}, $\langle \pi \rangle = \$3,000 $}
        \label{fig:sub4}
    \end{subfigure}
    
    \vspace{5mm} 
    
    \begin{subfigure}{.5\textwidth}
        \centering
        \includegraphics[width=\linewidth]{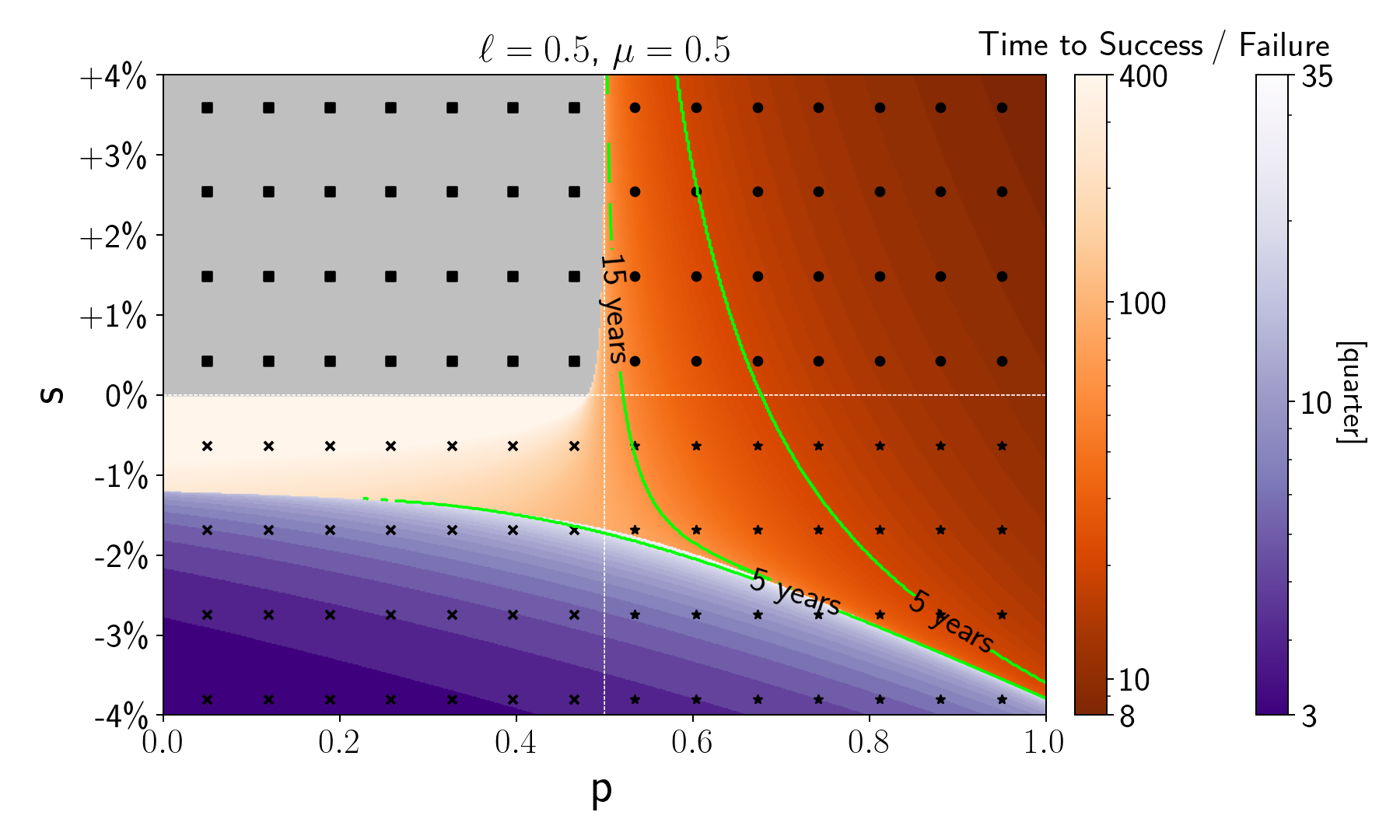}
        \subcaption{Case $3a$: $\langle E_0 \rangle = \$90,000 $; $\langle M_0 \rangle = \$900,000 $; \\ \textcolor{black}{$c_1=10$}, $\langle \pi \rangle = \$10,000 $}
        \label{fig:sub5}
    \end{subfigure}
    
    \caption{Phase diagrams for different initial conditions. 
    Area in orange represents where debt recycling is (either weakly or strongly) successful; area in purple represents regions in the plane where debt recycling fails; area in gray denotes the zone of permanent re-mortgaging. The intensity of the orange (purple) directly correlates with the speed at which the strategy succeeds (fails). The colorbar is measured in quarters. Lime lines are contour lines, and the markers on the plot represent the values of the eigenvalues of the process at that point, with the white lines marking the transition between one area and another. The markers follow the same legend described in Fig. \ref{fig:phase diagram 1}.}
    \label{fig:pd different initial conditions}
\end{figure}

In Sec. \ref{summary of results} we presented a phase diagram with the following parameters. Average initial equity \(\langle E_0 \rangle = \$30,000\); average initial mortgage \(\langle M_0 \rangle = \$300,000\); value of the installment of the repayment to the bank \(\pi^\star = \$3,000\); \(q = 1 \%\) chance of skipping an installment; $\ell=0.5$; $\mu=0.5$. 
Of course, under unfavorable investment parameters (\(p \sim 0\) and \(s \sim -4\%\)), the house is entirely mortgaged, and the strategy defaults; with favorable parameters (\(p \sim 1\) and \(s \sim +4\%\)), the strategy is strongly successful. We now analyze and present in Fig. \ref{fig:pd different initial conditions} how the phase diagram changes for a wider range of initial conditions (average equity $\langle E_0 \rangle$, average mortgage $\langle M_0 \rangle$, average installment of the repayment to the bank $\langle \pi \rangle$), but keeping the same value of the product of the LTV ratio and risk factor $\ell \mu$\footnote{The official data available on the \href{https://www.abs.gov.au/statistics/economy/price-indexes-and-inflation/total-value-dwellings/latest-release}{Australian Bureau of Statistics} (ABS), updated to December 2023, states that the mean price of residential dwellings is $\$ 933,800$. We chose to reference Australian data because this strategy is already in use there.}:
\begin{itemize}
    \item \textbf{Case $1a$}: $\langle E_0 \rangle = \$300,000 $ and $\langle M_0 \rangle = \$300,000 $ for houses of low value, with fixed $\langle \pi \rangle = \$3,000 $.
    \item \textbf{Case $1b$}: $\langle E_0 \rangle = \$800,000 $ and $\langle M_0 \rangle = \$800,000 $ for houses of high value, with fixed $\langle \pi \rangle = \$3,000 $.
    \item \textbf{Case $2a$}: $\langle E_0 \rangle = \$90,000 $ and $\langle M_0 \rangle = \$900,000 $ for houses of medium value, with fixed $\langle \pi \rangle = \$3,000 $.
    \item \textbf{Case $2b$}: $\langle E_0 \rangle = \$150,000 $ and $\langle M_0 \rangle = \$1,500,000 $ for houses of high value, with fixed $\langle \pi \rangle = \$3,000 $.
    \item \textbf{Case $3a$}:
    $\langle E_0 \rangle = \$90,000 $ and $\langle M_0 \rangle = \$900,000 $ for houses of medium value, with fixed $\langle \pi \rangle = \$10,000 $.
\end{itemize}
\textcolor{black}{\subsubsection{Impact of property value with initial mortgage and equity ratio}
We first analyze the impact of the property value on the strategy outcomes. We provide two examples (Case $1a$-$1b$ and Case $2a$-$2b$) where we compare - \textit{ceteris paribus} - debt recycling strategy outcomes for properties with lower ($a$ - panels) or higher ($b$ - panels).\\
The comparison between Case $1a$ and $1b$ shows the dependence on the initial conditions $\langle E_0 \rangle$ and $\langle M_0 \rangle$ separately, with fixed ratio $c_1=1$; the comparison between Case $2a$ and $2b$ shows the dependence on the initial conditions $\langle E_0 \rangle$ and $\langle M_0 \rangle$ separately, with fixed ratio $c_1=10$.\\
By comparing these cases, we see the effect of different property values (low vs. high) with a fixed mortgage-to-equity ratio (\(c_1\)). Although the overall structure of success and failure regions remains similar, there are subtle but important differences: 
\begin{itemize}
        \item \textbf{Time to outcome}: For higher property values (e.g., Case $1b$ and Case $2b$ compared with Case $1a$ and Case $2a$ respectively), the strategy requires more time to reach either success or failure. In the orange area, where the strategy is successful, an increase in the total value of the house $H$ ($H = E + M$) corresponds to a slightly longer time required to fully repay it, converting its entire value into equity. Meanwhile, again in the case of a higher house value, the time to reach failure in the area where the strategy fails (in purple) is shorter.  Where the strategy is successful, a higher house value means there is more capital to be managed and repaid. This results in a longer time period needed to transform all mortgage into equity, even though the strategy ultimately succeeds. In contrast, in scenarios where the strategy fails, the financial strain of maintaining large mortgage payments against a failing investment strategy becomes unsustainable more quickly, leading to a faster default.
        \item \textbf{Expansion of failure region}: The purple zone, where debt recycling fails, slightly expands with increased property values. This suggests that while debt recycling has potential rewards for high-value properties, it also entails heightened risks, as these properties amplify both gains and losses. In scenarios with unfavorable market conditions, this expanded failure zone underscores the potential for accelerated default due to increased financial strain.
    \end{itemize}
These observations imply that while higher property values can enhance the potential gains from successful debt recycling, they also increase the risks and consequences of failure, requiring a more extended commitment period.
\subsubsection{Impact of Mortgage-to-Equity Ratio}
The comparison between cases $1a-2a$ and $1b-2b$ highlights the effect of changing the mortgage-to-equity ratio \(c_1\) for low and high value properties, respectively.    
    \begin{itemize}
        \item \textbf{Risk of Default in High \(c_1\) Scenarios}: A higher mortgage-to-equity ratio ($c_1$), as seen in Cases $2a$ and $2b$ (independently of the initial house value, high vs. low respectively), results in an expanded default outcome space, extending well into the \MakeUppercase{\romannumeral 4} quadrant. This suggests that even with a good chance of investment gains ($p>0.5$), if the average market value trend is negative ($s<0$), the strategy on average still defaults when the initial debt burden is significantly higher relative to the property’s equity.
        \item \textbf{Implications for debt recycling feasibility}: As we will discuss in Sec. \ref{sec:Policy}, this sensitivity to the \(c_1\) ratio implies that debt recycling strategies are more likely to be successful when borrowers have a lower mortgage-to-equity ratio. Borrowers starting with a high debt relative to equity should be more cautious, as this increases the likelihood of default, even under relatively favorable investment probabilities.
    \end{itemize}
\subsubsection{Effect of Increased Repayment Size}
    The comparison between Case $2a$ and $3a$ shows, \textit{ceteris paribus}, the dependence on the value of repayment installments $\langle \pi \rangle$, increased in these scenarios from \$3,000 to \$10,000.
     \begin{itemize}
        \item \textbf{Enlarged success region}: Increasing the repayment significantly widens the orange (success) zone even in the \MakeUppercase{\romannumeral 3} quadrant where $s<0$ and $p<0.5$, indicating that higher repayments improve the likelihood of successful debt recycling. This is because larger repayments reduce the outstanding mortgage more quickly, thereby increasing equity and reducing debt exposure over time.
        \item \textbf{Reduced default risk}: The purple (failure) zone shrinks when the repayment is increased, as the faster reduction of debt mitigates the risk of default under a range of market conditions. Financially, this suggests that borrowers with the capacity for higher repayments are more likely to sustain a debt recycling strategy successfully, as they can more readily manage and reduce debt exposure.
    \end{itemize}
To summarize, the most influential parameters are the mortgage-to-equity initial ratio, rather than the individual values of mortgage and equity themselves -- meaning the strategy's outcome does not depend strongly on whether we are dealing with a low or high-value house -- and the size of the monthly repayments. The implications for policymakers and consideration on feasibility for borrowers will be discussed in Sec. \ref{sec:Policy}.}

\textcolor{black}{\subsubsection{Dependence on risk factor and LTV ratio}}
Now, we analyze the phase diagrams for different values of $\ell$ and $\mu$. As showed in Sec. \ref{subsec:average}, the product $\ell \mu \in [0,1]$ is the critical factor influencing the amplitude of the fluctuations of $E_t$ and $M_t$. 

\begin{figure}[H]
    \centering
    
    \begin{subfigure}{.5\textwidth}
        \centering \includegraphics[width=\linewidth]{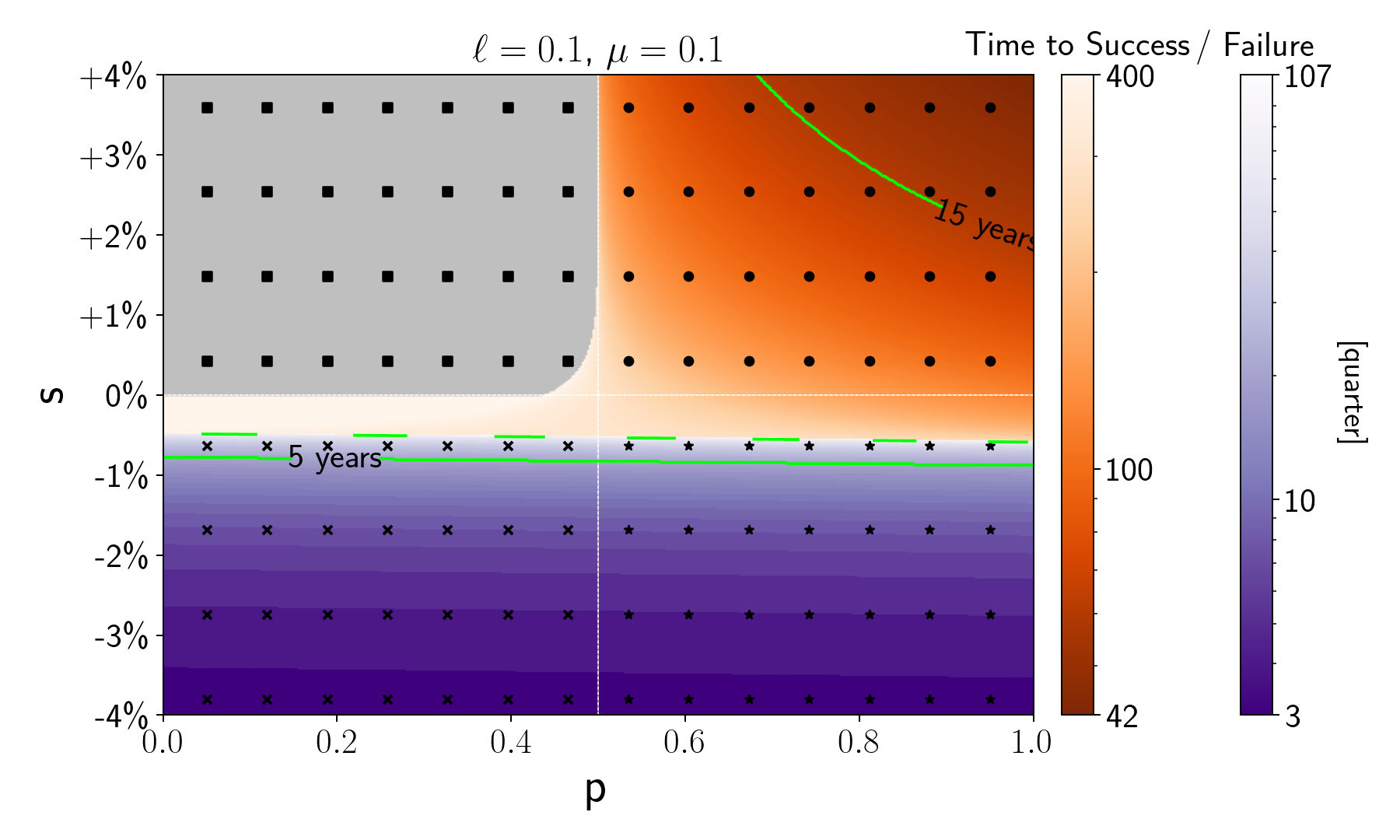}
        \subcaption{Case $4a$: $\ell = 0.1$; $\mu = 0.1$}
        \label{fig:sub6}
    \end{subfigure}\hfill
    \begin{subfigure}{.5\textwidth}
        \centering
        \includegraphics[width=\linewidth]{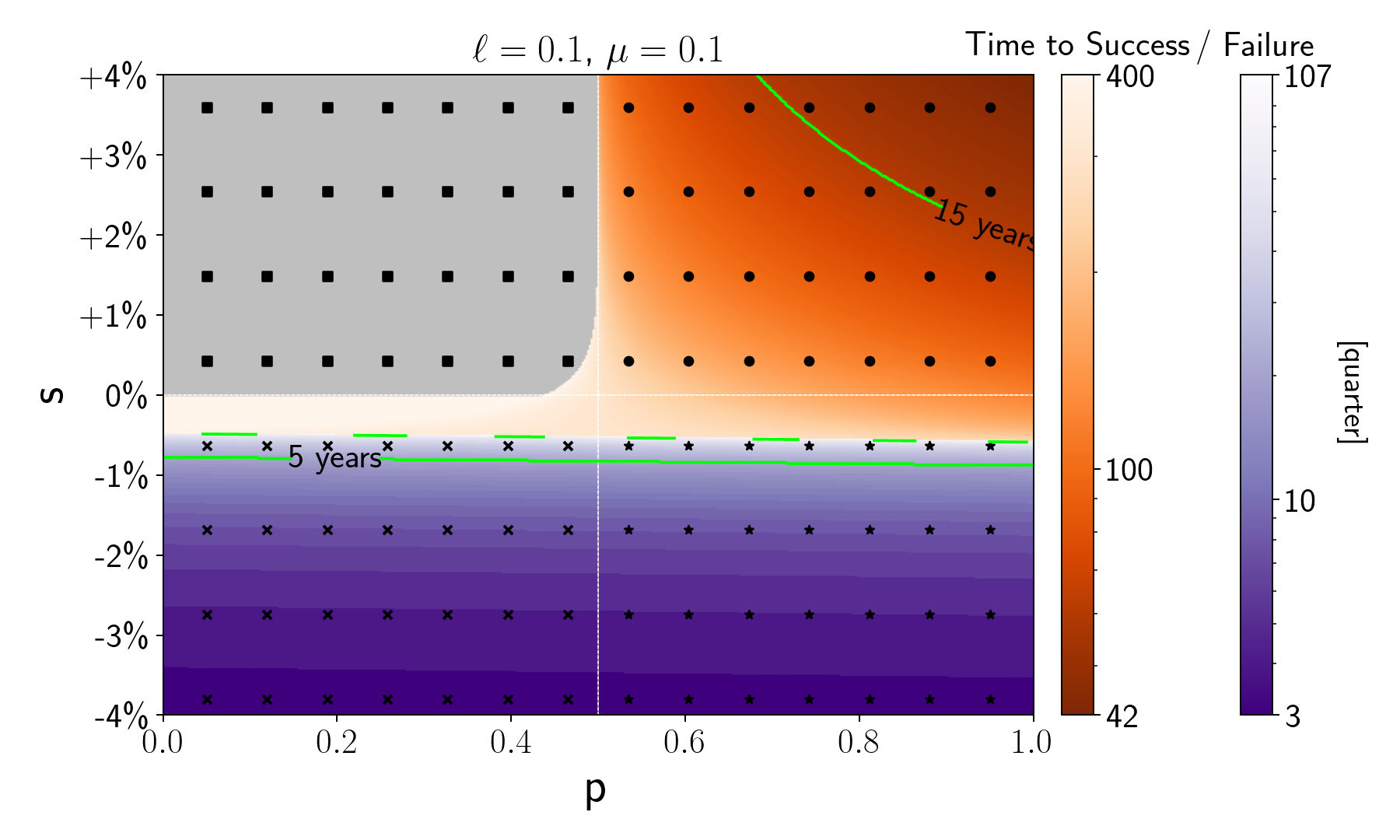}
        \subcaption{Case $4b$: $\ell = 0.9$; $\mu = 0.9$}
        \label{fig:sub7}
    \end{subfigure}
    
    \vspace{5mm} 
    
    \begin{subfigure}{.5\textwidth}
        \centering
        \includegraphics[width=\linewidth]{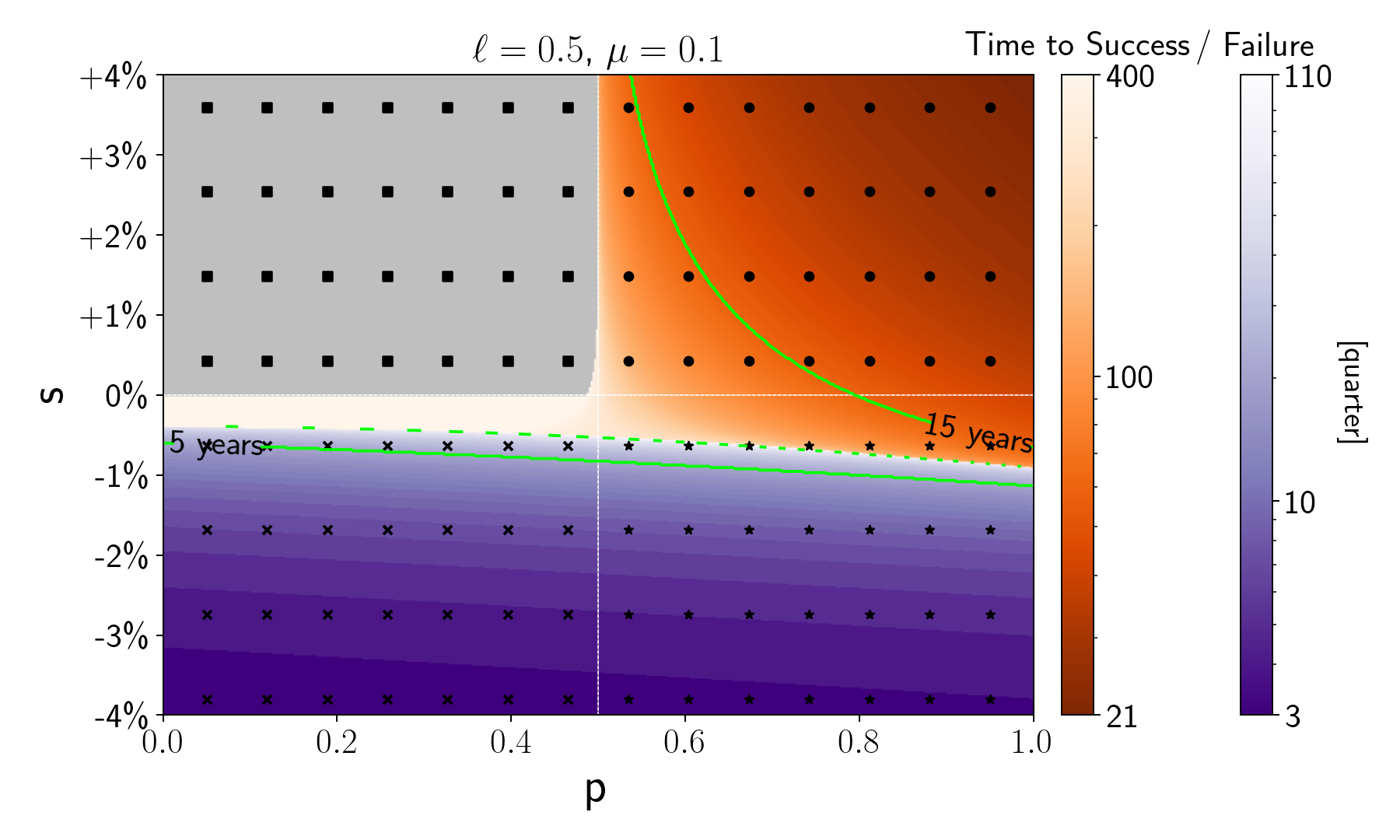}
        \subcaption{Case $5a$: $\ell = 0.5$; $\mu = 0.1$}
        \label{fig:sub8}
    \end{subfigure}\hfill
    \begin{subfigure}{.5\textwidth}
        \centering
        \includegraphics[width=\linewidth]{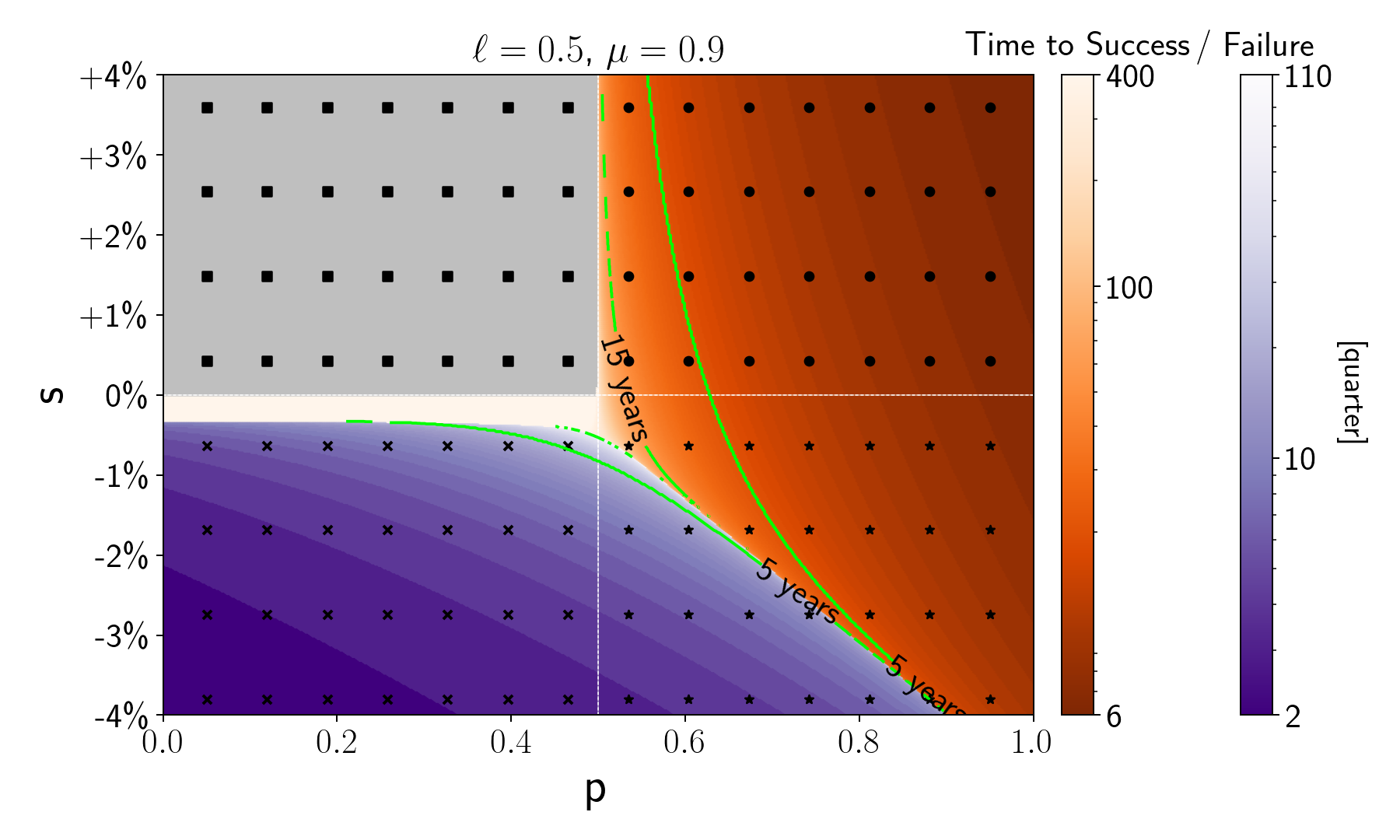}
        \subcaption{Case $5b$: $\ell = 0.5$; $\mu = 0.9$}
        \label{fig:sub9}
    \end{subfigure}
    
    \caption{Phase diagrams for different values of LTV ratio $\ell$ and risk factor $\mu$. Area in orange represents where debt recycling is (either weakly or strongly) successful; area in purple represents regions in the plane where debt recycling fails; area in gray denotes the zone of permanent re-mortgaging. The intensity of the orange (purple) directly correlates with the speed at which the strategy succeeds (fails). The colorbar is measured in quarters. Lime lines are contour lines, and the markers on the plot represent the values of the eigenvalues of the process at that point, with the white lines delineating the transition between one area and another. The markers follow the same legend described in Fig. \ref{fig:phase diagram 1}.}
    \label{fig:pd different l u}
\end{figure}

For the following phase diagrams (see Fig. \ref{fig:pd different l u}) we fix the values of the initial conditions to a house of medium value with $\langle E_0 \rangle = \$90,000 $, $\langle M_0 \rangle = \$900,000 $, $\langle \pi \rangle = \$3,000 $, and we explore the following combinations:
\begin{itemize}
    \item \textbf{Case $4a$}: $\ell = 0.1$ and $\mu = 0.1$ $\rightarrow\ell \mu = 0.01$.
    \item \textbf{Case $4b$}: $\ell = 0.9$ and $\mu = 0.9$ $\rightarrow\ell \mu = 0.81$
    \item \textbf{Case $5a$}: $\ell = 0.5$ and $\mu = 0.1$ $\rightarrow\ell \mu = 0.05$.
    \item \textbf{Case $5b$}: $\ell = 0.5$ and $\mu = 0.9$ $\rightarrow\ell \mu = 0.45$.
\end{itemize}
The comparison between Case $4a$ and $4b$ shows the dependence on the parameter $\ell \mu$ combined; the comparison between Case $5a$ and $5b$ shows instead the dependence on the risk factor only when we decide to keep the other one (LTV ratio) constant.

Comparing Case $4a$ with Case $4b$, we observe that increasing the value of the product \(\ell \mu\) yields two effects: it reduces the default zone in the \MakeUppercase{\romannumeral 4} quadrant and accelerates the strategy outcome. With higher \(\ell \mu\), the system becomes more sensitive to changes in parameters $p$ and $s$. Small variations in these parameters can lead to significant changes in the outcome, as indicated by the steeper gradients in Case $4b$. While higher \(\ell \mu\) values reduce the risk of default and accelerate success, they also imply that rapid changes of the outcome of the strategy are more likely.

Additionally, one may note that only low values of \(\ell \mu\) may lead to a weakly successful strategy in the \MakeUppercase{\romannumeral 2} quadrant, as outlined in Tab. \ref{tab:dynamic outcomes}.

The straightforward comparison between Case $5a$ and Case $5b$, where only the risk factor \(\mu\) increases, leads to qualitatively similar conclusions.

We can either increase the LTV ratio parameter, the risk parameter, or both simultaneously for an amplified effect: the result is a faster process, and in the \MakeUppercase{\romannumeral 4} quadrant, default is shifted back in favor of success. This effect is amplified as the total value of the house decreases.

\subsubsection{Velocity of the outcome}

The phase diagrams are color-coded to indicate whether debt recycling represents a strong/weak winning or losing strategy, with the intensity of the color reflecting the speed at which the average equity/mortgage process reaches the absorbing boundary first. In this section we present slice plots, an analysis of the phase diagrams by fixing the variable $p$ ($s$) on the $x-$axis ($y-$axis) and examining the variations in first hitting time along the other dimension. 

\begin{figure}[h]
    \centering
    \includegraphics[width=0.7\textwidth]{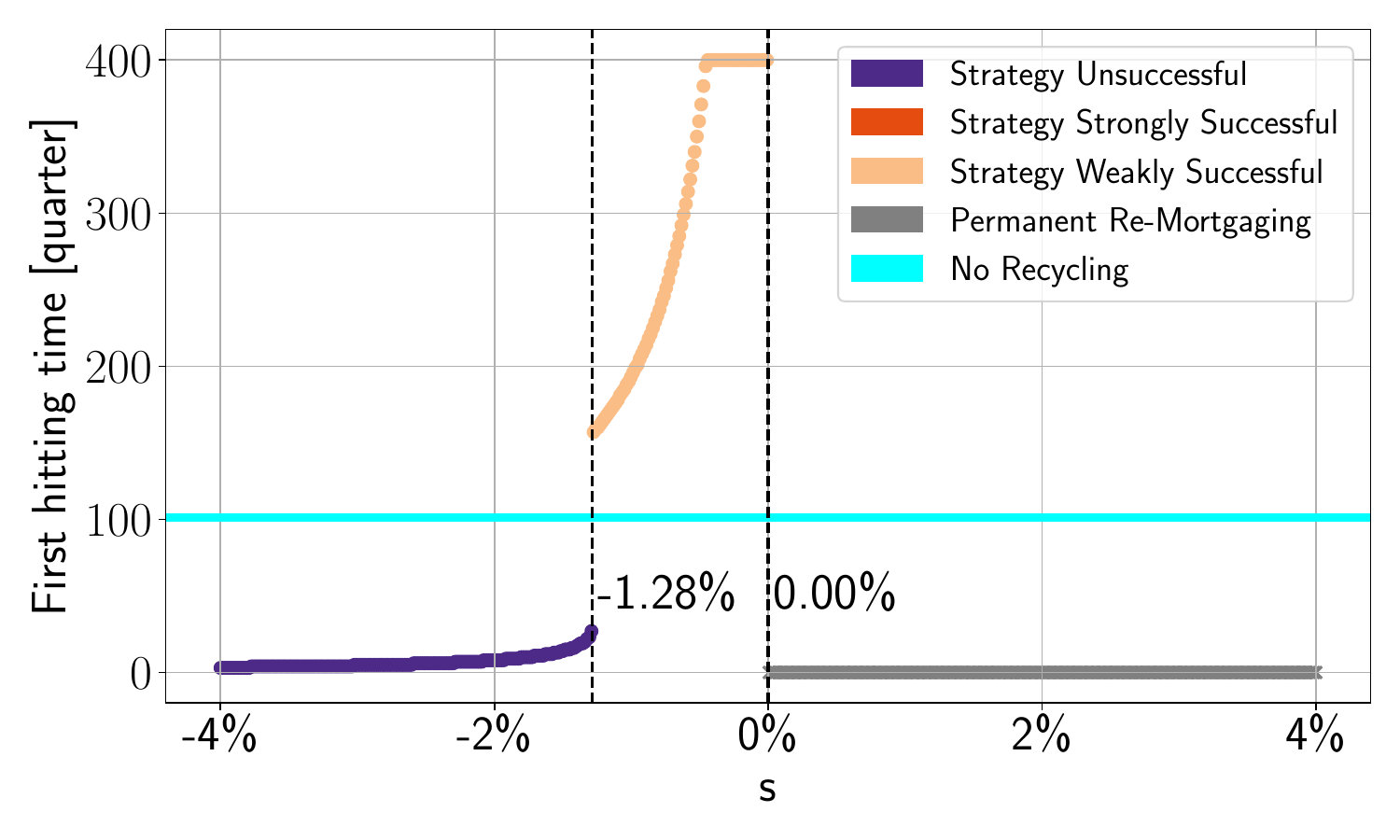}
    \caption{First hitting time varying $s$, with $\ell =0.5$, $\mu =0.5$, $p=0.4$.}
    \label{fig:compare_t_1}
\end{figure}

We show in Fig. \ref{fig:compare_t_1} the plot of the first hitting time for $\ell =0.5$, $\mu =0.5$, and $p=0.4$, the same parameters chosen for the plot in Fig. \ref{fig:translation s}.
Taking Fig. \ref{fig:phase diagram 1} as a reference, we travel along the vertical line at $p=0.4$ from bottom to top, increasing the value of $s$. At first, we are deeply into the default (purple) region, but increasing $s$ the time it takes for the average equity to hit zero increases. This is in clear agreement with the housing market becoming progressively more favorable, as also evidenced by the first hitting root of the average equity curves in Fig. \ref{fig:translation s}.
At \(s = -1.28\%\), we cross the purple-orange boundary in the phase diagram in Fig. \ref{fig:phase diagram 1}, meaning that the strategy turns into a (weakly) successful one (light orange). The time required to gain ownership of the house indeed turns out to be longer than without any strategy -- just paying back the lender on a monthly basis (cyan line). The time required to repay the debt without applying the strategy is computed assuming a constant repayment amount \(\pi^\star\) to the bank at each time step. With an initial mortgage \(M_0\) of \$300,000 and \(\pi^\star\) set at \$3,000, the required repayment period is $100$ time steps (or $25$ years). For visual clarity, in the phase diagram we have limited the maximum color intensity to \(t = 400\) quarters; instances where debt recycling is effective beyond $100$ years are here represented at \(t = 400\), as well. At $s=0\%$ we further cross the orange-gray boundary and enter the phase of permanent re-mortgaging.

\begin{figure}[h]
    \begin{minipage}{0.5\textwidth}
        \centering \includegraphics[width=\textwidth]{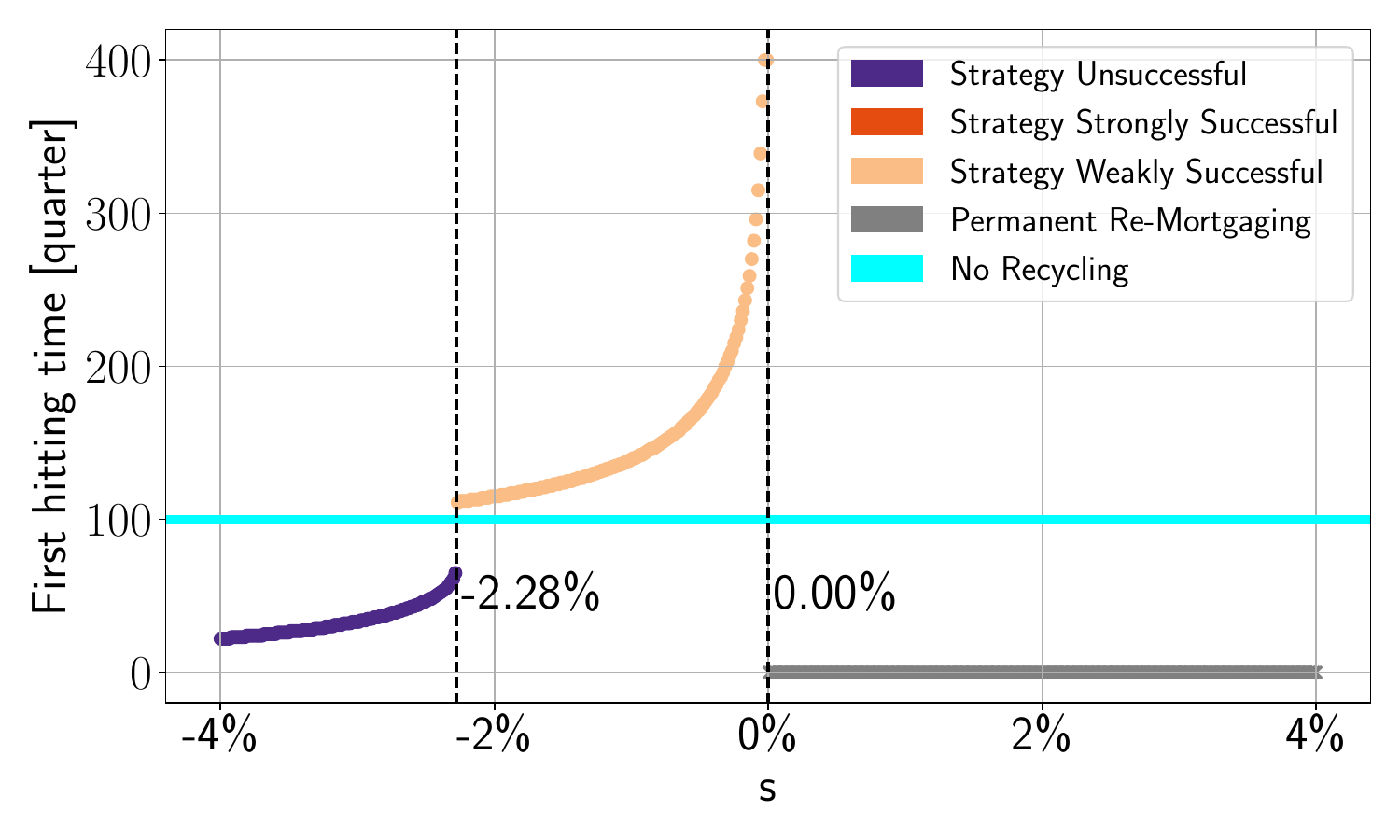} \subcaption{First hitting time of Case $1a$, with $p=0.49$.} \label{fig:compare_2}
    \end{minipage}
    \begin{minipage}{0.5\textwidth}
        \centering \includegraphics[width=\textwidth]{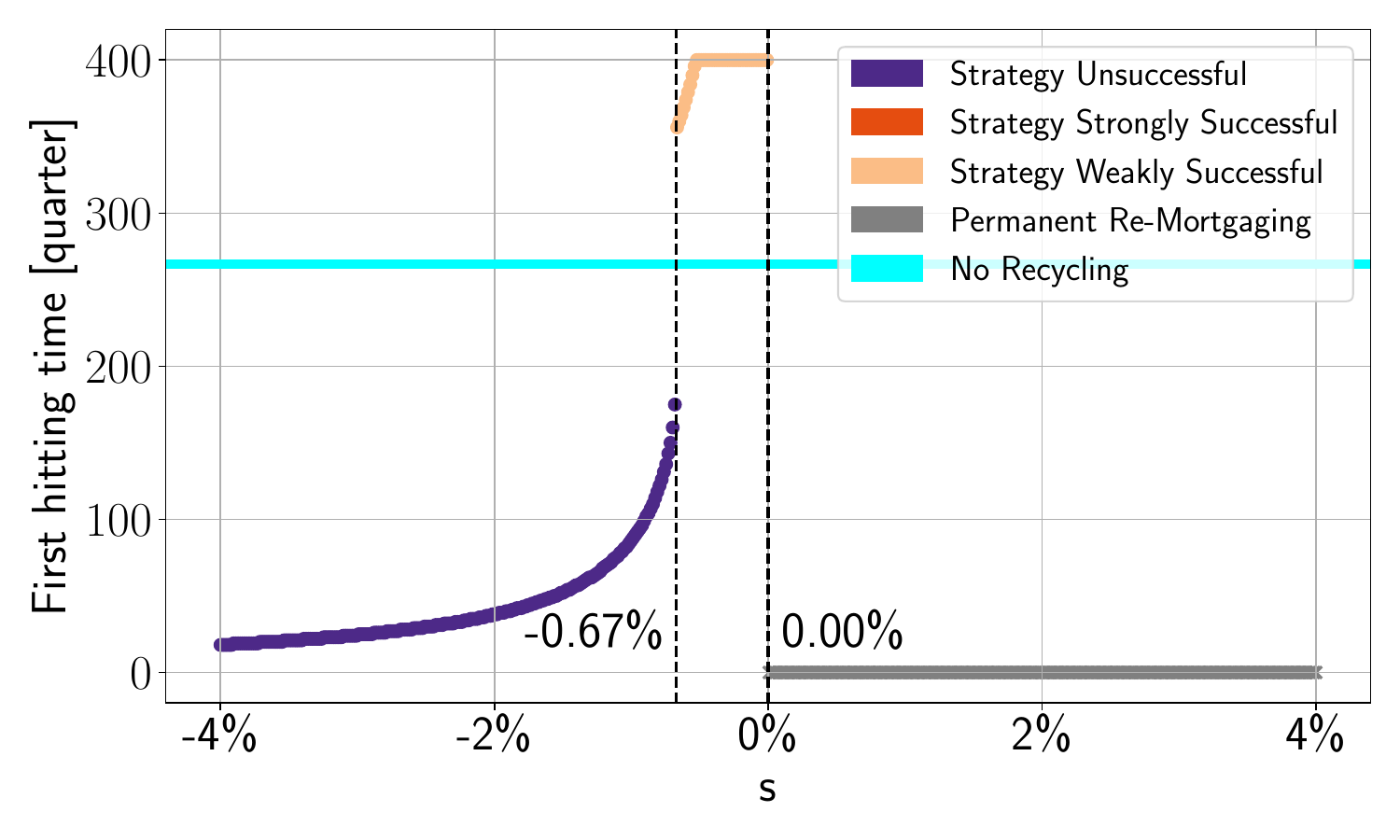} \subcaption{First hitting time of Case $1b$, with $p=0.49$.}\label{fig:compare_3}
    \end{minipage}
    \caption{First hitting time varying $\langle E_0 \rangle$ and $\langle M_0 \rangle$.} 
    \label{fig: compare_4}
\end{figure}

To further validate the findings presented in Section \ref{subsec:phase diagram}, we show in Fig. \ref{fig: compare_4} a comparison of the process speed between Cases $1a$ and $1b$, reproduced here for ease of reference:
\begin{itemize}
    \item \textbf{Case $1a$}: $\langle E_0 \rangle = \$300,000 $ and $\langle M_0 \rangle = \$300,000 $ for houses of low value, with fixed $\langle \pi \rangle = \$3,000 $.
    \item \textbf{Case $1b$}: $\langle E_0 \rangle = \$800,000 $ and $\langle M_0 \rangle = \$800,000 $ for houses of high value, with fixed $\langle \pi \rangle = \$3,000 $.
\end{itemize}

The mortgage-to-equity ratio \(c_1\) is held constant but we have different initial conditions of equity and mortgage. Setting $p=0.49$ in Fig. \ref{fig:sub1} and \ref{fig:sub2} and increasing $s$, we confirm that the purple region (strategy failing) is stretched out for higher-value houses, with a purple-to-orange threshold moving up from \(s=-2.28\%\) to \(s=-0.67\%\)). It is noteworthy that in both cases, the strategy past the threshold is only weakly successful, corroborating the result found in Section \ref{summary of results} that \(p=0.5\) is the threshold between weak and strong success.

\begin{figure}[h]
    \begin{minipage}{0.5\textwidth}
        \centering \includegraphics[width=\textwidth]{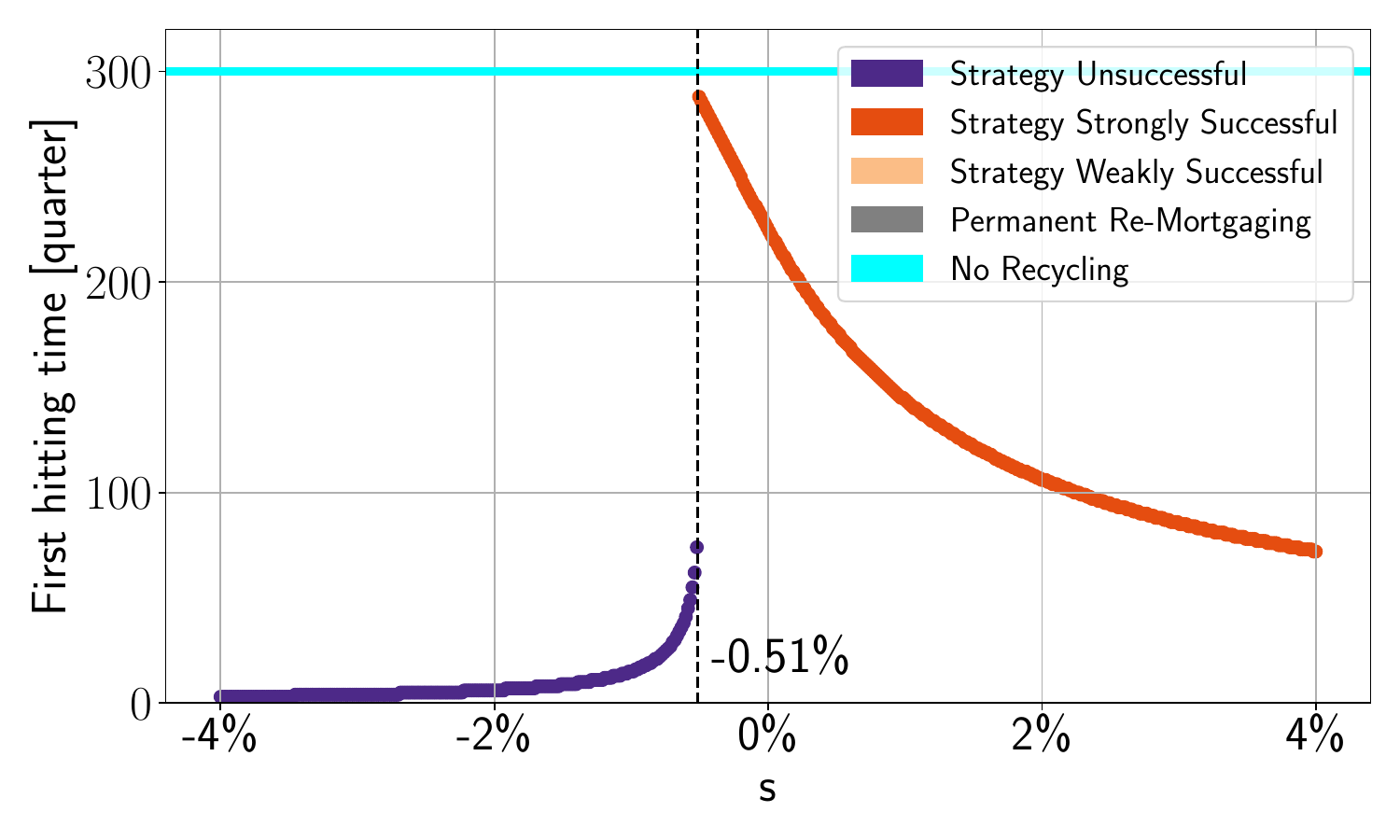} \subcaption{First hitting time of Case $4a$, with $p=0.6$.} \label{fig:compare_5}
    \end{minipage}
    \begin{minipage}{0.5\textwidth}
        \centering \includegraphics[width=\textwidth]{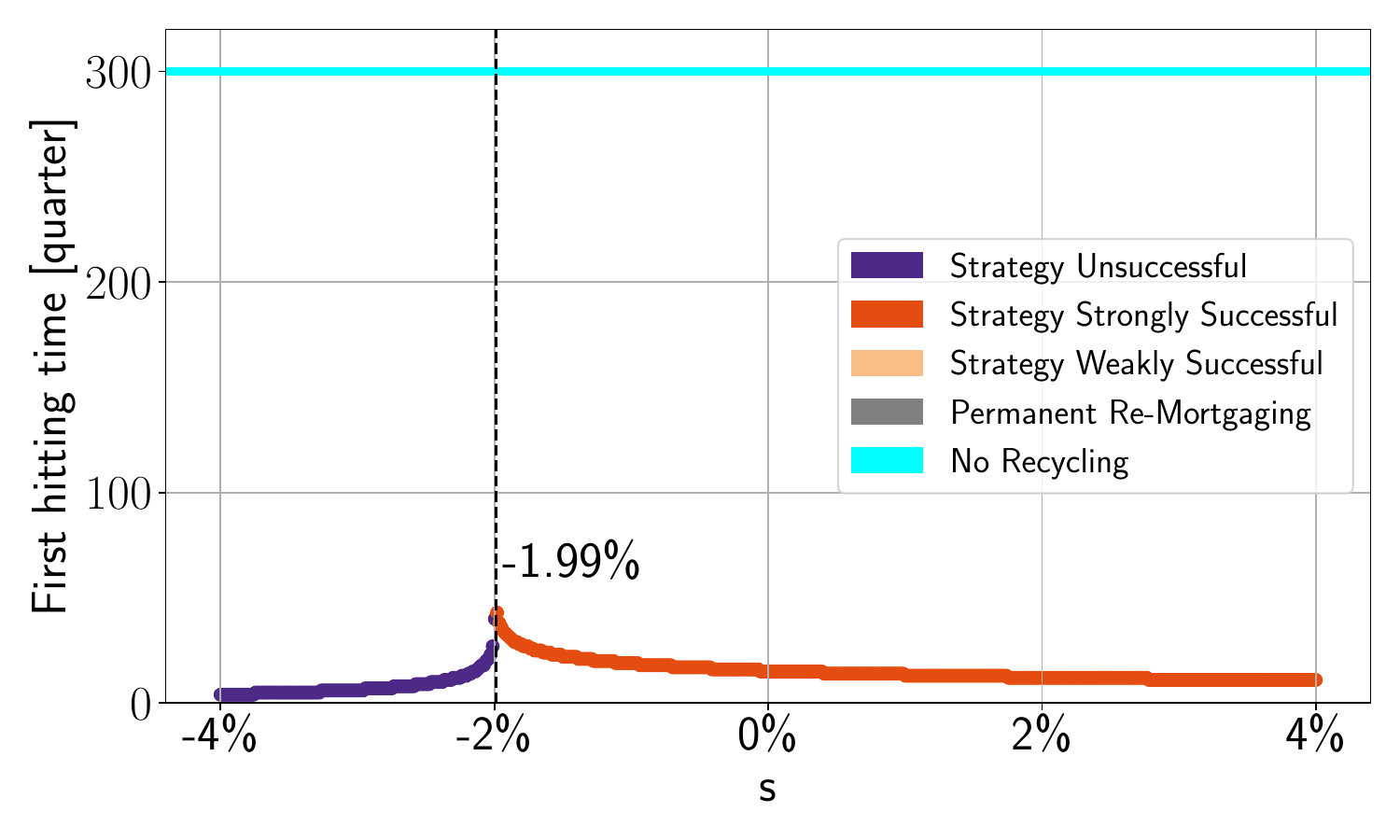} \subcaption{First hitting time of Case $4b$, with $p=0.6$.}\label{fig:compare_6}
    \end{minipage}
    \caption{First hitting time varying $\ell \mu$. 
    } 
    \label{fig: compare_7}
\end{figure}

In Fig. \ref{fig: compare_7}, we present a comparison of the process speed between Cases $4a$ and $4b$ (see Fig. \ref{fig:sub6} and \ref{fig:sub7} for reference), where $p=0.6$. As expected, an increase in the product \(\ell \mu\) results in a reduction of the default region (transitioning from a threshold of \(s=-0.51\%\) to \(s=-1.99\%\)) and a noticeable acceleration of the outcome. In this scenario, the strategy is strongly successful beyond the threshold, as gaining ownership via regular monthly repayments would take much longer (see the cyan constant line). 

\begin{figure}[h]
    \centering
    \includegraphics[width=0.7\textwidth]{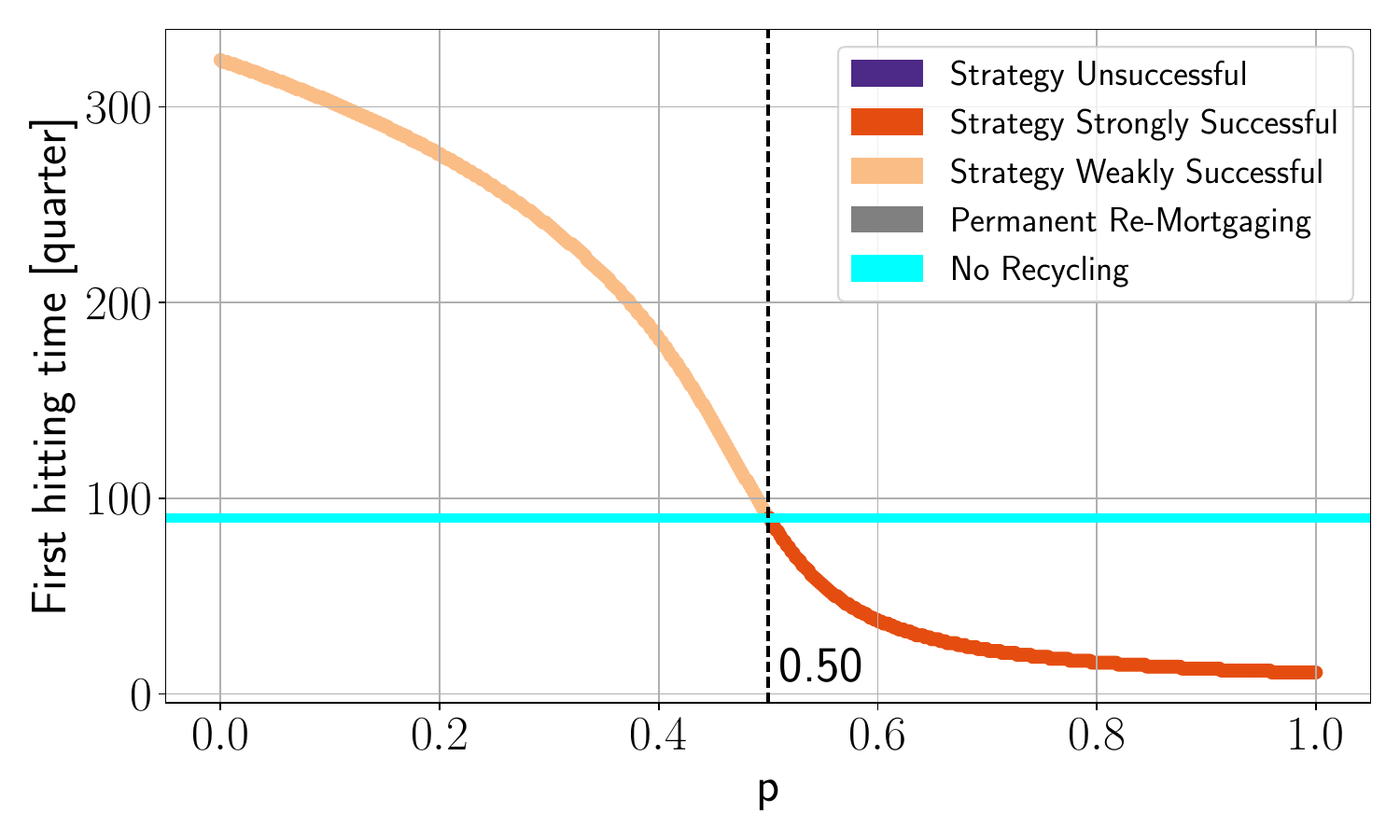}
    \caption{First hitting time of Case $3a$, with $s=-1\%$.}
    \label{fig:compare_9}
\end{figure}

Finally, to show that \(p=0.5\) is the threshold between weak and strong success in the majority of cases, we show in Fig. \ref{fig:compare_9} a phase diagram for Case $3a$, focusing just on the success area. This case outlines a realistic and not too risky situation for a mid-value house: \(\langle E_0 \rangle = \$90,000\); \(\langle M_0 \rangle = \$900,000\); \(\langle \pi \rangle = \$10,000\); \(\ell =0.5\); \(\mu =0.5\); \(s=-1\%\).

Below, in Fig. \ref{fig: compare_10}, we present two edge-case scenarios: in the Left Panel, a case where weak success occurs even for \(p>0.5\), and in the Right Panel, a scenario where strong success is achieved for \(p<0.5\).

\begin{figure}[h]
    \begin{minipage}{0.5\textwidth}
        \centering \includegraphics[width=\textwidth]{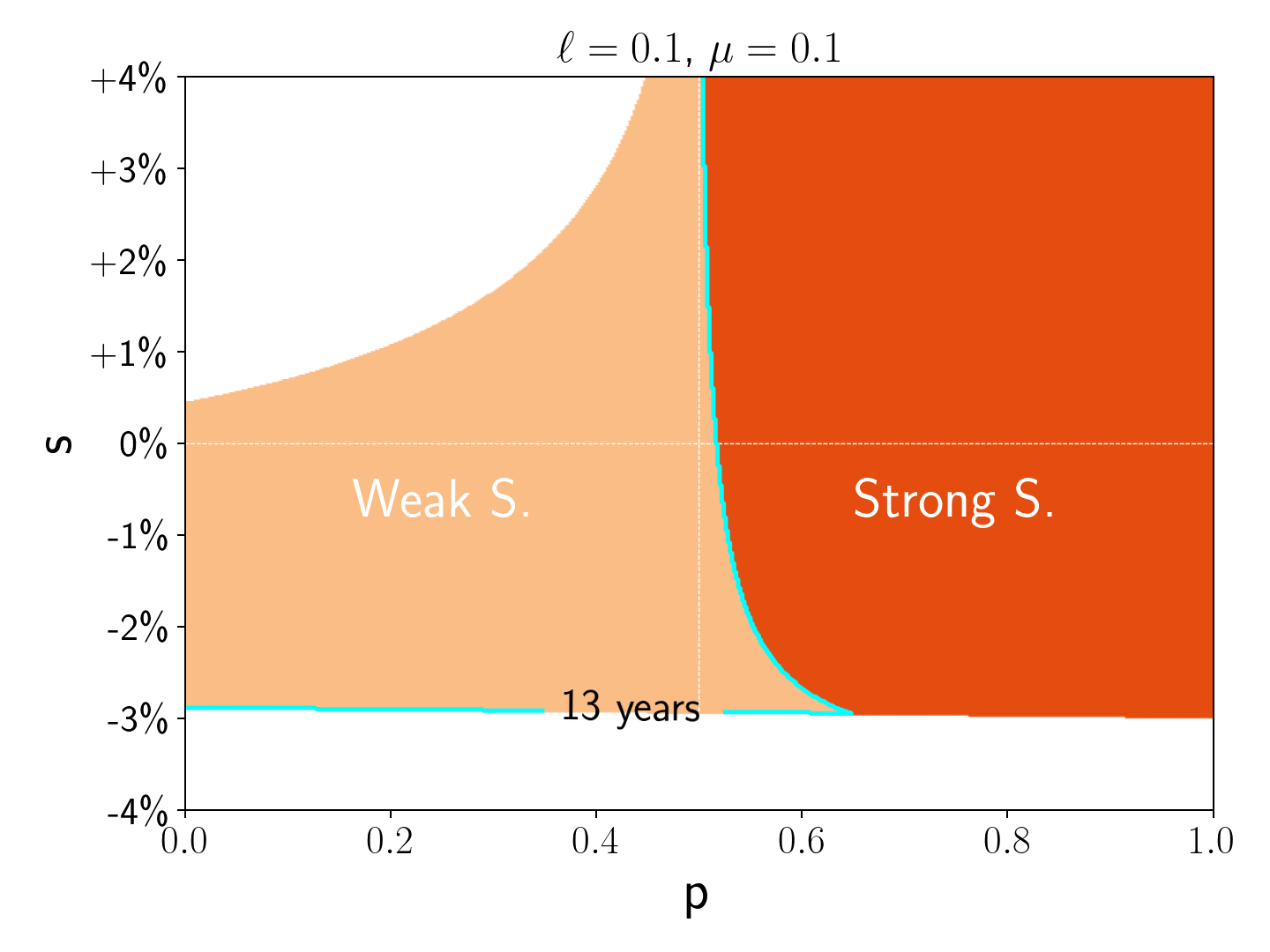}\subcaption{\(\langle E_0 \rangle = \$15,000\); \(\langle M_0 \rangle = \$150,000\). \\ In light orange, weak success; in dark orange, strong success. In cyan, the contour line $t^* = t_{\text{ no recycling}}$.} \label{fig:compare_11}
    \end{minipage}
    \hspace{10pt}
    \begin{minipage}{0.5\textwidth}
        \centering \includegraphics[width=\textwidth]{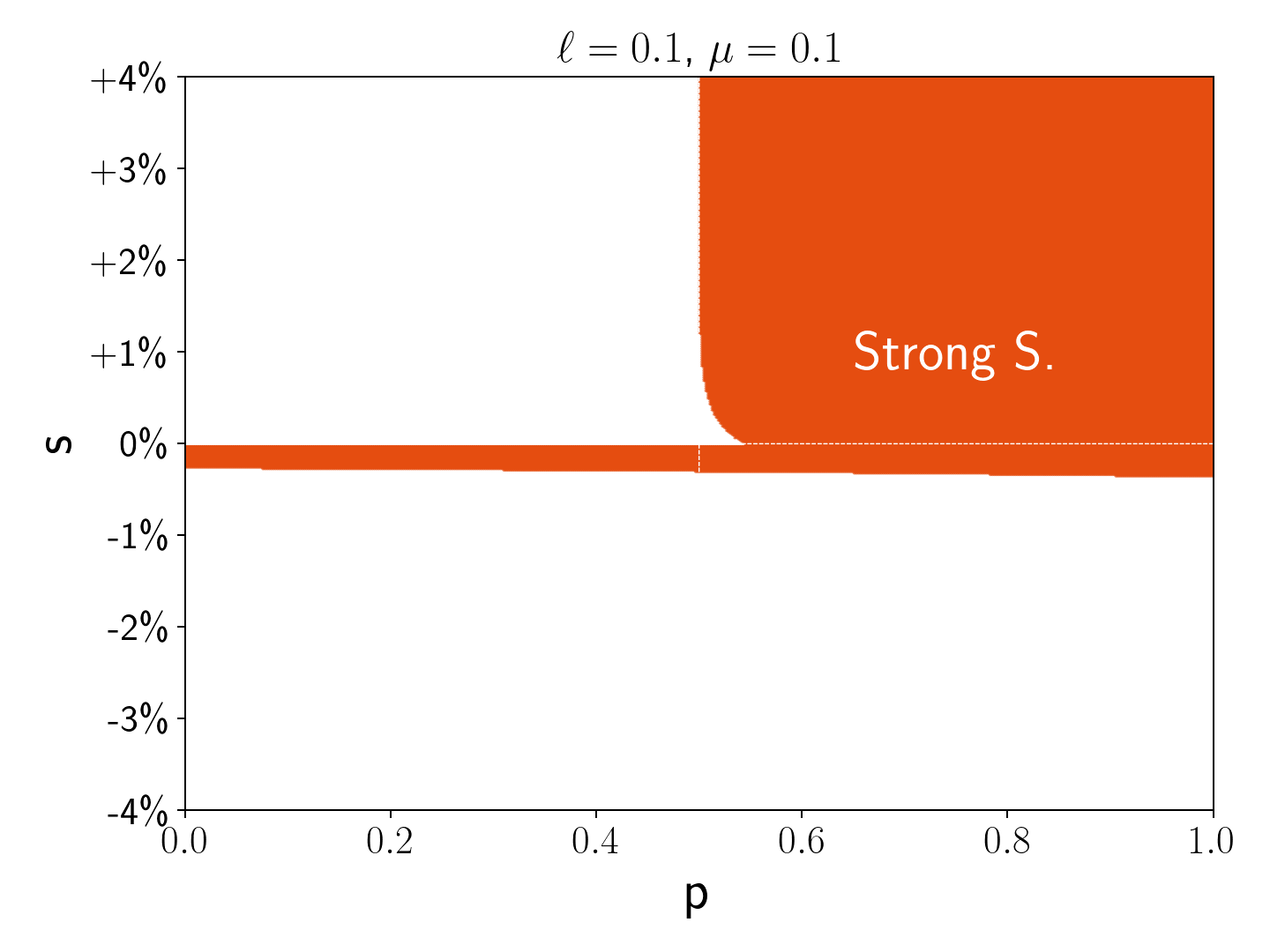} \subcaption{\(\langle E_0 \rangle = \$150,000\); \(\langle M_0 \rangle = \$1,500,000\). \\ In dark orange, strong success.}\label{fig:compare_12}
    \end{minipage}
    \caption{Scenarios where $p=0.5$ is not the threshold between strong and weak success.} 
    \label{fig: compare_10}
\end{figure}

The Left Panel is defined by parameters where the equity and mortgage values are relatively low: \(\langle E_0 \rangle = \$15,000\); \(\langle M_0 \rangle = \$150,000\); \(\pi^\star = \$3,000\); \(q=1\%\); \(\ell =0.1\); \(\mu=0.1\). In contrast, the Right Panel features significantly higher values: \(\langle E_0 \rangle = \$150,000\); \(\langle M_0 \rangle = \$1,500,000\); \(\pi^\star = \$3,000\); \(q=1\%\); \(\ell =0.1\); \(\mu=0.1\). Both scenarios represent over-collateralization of the house, as evidenced by the consistent mortgage-to-equity ratio \(c_1=10\) across both panels, and the product \(\ell \mu\) remains low in each. The key distinction between these cases lies in the total value of the house, which is on the order of \(10^5\) in the Left Panel and \(10^6\) in the Right Panel, reflecting the scale of financial involvement and potential risk exposure. In the Right Panel, monthly repayments should be properly rescaled (e.g., increased) to account and hedge for the increased risk of the debt recycling strategy -- in which case the two phase diagram would become more similar. The scale of the problem, therefore, is also an important parameter to consider to assess the success of the strategy, and a re-evaluation of all parameters' involved in the initial mortgage issuance (e.g., monthly repayment schedule) should be carried out.

\color{black}
\section{Policy Implications}\label{sec:Policy}

\textcolor{black}{Debt recycling, as analyzed through our model, presents a complex financial mechanism with considerable potential for both accelerating mortgage repayment and enhancing financial flexibility. However, its inherent risks necessitate a structured, adaptable regulatory approach that can both support viable use cases and provide robust safeguards against the heightened risk of default or unsustainable leverage. Policymakers and regulators are expected to play a central role in structuring the framework within which debt recycling can operate safely, balancing the incentives for growth with a robust system of warnings and checks to protect both individual borrowers and the financial system at large.}

\textcolor{black}{Our findings highlight the need for a tiered regulatory framework that dynamically responds to the market conditions mostly responsible for the success or failure of debt recycling strategies. During periods of housing market strength, indicated in our model by the parameter \(\lambda_1 > 1\), regulatory guidelines may be implemented and calibrated to support the strategy’s viability. In these conditions, the positive performance of real estate assets makes a favorable case for debt recycling, as equity can be leveraged efficiently and investments are more likely to yield returns that accelerate mortgage repayments. Regulatory leniency and restraint in such periods could for instance involve moderate LTV ratio allowances, with an emphasis on supporting investment-backed debt products, provided that comprehensive risk disclosures are mandated. Borrowers should in any case be required to undertake ``stress test'' exercises that assess their ability to manage both primary and investment-linked debt obligations, ensuring that any downturn in either the housing or investment markets does not directly lead to insolvency. Disclosure requirements should be structured to make the correlation between market risks fully transparent to borrowers, helping them understand how adverse scenarios may impact their financial health in the longer run.}

\textcolor{black}{In contrast, during market downturns or periods where \(\lambda_1 < 1\) signals declining home values, regulatory oversight must tighten to mitigate risks of over-leverage and potential default. For debt recycling to remain viable under these conditions, stricter borrowing criteria and higher LTV ratio caps should be applied.}

\textcolor{black}{It should be noted, however, that such approaches are inherently procyclical, with the potential to amplify both market upswings and downturns, particularly in larger or highly interconnected markets, which could pose systemic challenges.}
\textcolor{black}{Increasing capital requirements for financial institutions offering debt recycling products can reduce systemic risk, as these higher reserves would act as a buffer against widespread market shocks. More frequent portfolio reviews should be required for both borrowers and lending institutions, along with restrictions or phased limits on further equity extraction in volatile conditions, thus reducing the likelihood of excessive leveraging during unfavorable market trends.}

\textcolor{black}{Several key policy levers within the debt recycling framework may enhance the strategy’s resilience across different economic conditions. These include the LTV ratio, the risk factor associated with the underlying investment, and the repayment schedule. Adjusting the LTV ratio dynamically -- allowing greater flexibility in favorable markets while capping ratios in downturns -- may prevent borrowers from becoming overly exposed to declining equity. Our model suggests that in declining markets, policymakers should restrict LTV ratios to lower levels, which would help contain default risk by reducing the amount of equity that can be extracted as collateral. Furthermore, policy interventions should establish diversified and standardized risk assessment frameworks for the investment assets used in debt recycling, especially for high-risk ventures. Lenders should probably be required or induced to diversify the portfolios of investments that back these equity-extraction strategies, which would help distribute and mitigate exposure to risk. Standardizing investment categories with differential risk weightings would also ensure that borrowers only access debt recycling for investments that meet approved risk criteria, thereby minimizing individual exposure to high-volatility assets. A closer look -- also from the theoretical side -- ought to be paid to likely correlations and feedback loops arising between the stock and the housing markets, where virtuous and vicious cycles are likely to come hand in hand, requiring a more holistic regulatory approach.}

\textcolor{black}{The larger ``risk appetite'' of borrowers and institutions in Australia is, indeed, responsible for debt recycling being regarded as a particularly attractive strategy there. Indeed, LTV ratios in the UK are typically in the range of 60\% to 90\%, with very few loans with an LTV up to the maximum of 95\% available. Similarly, the US is also rather risk averse, with LTV generally capped at 80\% and private mortgage insurance being generally required if the LTV ratio goes above that. New Zealand is even more risk averse, with very few loans with LTV ratios above 80\%.  In contrast, Australia is very risk-prone, and regularly allows and issues loans with LTV ratios of 100\% and over \cite{FSBEU,BOE}. This creates an economic climate ripe for debt recycling, as shown by our analysis, and may help explain why Australia is currently the only country where this strategy is seen as generally viable and regularly offered as an option.}

\textcolor{black}{Monthly repayment schedules also serve as a powerful tool for modulating the risk level involved in debt recycling strategies, this time down to the 'individual borrower' level. Policymakers could impose minimum repayment requirements \cite{payment size} that adjust based on prevailing market conditions and to the borrower’s debt-to-income (DTI) ratio \cite{Debelle}, thus preventing borrowers from deferring payments and accumulating debt burdens that may become unsustainable. Minimum repayment requirements would ensure that borrowers maintain a stable debt service capability, which our model suggests is crucial for avoiding default scenarios, particularly during the ``permanent re-mortgaging'' phase. This repayment flexibility should be embedded into the regulatory framework to allow for adjustments in repayment terms that respond to significant shifts in the borrower’s financial situation or the housing market, such as permitting lower payments in economic downturns.}

\textcolor{black}{Beyond regulatory controls, systemic risk management mechanisms are vital for preventing cyclical or permanent debt dependencies that may arise from repeated re-mortgaging phases. To this end, our model identifies specific phase transition lines and discontinuities that policymakers should monitor closely to preemptively intervene when borrowers are likely to enter the cycle of continuous re-mortgaging, particularly in the second quadrant where market conditions are ambivalent. ``Circuit breakers'', such as imposing limits on the duration or frequency of refinancing, could prevent debt recycling from devolving into an endless loop of re-mortgaging phases. Mandatory periods for equity building between refinancing rounds would ensure that the strategy leads to real progress toward ownership rather than reliance on recurrent refinancing. Progressive restrictions on LTV caps for repeated debt recycling engagements could also help mitigate the risk of default by enforcing stricter limits each time the borrower opts to extract additional equity.}

\textcolor{black}{In addition to market-driven regulatory interventions, consumer protection measures are essential to ensure that borrowers fully understand the complexities and risks associated with debt recycling. Mandatory financial literacy programs would equip prospective participants in the scheme with a realistic understanding of how debt recycling strategies behave across different market phases. Disclosure guidelines should be crafted to transparently communicate how the phases outlined in our model -- success (weak and strong), default, and permanent re-mortgaging -- apply to the borrower’s specific initial circumstances, ensuring that participants are aware of the potential for rapid shifts between these states based on the prevailing economic environment. Suitability requirements could involve a more granular assessment of borrower profiles, ensuring that debt recycling is only offered to individuals with the financial capability and risk tolerance to manage such complex arrangements. Regular reassessment of borrower circumstances would allow for adjustments in strategy recommendations, encouraging borrowers to exit the strategy in deteriorating market conditions or if they experience financial hardships.}

\textcolor{black}{Finally, tax policy adjustments (currently not considered in our model) could support debt recycling as a sustainable financial instrument, while maintaining overall market stability. Calibrating deductibility rates for investment-related interest expenses based on holding periods may incentivize longer-term investments rather than speculative or short-term ventures. Progressive tax treatments for recurring equity extractions would prevent borrowers from relying excessively on debt recycling, while preserving its benefits for strategic, growth-oriented use. Tax incentives could also be structured to favor responsible LTV ratios, with diminished tax advantages for higher-risk investment categories, thereby encouraging more prudent financial decisions.}

\textcolor{black}{Overall, debt recycling can serve as a valuable financial strategy under the right regulatory framework, offering significant benefits when managed carefully. However, its successful implementation requires a fine-tuned combination of flexible regulations, risk assessment frameworks, and consumer protection mechanisms. Policymakers should focus on creating conditions that enhance the probability of achieving a ``strongly successful'' outcome, as indicated in our model, while maintaining stringent safeguards against default scenarios. By aligning policy levers with the specific parameters that our model identifies as critical -- such as the LTV ratio, investment risk, repayment schedules, and consumer risk-proneness -- regulators can turn debt recycling into a safer, more productive element within the financial landscape, allowing it to flourish outside its currently geographically constrained situation.}

\color{black}

\section{Conclusions and Outlook}\label{sec:concl}

We have considered a simple dynamical model for the joint time evolution of equity $E_t$ and mortgage balance $M_t$ in the case where an aggressive ``debt recycling'' strategy is employed. Debt recycling consists in a mortgage holder investing in an income-producing asset backed by their equity (the current market value of the fraction of the house they have already repaid), with the aim to hopefully pay off their home loan quicker. When interest rates rise, the immediate benefit of paying down a mortgage becomes evident due to the higher cost of borrowing. However, debt recycling emerges as a powerful strategy, leveraging higher interest rates to maximize tax deductions associated with mortgage interest payments.
While debt recycling offers significant advantages, especially in higher interest environments, its success strongly depends on a proper execution and continuous management. The strategy can outperform regular mortgage repayment when executed with precision and supported by favorable house market and investment conditions. However, the risk associated with investments must be managed to prevent potential setbacks from market volatility. The choice between paying down a mortgage directly or engaging in debt recycling should be informed by an understanding of both financial strategies and personal financial goals, however this hoped-for level of awareness is hindered by the current lack of workable models for the joint evolution of equity and mortgage burden. 

Our paper aims at filling this gap, by incorporating a risk factor $\mu$ of the investment in our model, as well as a stochastic trajectory $\bm\sigma$ of positive or negative returns $\sigma_t\in\{\pm 1\}$ of the investment performance. The mortgage is assumed to decrease in time due to regular repayments, but also due to a larger influx of cash coming from a rewarding investment. The flip side is of course that the mortgage holder's debt or exposure can actually increase in case of a bad investment, or due to shrinking available equity in a contracting house market. These interacting effects are modeled by the set of coupled dynamical equations in \eqref{eq:processmatrix}. Under reasonable assumptions about the parameter ranges, we found that the average processes can be characterized analytically in terms of the eigenvalues $\lambda_1=s+1$ and $\lambda_2=\ell \mu (2p-1) +1$ of the average matrix given in Eq. \eqref{eq:A avg}. The eigenvalues depend on four parameters: $\ell$ (LTV ratio), $\mu$ (risk factor), $p$ (probability of gaining from the investment), and $s$ (average value of the market's fluctuations).

Depending on these combinations of parameters, we find three outcomes for the strategy: \textit{successful}, \textit{unsuccessful}, \textit{permanent re-mortgaging}. 
The strategy is \textit{successful} when the average mortgage process $\langle M_t \rangle$ hits the absorbing boundary $\langle M_t \rangle =0$ first. In this scenario, we distinguish between \textit{strong success}, where the time to achieve house ownership through debt recycling is shorter than without any strategy, and \textit{weak success}, where the strategy requires more time than the conventional monthly repayment approach, suggesting the latter as the more advisable option. In either case, though, the house is eventually secured.
Conversely, the strategy is deemed \textit{unsuccessful} when it is the average equity process $\langle E_t \rangle$ that hits the absorbing boundary $\langle E_t \rangle =0$ first. In this case the homeowner owns far less equity (seen as portion of ownership of the house) than they would have had they not engaged in debt recycling, meaning that the value of the house is now entirely mortgaged, and only monthly repayments could gain the house back.
The strategy may also lead to a state of \textit{permanent re-mortgaging} where neither process ever hits the absorbing boundary: as the house value increases, it is theoretically possible to keep securing additional bank funding for investment that feeds a cycle of equity extraction. Without regulatory intervention to cap bank lending, individuals could in principle exploit this strategy indefinitely, potentially leading to financial malpractice.

The debt recycling strategy is overall strongly successful with favorable parameters (\(p>0.5\), \(s>0\)); however, it leads to a state of permanent re-mortgaging when \(p<0.5\), \(s>0\). With \(s<0\), there is a potential for either default or success. The phase transition between default and success marks a critical region within the parameter space: small variations in \(s\) or \(p\) can lead to drastically different outcomes. In this region, the time it takes for the strategy to lead to a final outcome experiences a discontinuous jump, resembling a first-order transition.

The outcome of the strategy is sensitive to various parameters with different effects: a higher percentage of the mortgage yet to be repaid relative to the already owned equity increases the likelihood of default, expanding the region where the strategy is expected to fail, especially at high \(p\) (probability of successful investment). However, increasing the scheduled monthly repayment to the bank expands the region where the strategy is expected to be successful, even for \(p<0.5\). The factors \(\ell\) (LTV ratio) and \(\mu\) (risk factor) have the same influence on the outcome of the strategy: the higher their values, the more they expand the region where the strategy is expected to be successful (for \(p>0.5\) and \(s<0\)). \textcolor{black}{A thorough calculation of the level of fluctuations expected for the processes at each time reveals that the conclusions drawn from the average processes are very robust, and the predicted outcome (success/default/permanent remortgaging) is not easily subverted even taking into account fluctuations within the accuracy of one standard deviation. }

Future research could focus on the following issues:
\begin{enumerate}
\item Microscopic modeling of taxation issues. In our current model, we have ignored the interplay between deductible and non-deductible debt, and subtleties related to different taxation regimes for locked-in equity and investment-generated wealth. It would be interesting to incorporate this further complexity layer in our model. Also, our model does not currently allow for a finer sub-splitting of the primary debt (mortgage) and the locked-in equity. 
\item Determining the optimal parameters for different phases of the mortgage's life. For example, how should the monthly repayment amounts or optimal investment strategies be scaled to ensure that the mortgage is repaid on average in 20 years?
\item A potential diversification of the asset portfolio is another research avenue worth pursuing. We have currently limited our setting to a single risky investment for simplicity, but extending to the setting to multiple assets being traded is potentially very interesting.
\item It will be also interesting to include potential correlations between the house market and the return/volatility performance of the risky asset, which are considered for simplicity independent of each other at the moment.

\item It will be also interesting to add some mechanism to the model to suppress the permanent re-mortgaging phase, for example a time cap to the propensity of the lender to provide a line of credit.

\item In traditional banking, mortgages serve as collateral where the financed property secures the loan. In crypto-backed lending, cryptocurrencies act as collateral, allowing borrowers to access liquidity without selling their assets. This enables investment in additional assets, leveraging potential returns. However, the volatile nature of cryptocurrency markets can significantly affect the leverage or loan value, with interest rates also varying based on the lending platform and the collateral used \cite{defi}.
For instance, an investor might use Bitcoin to secure a loan in USDC through a decentralized finance protocol, then reinvest these funds in yield-generating activities. Given the liquidity and potential high returns of cryptocurrency investments, this could potentially accelerate the benefits of debt recycling. 
Combining debt recycling with the peculiar features of crypto markets is another exciting avenue for further research.
\end{enumerate}

\section*{Acknowledgments} P.V. acknowledges support from UKRI Future Leaders Fellowship Scheme
(No. MR/X023028/1).

\newpage
\appendix

\section{Extended Derivation of Main Equations for the Average Process}\label{app:Calculation Methods}

We start from Eq. \eqref{eq:master E}:
\begin{equation}
\frac{\langle E_t \rangle}{\langle E_0 \rangle} = (U \Lambda^t U^{-1})_{11} + c_1  (U \Lambda^t U^{-1})_{12} + c_2 \sum_{k=1}^{t} \left[ (U \Lambda^{t-k} U^{-1})_{11} - (U \Lambda^{t-k} U^{-1})_{12} \right]\ ,
\end{equation}
with
\begin{equation}
c_1 = \frac{\langle M_0 \rangle}{\langle E_0 \rangle},\qquad c_2 = \frac{\langle \pi \rangle}{\langle E_0 \rangle}\ .
\end{equation}

Writing the matrix entries explicitly, we have

\begin{align}
\nonumber\frac{\langle E_t \rangle}{\langle E_0 \rangle} &= \sum_{l=1}^{2} U_{1l} \lambda_l ^t (U^{-1})_{l1} + c_1 \sum_{l=1}^{2} U_{1l} \lambda_l ^t (U^{-1})_{l2} \\
\nonumber &+ c_2 \sum_{k=1}^{t} \sum_{l=1}^{2} \left[ U_{1l} \lambda_l ^{t-k} (U^{-1})_{l1} - U_{1l} \lambda_l ^{t-k} (U^{-1})_{l2}\right] = \\
\nonumber &= \sum_{l=1}^{2} \phi_l \lambda_l ^t + c_1 \sum_{l=1} ^2 \psi_l \lambda_l ^t + c_2 \sum_{l=1}^{2} \phi_l \sum_{k=1} ^t \lambda_l ^{t-k} - c_2 \sum_{l=1} ^2 \psi_l \sum_{k=1}^t \lambda_l ^{t-k} = \\
&= \sum_{l=1}^2 \phi_l \lambda_l ^t + c_2 \sum_{l=1}^2 \phi_l \sum_{k=1}^t \lambda_l ^{t-k} + c_1 \sum_{l=1}^2 \psi_l \lambda_l ^t - c_2 \sum_{l=1}^2 \psi_l \sum_{k=1}^t \lambda_l ^{t-k}
\end{align}
where we have defined for $l=1,2$
\begin{align}
\phi_l &\equiv U_{1l} (U^{-1})_{l1}\\
\psi_l &\equiv U_{1l} (U^{-1})_{l2}\ ,
\end{align}
with the $2\times 2$ matrices $U$ and $U^{-1}$ given in Eqs. \eqref{U} and \eqref{Uminusone}.

Using the geometric sum 
\begin{equation}
\sum_{k=1}^t \lambda_l ^{t-k} = \frac{1 - \lambda_l ^t}{1 - \lambda_l}
\end{equation}
we can write
\begin{align}
\nonumber &\frac{\langle E_t \rangle}{\langle E_0 \rangle} = \sum_{l=1}^2 \phi_l \lambda_l ^t + c_2 \sum_{l=1}^2 \phi_l \frac{1 - \lambda_l ^t}{1 - \lambda_l} + c_1 \sum_{l=1}^2 \psi_l \lambda_l ^t - c_2 \sum_{l=1}^2 \psi_l \frac{1 - \lambda_l ^t}{1 - \lambda_l} = \\
\nonumber &= \phi_1 \lambda_1 ^t + \phi_2 \lambda_2 ^t + c_1 \psi_1 \lambda_1 ^t + c_1 \psi_2 \lambda_2 ^t + c_2 \left[\phi_1 \frac{1- \lambda_1 ^t}{1-\lambda_1} + \phi_2 \frac{1- \lambda_2 ^t}{1- \lambda_2} \right.\\
&\left. - \psi_1 \frac{1- \lambda_1 ^t}{1-\lambda_1} - \psi_2 \frac{1- \lambda_2 ^t}{1- \lambda_2} \right]\ .
\end{align}
Collecting now the coefficients of $\lambda_1^t$ and $\lambda_2^t$, we obtain the main result in Eq. \eqref{avproc1} in the main text, with

\begin{align}
\mathcal{A} &\equiv \phi_1 + c_1 \psi_1 - \frac{c_2 \phi_1}{1-\lambda_1} + \frac{c_2 \psi_1}{1-\lambda_1} \\
\mathcal{B} &\equiv \phi_2 + c_1 \psi_2 - \frac{c_2 \phi_2}{1-\lambda_2} + \frac{c_2 \psi_2}{1-\lambda_2}\\
\mathcal{C} &\equiv \frac{c_2 \phi_1}{1- \lambda_1} + \frac{c_2 \phi_2}{1- \lambda_2} - \frac{c_2 \psi_1}{1- \lambda_1} - \frac{c_2 \psi_2}{1- \lambda_2}\ , 
\end{align}
which reduce to the expressions in Eqs. \eqref{eq: A},\eqref{eq: B}, \eqref{eq: C} once the values are substituted in. The derivation for the average mortgage process is entirely analogous.

\section{\textcolor{black}{Fluctuations of the $(E_t, M_t)$ Processes}}\label{app:variance}

\textcolor{black}{Starting from the $(E_t, M_t)$ processes in Eq. \ref{eq:processmatrix} and the introduction of the matrix $A_t$ in Sec. \ref{subsec:average}, we can write
\begin{equation}
\begin{pmatrix}
 E_t \\ M_t
\end{pmatrix}
=
A_t
\begin{pmatrix}
E_{t-1} \\ M_{t-1}
\end{pmatrix}
+
\begin{pmatrix}
\pi_t \\ -\pi_t  
\end{pmatrix}\ . 
\label{eq: b1}
\end{equation}
Our objective is to derive the equations for the evolution of \( (E_t^2, M_t^2) \), enabling us to compute the theoretical variance of the equity and mortgage processes:
\begin{equation}\label{var_Et}
\text{Var}(E_t) = \langle E_t^2 \rangle - \langle E_t \rangle^2
\end{equation}
\begin{equation}\label{var_Mt}
    \text{Var}(M_t) = \langle M_t^2 \rangle - \langle M_t \rangle^2\ .
\end{equation}}

\textcolor{black}{We can start decomposing Eq. \ref{eq: b1} for the two components, obtaining:
\begin{equation}
\begin{cases} 
E_t = (A_t)_{11} E_{t-1} + (A_t)_{12} M_{t-1} + \pi_t\\ 
M_t = (A_t)_{21} E_{t-1} + (A_t)_{22} M_{t-1} - \pi_t\ .
\end{cases}
\end{equation}
By squaring both the left-hand side (LHS) and right-hand side (RHS) of each equation, and then multiplying those, we obtain the following three equations:
\begin{equation}
    E_t^2 = (A_t)_{11}^2 E_{t-1}^2 + (A_t)_{12}^2 M_{t-1}^2 + \pi_t^2 + 2 (A_t)_{11} (A_t)_{12} E_{t-1} M_{t-1} + 2 (A_t)_{11} E_{t-1} \pi_t + 2 (A_t)_{12} M_{t-1} \pi_t
\end{equation}
\begin{equation}
    M_t^2 = (A_t)_{21}^2 E_{t-1}^2 + (A_t)_{22}^2 M_{t-1}^2 + \pi_t^2 + 2 (A_t)_{21} (A_t)_{22} E_{t-1} M_{t-1} - 2 (A_t)_{21} E_{t-1} \pi_t - 2 (A_t)_{22} M_{t-1} \pi_t
\end{equation}
\begin{align}
\nonumber E_t M_t 
&=  \ (A_t)_{11} (A_t)_{21} E_{t-1}^2 + (A_t)_{11} (A_t)_{22} E_{t-1} M_{t-1} - (A_t)_{11} E_{t-1} \pi_t \\
\nonumber & + (A_t)_{12} (A_t)_{21} M_{t-1} E_{t-1} + (A_t)_{12} (A_t)_{22} M_{t-1}^2 - (A_t)_{12} M_{t-1} \pi_t \\
& + \pi_t (A_t)_{21} E_{t-1} + \pi_t (A_t)_{22} M_{t-1} - \pi_t^2\ .
\end{align}}

\textcolor{black}{
These Eqs. can be written in a more compact way as
\begin{equation}
    W_t = M W_{t-1}+n_t\ ,\label{WtIteration}
\end{equation}
where $M$ is the $5\times 5$ matrix
\begin{equation}
    M=\begin{pmatrix}
(A_t)_{11} ^2 & (A_t)_{12} ^2 & 2 (A_t)_{11} \pi_t & 2 (A_t)_{12} \pi_t & 2 (A_t)_{11} (A_t)_{12} \\
(A_t)_{21} ^2 & (A_t)_{22} ^2 & - 2 (A_t)_{21} \pi_t & - 2 (A_t)_{22} \pi_t & 2 (A_t)_{21} (A_t)_{22} \\
0 & 0 & (A_t)_{11} & (A_t)_{12} & 0 \\
0 & 0 & (A_t)_{21} & (A_t)_{22} & 0 \\
(A_t)_{11} (A_t)_{21} & (A_t)_{12} (A_t)_{22} & \pi_t [(A_t)_{21} - (A_t)_{11}] & \pi_t [(A_t)_{22} - (A_t)_{12}] & (A_t)_{11} (A_t)_{22} + (A_t)_{12} (A_t)_{21}\\
\end{pmatrix}
\end{equation}
and the column vectors are
\begin{equation}
    W_t = \begin{pmatrix}
E_t^2 \\
M_t^2 \\
E_t \\
M_t \\
E_t M_t
\end{pmatrix}\qquad n_t = \begin{pmatrix}
\pi_t^2 \\
\pi_t^2 \\
\pi_t \\
-\pi_t \\
-\pi_t^2
\end{pmatrix}\ .
\end{equation}
}

\textcolor{black}{From this point onward, we can apply the same approach as outlined in the main text to deal with Eq. \eqref{eq:processmatrix} for computing the average values.}

\textcolor{black}{Iterating Eq. \eqref{WtIteration}, we obtain:
\begin{equation}
    W_t = M_t M_{t-1} \dots M_1 W_0 + M_t M_{t-1} \dots M_2 n_1 + M_t M_{t-1} \dots M_3 n_2 + \dots + n_t 
\end{equation}
and, averaging on multiple realizations of the random variables involved:
\begin{equation}\label{eq:b7}
    \langle W_t \rangle = \overline{M}^t \langle W_0 \rangle + \overline{M}^{t-1} \langle n_1 \rangle  
    + \overline{M}^{t-2} \langle n_2 \rangle + \dots + \langle n_t \rangle\ .
\end{equation}
Below, we present the forms that $\overline{M}$, $\langle W_0 \rangle$, and $\langle n_k \rangle$  take after some algebraic manipulation:
\tiny
\begin{equation}
\overline{M} =
\begin{pmatrix}
1 + \phi^2 + s^2 + \ell^2 \mu^2 + 2 \ell \mu (2p-1)(1 + s) + 2s
 & \phi^2 + s^2 & 2 \langle \pi \rangle (1+\ell \mu (2p-1) +s)  & 2 \langle \pi \rangle s & 2 [s+\ell \mu s (2p-1) + s^2 + \phi^2] \\
\ell^2 \mu^2 & 1 & 2 \langle \pi \rangle \ell \mu (2p-1) & -2 \langle \pi \rangle & -2 \ell \mu (2p-1) \\
0 & 0 & 1+\ell \mu (2p-1) +s & s & 0 \\
0 & 0 & -\ell \mu (2p-1) & 1 & 0 \\
- \ell \mu (2p-1) (1+ s) - \ell^2 \mu^2 & s & - \langle \pi \rangle [1+2 \ell \mu (2p-1) +s] & \langle \pi \rangle (1-s) & 1+ \ell \mu (2p-1) (1-s) +s \\
\end{pmatrix}
\end{equation}
\normalsize  
\begin{equation}
\langle W_0 \rangle =
\begin{pmatrix}
\langle E_{0} \rangle ^2 \\
\langle M_{0} \rangle ^2 \\
\langle E_{0} \rangle \\
\langle M_{0} \rangle \\
\langle E_{0} \rangle \langle M_{0} \rangle
\end{pmatrix}
\end{equation}
\begin{equation}
\langle n_k \rangle =
\begin{pmatrix}
(1-q) (\pi^\star)^2 \\
(1-q) (\pi^\star)^2 \\
(1-q) \pi^\star \\
- (1-q) \pi^\star \\
- (1-q) (\pi^\star)^2 
\end{pmatrix}\ ,
\end{equation}
where $\langle\pi\rangle=(1-q)\pi^\star$ is the average monthly repayment, while $\phi$ is the standard deviation of the fluctuations in the house market value.}
\normalsize

\textcolor{black}{We set \( \langle E_0^2 \rangle \approx \langle E_0 \rangle^2 \), assuming \( \text{var}(E_0) = 0 \), since there is no uncertainty at the initial step of the process. The same assumption holds for \( \langle M_0^2 \rangle \) and \( \langle E_0 M_0 \rangle \). Additionally, note that \( \langle n_k \rangle \) is time-independent, so \( \langle n_k \rangle \equiv \langle n \rangle \) and we use $\langle\pi^2\rangle=(1-q)(\pi^\star)^2$ from Eq. \eqref{pofpi}.}

\textcolor{black}{At this point, as done in the main text, we decompose the average matrix $\big( \overline{M} = V \Gamma V^{-1} \big)$, obtaining from Eq. \eqref{eq:b7}:
\begin{equation}\label{eq:b11}
\langle W_t \rangle = \big( V \Gamma V^{-1} \big)^t \langle W_0 \rangle + \sum_{k=1}^t \big( V \Gamma V^{-1} \big)^{t-k} \langle n \rangle\ ,
\end{equation}
where $\Gamma$ is the $(5 \times 5)$ diagonal matrix of (distinct) eigenvalues $\{\gamma_1,\ldots,\gamma_5\}$
\begin{equation}
\Gamma =
\left(
\begin{array}{ccccc}
 s+1 & 0 & 0 & 0 & 0 \\
 0 & \ell \mu  (2 p-1)+1 & 0 & 0 & 0 \\
 0 & 0 & (s+1) (\ell \mu  (2 p-1)+1) & 0 & 0 \\
 0 & 0 & 0 & \ell \mu  (\ell \mu + 4 p -2)+1 & 0 \\
 0 & 0 & 0 & 0 & (s+1)^2+\phi ^2 \\
\end{array}
\right)\ .
\end{equation}}

\textcolor{black}{We are interested in $\langle (W_t)_1
\rangle = \langle E_t^2 \rangle$ and $\langle (W_t)_2
\rangle = \langle M_t^2 \rangle$ and, following the structure of the results obtained in \ref{app:Calculation Methods}, we expect to obtain the following expressions:
\begin{equation}\label{eq:b12}
    \langle E_t^2 \rangle = \mathcal{G} \gamma_1^t + \mathcal{H} \gamma_2^t + \mathcal{I} \gamma_3^t + \mathcal{J} \gamma_4^t + \mathcal{K} \gamma_5^t + \mathcal{L}\ ,
\end{equation}
\begin{equation}\label{eq:b13}
    \langle M_t^2 \rangle = \mathcal{M} \gamma_1^t + \mathcal{N} \gamma_2^t + \mathcal{O} \gamma_3^t + \mathcal{P} \gamma_4^t + \mathcal{Q} \gamma_5^t + \mathcal{R}\ ,
\end{equation}
with the coefficients to be determined.}

\textcolor{black}{We will provide a proof of Eqs. \eqref{eq:b12} and \eqref{eq:b13} at the end of this Appendix. }\\

\begin{figure}[H]
    \begin{minipage}{0.5\textwidth}
        \centering \includegraphics[width=\textwidth]{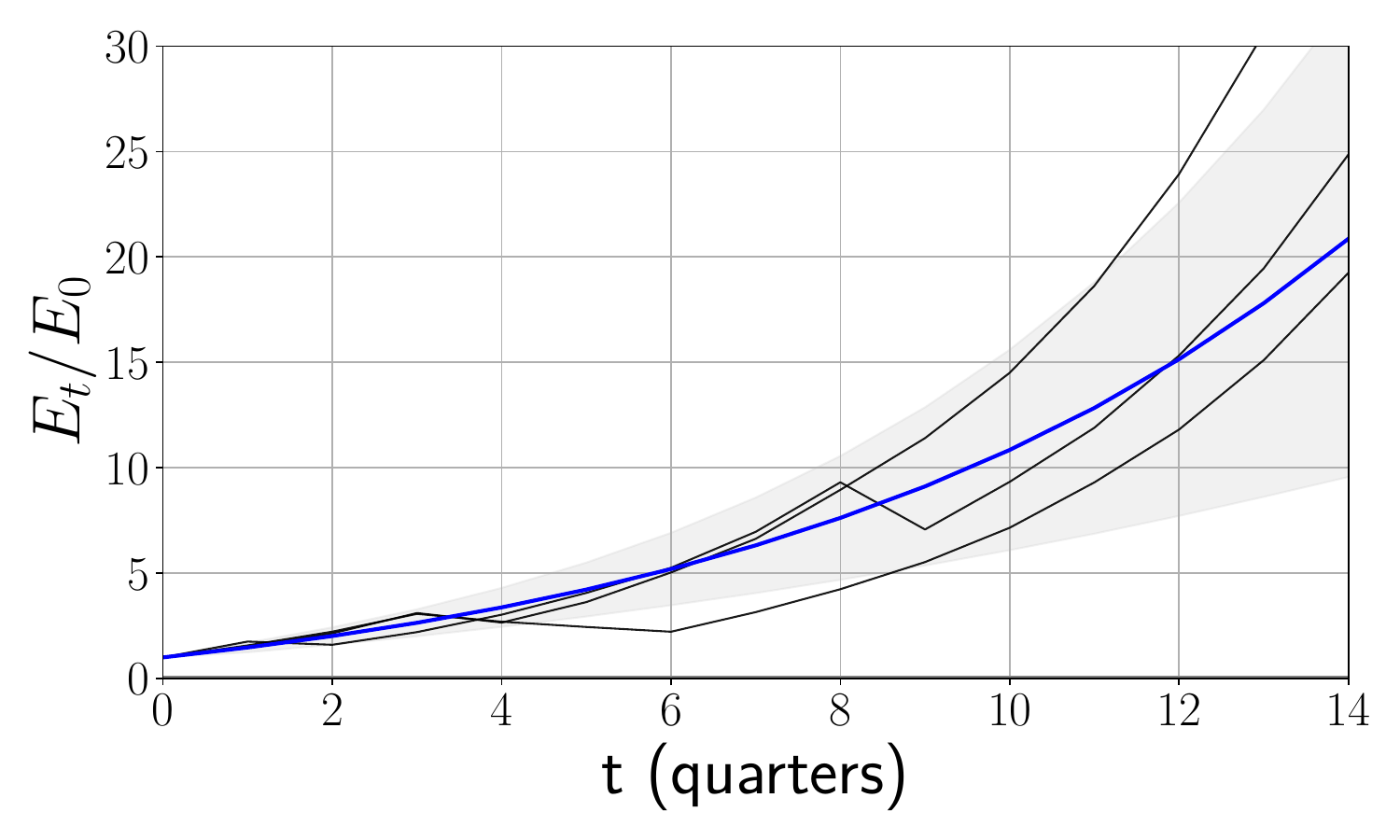}
        \subcaption{\textcolor{black}{Numerical simulations of the equity process in Eq. \eqref{eq:processmatrix} (solid black lines). In blue, the empirical mean (average over $10^5$ simulations). The gray area corresponds to the standard deviation band computed over all simulations.}} \label{fig:fluttuaz1}
    \end{minipage}
    \hspace{10pt}
    \begin{minipage}{0.5\textwidth}
        \centering \includegraphics[width=\textwidth]{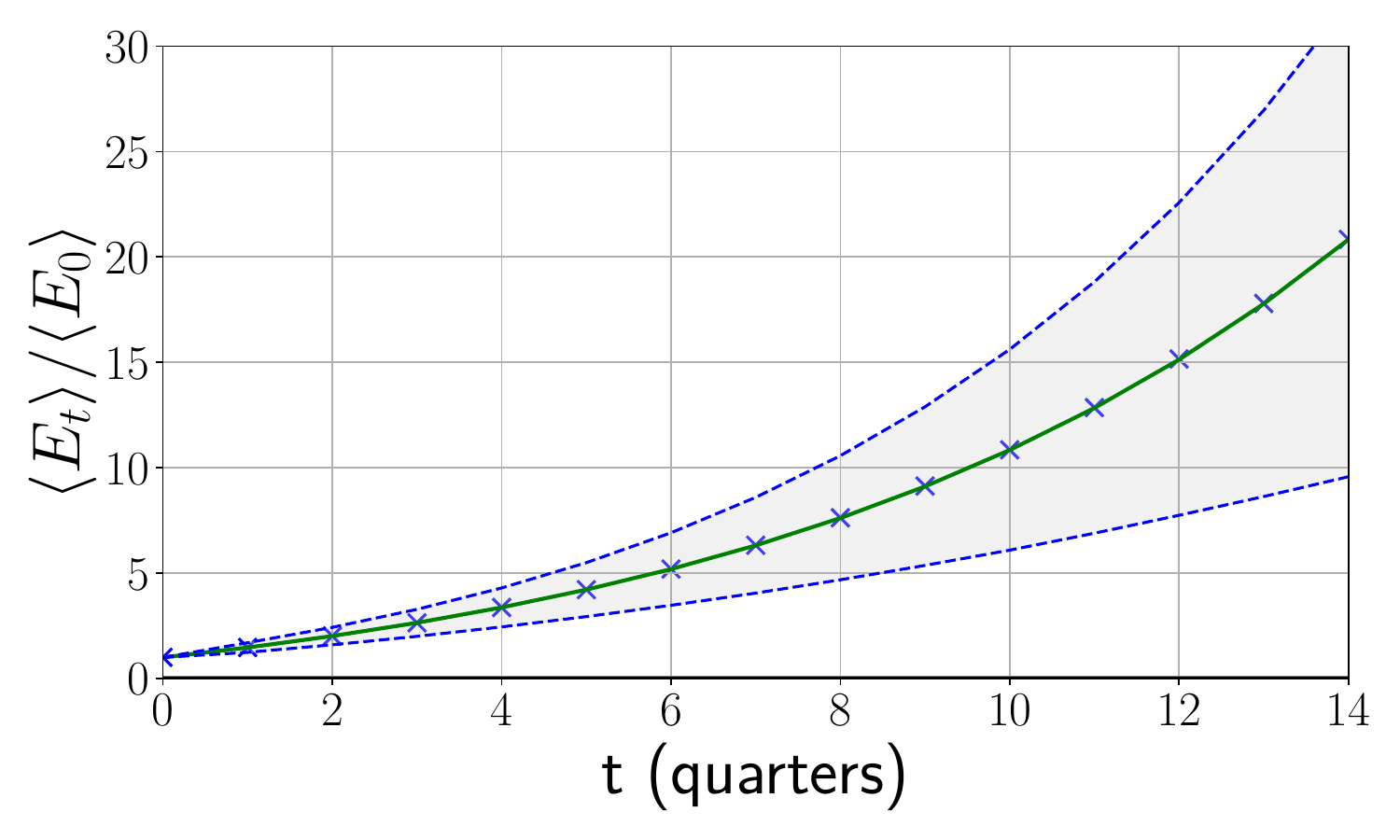}
        \subcaption{\textcolor{black}{Blue crosses ($\times$) mark the empirical mean values at each time, computed over $10^5$ realizations of the equity process in Eq. \eqref{eq:processmatrix}. They fall on top of the theoretical result in Eq. \eqref{avproc1} (solid green line). The gray band shows the theoretical standard deviation over time, given by Eq. \eqref{var_Et}. The dashed blue lines represent the empirical standard deviation bounds around the mean values, again computed over $10^5$ realizations of the process.}
       }\label{fig:fluttuaz2}
    \end{minipage}
    \caption{\textcolor{black}{Empirical and theoretical analysis of average equity over time, using the same parameters as in Fig. \ref{fig:trajectory1}: \( (\ell, \mu, p ,s) = (0.5, 0.5, 0.8, 2\%) \); \( (q, \pi^*) = (1\%, \$3,000) \); \( \phi^2 = 1\% \). Note the excellent agreement between the numerical simulations and the theory.}} 
    \label{fig: fluttuaz_tot1}
\end{figure}

\textcolor{black}{In Fig. \ref{fig: fluttuaz_tot1}, we provide a more detailed analysis of the equity trend previously presented in Fig. \ref{fig:trajectory1}, including this time the process fluctuations as well.}

\begin{figure}[H]
    \begin{minipage}{0.5\textwidth}
        \centering \includegraphics[width=\textwidth]{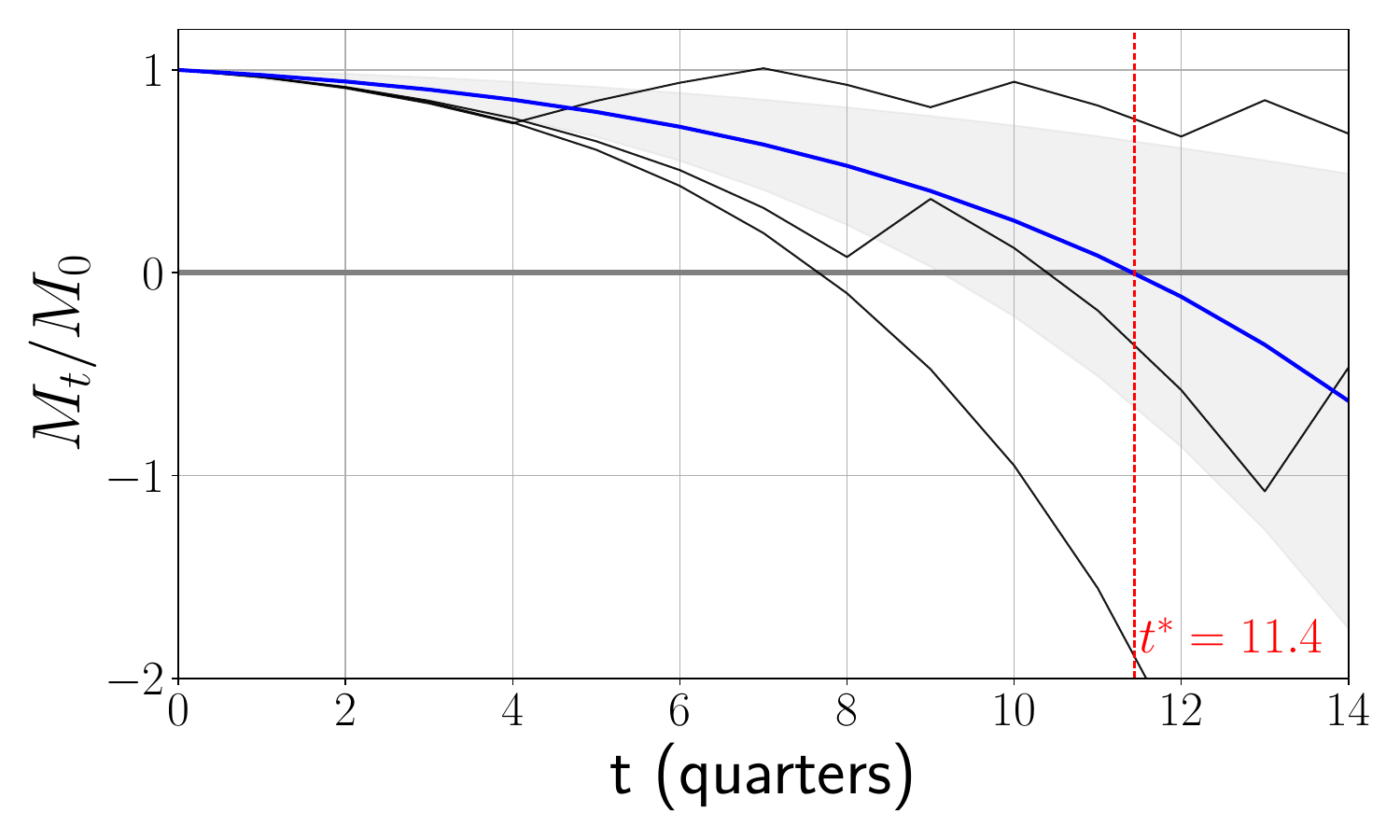}
        \subcaption{\textcolor{black}{Numerical simulations of the equity process in Eq. \eqref{eq:processmatrix} (solid black lines). In blue, the empirical mean (average over $10^5$ simulations). The gray area corresponds to the standard deviation band computed over all simulations. 
        The vertical line at $t^\star=11.4$ signals the moment the average mortgage balance hits the absorbing wall at zero.}} \label{fig:fluttuaz3}
    \end{minipage}
    \hspace{10pt}
    \begin{minipage}{0.5\textwidth}
        \centering \includegraphics[width=\textwidth]{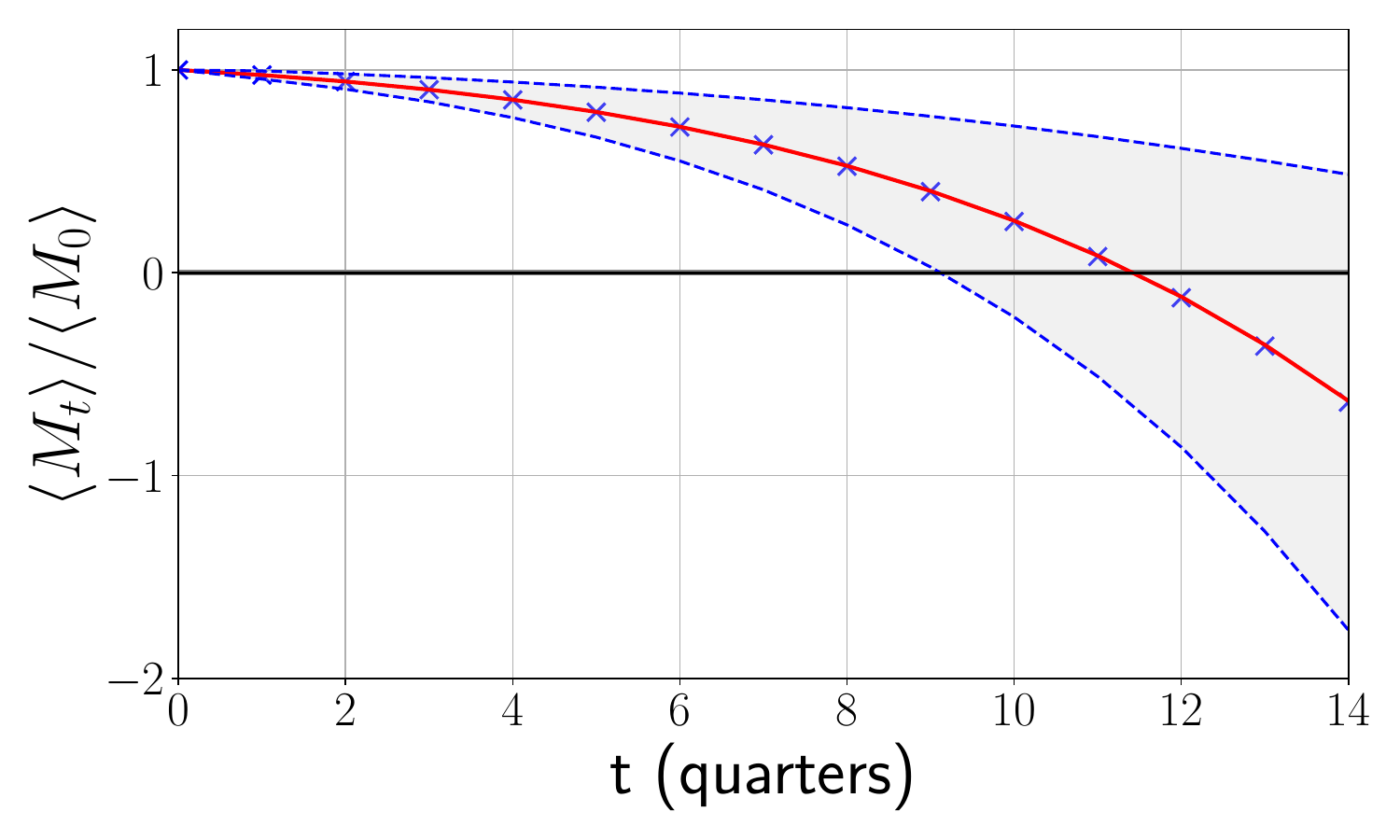} 
        \subcaption{\textcolor{black}{Blue crosses ($\times$) mark the empirical mean values at each time, computed over $10^5$ realizations of the equity process in Eq. \eqref{eq:processmatrix}. They fall on top of the theoretical result in Eq. \eqref{avproc2} (solid red line). The gray band shows the theoretical standard deviation over time, given by Eq. \eqref{var_Mt}. The dashed blue lines represent the empirical standard deviation bounds around the mean values, again computed over $10^5$ realizations of the process.}}\label{fig:fluttuaz4}
    \end{minipage}
    \caption{\textcolor{black}{Empirical and theoretical analysis of average mortgage value over time, using the same parameters as in Fig. \ref{fig:trajectory2}: \( (\ell, \mu, p ,s) = (0.5, 0.5, 0.8, 2\%) \); \( (q, \pi^*) = (1\%, \$3,000) \); \( \phi^2 = 1\% \). As before, note the excellent agreement between the numerical simulations and the theory.}} 
    \label{fig: fluttuaz_tot2}
\end{figure}

\textcolor{black}{In Fig. \ref{fig: fluttuaz_tot2}, we provide a similar analysis for the mortgage variable, previously presented in Fig. \ref{fig:trajectory2}.}

\begin{figure}[h]
    \centering
    \includegraphics[width=\textwidth]{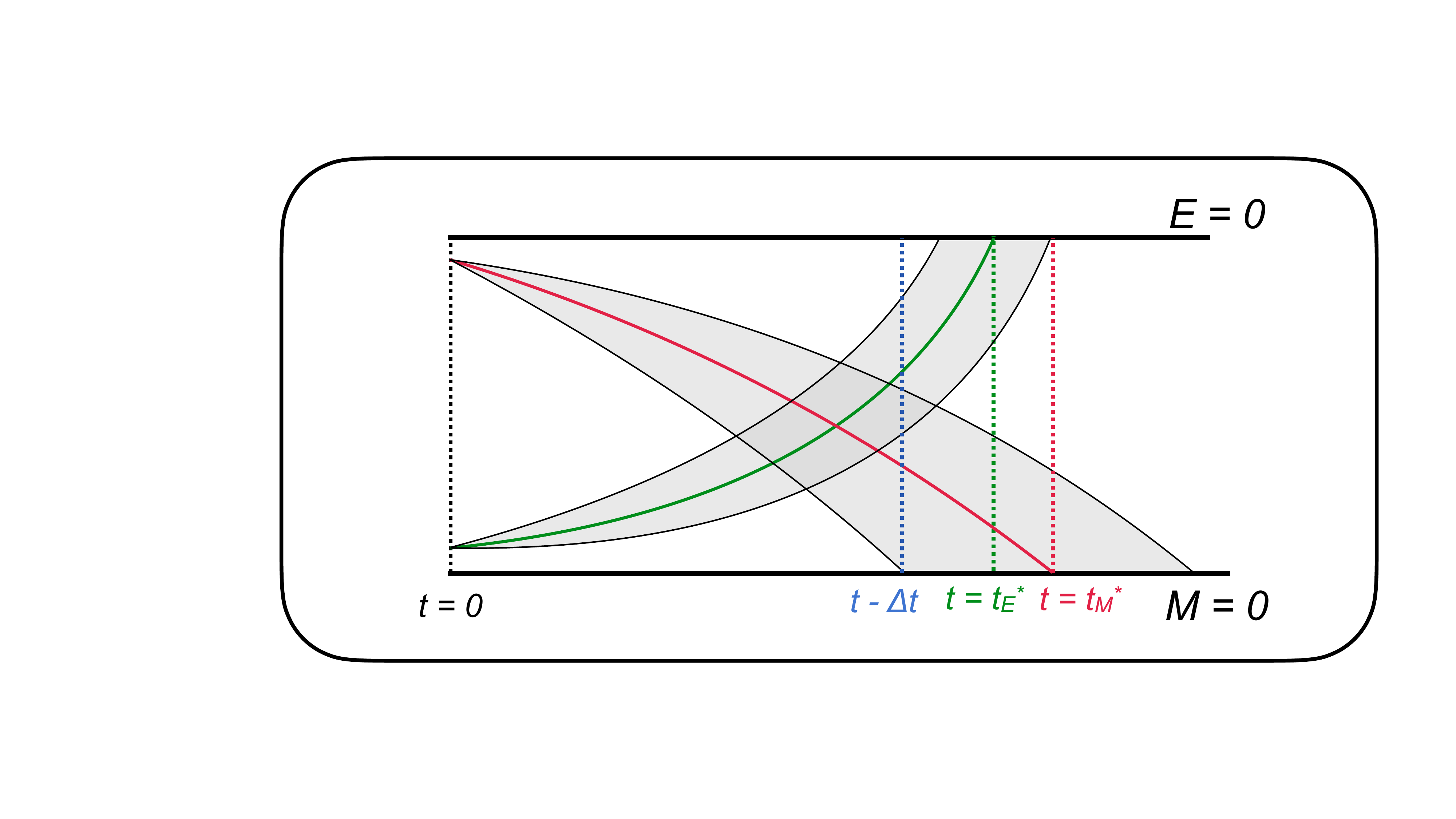}
    \caption{\textcolor{black}{Schematic representation of a potential situation where the average equity process (green line) reaches the absorbing boundary $E=0$ at $t=t_E^\star$, before the average mortgage process (red line) reaches the absorbing boundary $M=0$ at $t_M^\star$. However, the fluctuations around the average mortgage process (gray area) are such that this process (within the accuracy of one standard deviation) could actually hit the boundary anywhere between $t-\Delta t$ and $t_E^\star$, leading to the opposite call we would make based on the hitting times of the average processes alone. We have found that -- within a one standard deviation precision -- this situation never materializes in our model.}}
    \label{fig:schematicCrossing}
\end{figure}

\textcolor{black}{We conclude our analysis by investigating whether fluctuations can alter the outcome of the average process. Specifically, we examine if, at any time \( t \), a fluctuation-induced root in one variable may occur \textit{before} the root of the average process in the other variable: this may cast some doubt in the called outcome (success/default) of the strategy, as the \emph{opposite} outcome would still be possible within the error bounds (see Fig. \ref{fig:schematicCrossing} for a schematic representation). 
However, our analysis show that there is no region in the phase diagram where considering fluctuations around the average process may lead to calling a different strategic outcome. 
The only potential area where this effect could be seen is the one marked with \( (\times)\) in the phase diagram, representing the range where \( 0 < \lambda_{1,2} < 1 \) (see Table \ref{tab:dynamic outcomes}). In this region, both \( E_t \) and \( M_t \) exhibit a downward trend in the initial time steps -- consider that both processes always start from a value of \( 1 \), and we are interested in the first root (i.e. the time at which either mortgage or equity hits zero). Despite fluctuations, the outcome remains consistent because the decay rates are markedly different, leading to very different timescales at which each variable approaches zero. This case is illustrated in the main text in Fig. \ref{fig: trajectories2}. Furthermore, we stress that the model lacks external shocks; therefore, it is designed for orderly market conditions, making it reasonable to expect that fluctuations would not significantly disrupt the outcomes predicted on the basis of the average process alone.}

\subsection{\textcolor{black}{Proof Eq. \ref{eq:b12}}}

\textcolor{black}{Let us write explicitly Eq. \ref{eq:b11} for the first component:
\begin{align}
\nonumber \langle E_t^2 \rangle &= \big( V \Gamma^t V^{-1} \big)_{11} \langle E_0 \rangle^2 +
\big( V \Gamma^t V^{-1} \big)_{12} \langle M_0 \rangle^2 + 
\big( V \Gamma^t V^{-1} \big)_{13} \langle E_0 \rangle +
\big( V \Gamma^t V^{-1} \big)_{14} \langle M_0 \rangle  \\
\nonumber &+ \big( V \Gamma^t V^{-1} \big)_{15} \langle E_0 \rangle \langle M_0 \rangle + 
\sum_{l=1}^5 \sum_{k=1}^t \big( V \Gamma^{t-k} V^{-1} \big)_{1l} \langle n \rangle_l \\
\nonumber &= \langle E_0 \rangle^2 \sum_{j=1}^5 V_{1j} \big( V^{-1} \big)_{j1} \gamma_j^t
+ \langle M_0 \rangle^2 \sum_{j=1}^5 V_{1j} \big( V^{-1} \big)_{j2} \gamma_j^t
+ \langle E_0 \rangle \sum_{j=1}^5 V_{1j} \big( V^{-1} \big)_{j3} \gamma_j^t 
+ \langle M_0 \rangle \sum_{j=1}^5 V_{1j} \big( V^{-1} \big)_{j4} \gamma_j^t  \\
& +\langle E_0 \rangle \langle M_0 \rangle \sum_{j=1}^5 V_{1j} \big( V^{-1} \big)_{j5} \gamma_j^t +
\sum_{l=1}^5 \langle n \rangle_l \sum_{k=1}^t \sum_{j=1}^5 V_{1j} \big( V^{-1} \big)_{jl} \gamma_j^{t-k} 
\end{align}
Introducing
\begin{equation}
    \theta_{xy} \equiv V_{1x} (V^{-1})_{xy}, \quad \text{for } x,y = 1, \dots, 5\ ,
\end{equation}
we obtain:
\begin{align}
\nonumber\langle E_t^2 \rangle &= \langle E_0 \rangle^2 \sum_{j=1}^5 \theta_{j1} \gamma_j^t
+ \langle M_0 \rangle^2 \sum_{j=1}^5 \theta_{j2} \gamma_j^t
+ \langle E_0 \rangle \sum_{j=1}^5 \theta_{j3} \gamma_j^t 
+ \langle M_0 \rangle \sum_{j=1}^5 \theta_{j4} \gamma_j^t \\
& +\langle E_0 \rangle \langle M_0 \rangle \sum_{j=1}^5 \theta_{j5} \gamma_j^t +
\sum_{l=1}^5 \langle n \rangle_l \sum_{k=1}^t \sum_{j=1}^5 \theta_{jl} \gamma_j^{t-k} 
\end{align}
Using the geometric sum 
\begin{equation}
\sum_{k=1}^t \gamma_j ^{t-k} = \frac{1 - \gamma_j ^t}{1 - \gamma_j}\ ,
\end{equation}
we can write
\begin{align}
\nonumber\langle E_t^2 \rangle &= \langle E_0 \rangle^2 \sum_{j=1}^5 \theta_{j1} \gamma_j^t
+ \langle M_0 \rangle^2 \sum_{j=1}^5 \theta_{j2} \gamma_j^t
+ \langle E_0 \rangle \sum_{j=1}^5 \theta_{j3} \gamma_j^t 
+ \langle M_0 \rangle \sum_{j=1}^5 \theta_{j4} \gamma_j^t  \\
& +\langle E_0 \rangle \langle M_0 \rangle \sum_{j=1}^5 \theta_{j5} \gamma_j^t +
\sum_{l=1}^5 \langle n \rangle_l \sum_{j=1}^5 \theta_{jl} \frac{1 - \gamma_j ^t}{1 - \gamma_j}\ . 
\end{align}
Rearranging terms, we arrive at Eq. \eqref{eq:b12}, upon defining the following coefficients:
\begin{align}
\mathcal{G} &\equiv
\left\langle E_0\right\rangle^2 \theta_{11}+\left\langle M_0\right\rangle^2 \theta_{12}+\left\langle E_0\right\rangle \theta_{13}+\left\langle M_0\right\rangle \theta_{14}+\left\langle E_0\right\rangle\left\langle M_0\right\rangle \theta_{15} -\frac{\sum_{l=1}^5\langle n\rangle_l \theta_{1 l}}{1-\gamma_1} \\
\mathcal{H} &= \left\langle E_0\right\rangle^2 \theta_{21}+\left\langle M_0\right\rangle^2 \theta_{22}+\left\langle E_0\right\rangle \theta_{23}+\left\langle M_0\right\rangle \theta_{24}+\left\langle E_0\right\rangle\left\langle M_0\right\rangle \theta_{25} -\frac{\sum_{l=1}^5\langle n\rangle_l \theta_{2 l}}{1-\gamma_2} \\
\mathcal{I} &= \left\langle E_0\right\rangle^2 \theta_{31}+\left\langle M_0\right\rangle^2 \theta_{32}+\left\langle E_0\right\rangle \theta_{33}+\left\langle M_0\right\rangle \theta_{34}+\left\langle E_0\right\rangle\left\langle M_0\right\rangle \theta_{35} -\frac{\sum_{l=1}^5\langle n\rangle_l \theta_{3 l}}{1-\gamma_3} \\
\mathcal{J} &= \left\langle E_0\right\rangle^2 \theta_{41}+\left\langle M_0\right\rangle^2 \theta_{42}+\left\langle E_0\right\rangle \theta_{43}+\left\langle M_0\right\rangle \theta_{44}+\left\langle E_0\right\rangle\left\langle M_0\right\rangle \theta_{45} -\frac{\sum_{l=1}^5\langle n\rangle_l \theta_{4 l}}{1-\gamma_4} \\
\mathcal{K} &= \left\langle E_0\right\rangle^2 \theta_{51}+\left\langle M_0\right\rangle^2 \theta_{52}+\left\langle E_0\right\rangle \theta_{53}+\left\langle M_0\right\rangle \theta_{54}+\left\langle E_0\right\rangle\left\langle M_0\right\rangle \theta_{55} -\frac{\sum_{l=1}^5\langle n\rangle_l \theta_{5 l}}{1-\gamma_5} \\
\mathcal{L} &= \sum_{j=1}^5 \frac{\sum_{l=1}^5\langle n\rangle_l \theta_{j l}}{1-\gamma_j}\ .
\end{align}
The derivation of Eq. \eqref{eq:b13} is entirely analogous, once we define:
\begin{equation}
    \xi_{xy} \equiv V_{2x} (V^{-1})_{xy}, \quad \text{for } x,y = 1, \dots, 5\ .
\end{equation}}
\textcolor{black}{The coefficients $\mathcal{M},\mathcal{N},\mathcal{O},\mathcal{P},\mathcal{Q},\mathcal{R}$ are identical to $\mathcal{G},\mathcal{H},\mathcal{I},\mathcal{J},\mathcal{K},\mathcal{L}$ respectively, with each $\theta_{xy}$ replaced by $\xi_{xy}$.}

\end{document}